%% file: elsarticle-template-num.tex
\documentclass[preprint,12pt]{elsarticle}

\usepackage{natbib}
\usepackage{adjustbox}
\usepackage{amsmath}
\usepackage{amsfonts}
\usepackage{amsthm}
\usepackage[usenames]{color}
\usepackage[acronym,toc,shortcuts]{glossaries}
\usepackage{makeidx}
\usepackage{mathtools}
\usepackage{multirow,makecell}
\usepackage{enumerate} 
\usepackage{epstopdf}
\usepackage{epsfig}
\usepackage{eqparbox}
\usepackage{eurosym}
\usepackage{float}
\usepackage{graphicx}
\usepackage[utf8]{inputenc}
\usepackage{listings}
\usepackage{mdwmath}
\usepackage{mdwtab}
\usepackage[cmintegrals]{newtxmath}
\usepackage{placeins}
\usepackage{rotating}
\usepackage{spreadtab}
\usepackage{subfigure}
\usepackage{tabularx}
\usepackage{times}
\usepackage{xspace}
\usepackage{url}
\usepackage[dvipsnames]{xcolor}
\usepackage{nomencl}
\usepackage{adjustbox}

\immediate\write18{%
makeindex -s nomencl.ist -o \jobname.nls -t \jobname.nlg \jobname.nlo%
}

\input{src/acronyms.tex}

\makenomenclature

\begin{document}

\begin{frontmatter}

\title{An Extensive Survey on the Internet of Drones}

\author[inst1]{Pietro Boccadoro, IEEE Member}

\author[inst1]{Domenico Striccoli}

\author[inst1]{Luigi Alfredo Grieco, IEEE Member}

\affiliation[inst1]{organization={Dep. of Electrical and Information Engineering (DEI), Politecnico di Bari},%Department and Organization
	addressline={via Orabona 4}, 
	city={Bari},
	postcode={70126}, 
	state={Bari},
	country={Italy},
	email= {, email: name.surname@poliba.it}}

\begin{abstract}

The \ac{IoD} recently gained momentum due to its high adaptability to a wide variety of complex scenarios.
Indeed, \acp{UAV} can successfully be employed in different applications, such as agriculture, search and rescue missions, surveillance systems, mission-critical services, etc., thanks to some technological and practical advantages: high mobility, capability to extend wireless coverage areas, or ability to reach places inaccessible to humans.
Moreover, the employment of drones promisingly improves the performance parameters of different network architectures, i.e., reliability, connectivity, throughput, and delay, among others.
Nevertheless, the adoption of networks of drones gives rise to several issues related to the unreliability of the wireless medium, the duration of batteries, and the high mobility degree, which may cause frequent topology changes. Also security and privacy issues need to be properly investigated. This explains the very large number of works produced in the recent literature on \gls{IoD}-related topics.
With respect to other surveys on \gls{IoD}-related topics, the goal of the present work is to categorize the multifaceted aspects of \gls{IoD}, proposing a classification approach of the \gls{IoD} environment that develops along two main directions. At a macroscopic level, it follows the structure of the Internet protocol stack, starting from the physical layer and extending to the upper layers, without neglecting cross-layer and optimization approaches. At a finer level, all the most relevant works belonging to each layer of the stack are further classified, according to the different issues peculiar of the layer, and highlighting the most relevant differences with the other surveys present in literature.
To provide a deeper insight in the theme, the present work embraces many facets of the \gls{IoD}, including privacy and security considerations as well as the potential economic impact of the \gls{IoD}.
Finally, a discussion on the main research challenges and possible future directions is carried out, focusing on the open issues and the most promising technologies that deserve to be further developed in the \gls{IoD} field.

\end{abstract}




\end{frontmatter}


\input{src/1-intro.tex}
\input{src/3-related.tex}
\input{src/2-applications.tex}
\input{src/4-phy.tex}
\input{src/5-data.tex}
\input{src/6-network.tex}
\input{src/7-apps.tex}
\input{src/8-others.tex}
\input{src/9-misc.tex}
\input{src/10-disc.tex}

\input{src/11-conclusions.tex}

%\printglossary[type=acronym,style=tree,title=List Of Acronyms,nonumberlist]

\nomenclature{6G}{Sixth-Generation}
\nomenclature{6lo}{IPv6 over networks of resource-constrained nodes}
\nomenclature{6LoWPAN}{IPv6 over Low power Wireless Personal Area Networks}
\nomenclature{6top}{6tisch Operation Sublayer}
\nomenclature{6TiSCH}{IPv6 over the TSCH mode of IEEE 802.15.4}

\nomenclature{A2A}{Air-to-Air}
\nomenclature{A2G}{Air-to-Ground}
\nomenclature{ACODS}{Adaptive Computation Offloading Drone System}
\nomenclature{AGV}{Automated Guided Vehicle}
\nomenclature{AI}{Artificial Intelligence}
\nomenclature{AODV}{Ad-hoc On-demand Distance Vector}
\nomenclature{AP}{Allocation Policy} 
\nomenclature{AQ}{Autonomous Quadcopter}
\nomenclature{ASAP}{As-Soon-As-Possible Schedule Another Parent}
\nomenclature{ASN}{Absolute Slot Number}
\nomenclature{ASV}{Autonomous Surface Vehicle}
\nomenclature{AUV}{Autonomous Underwater Vehicle}

\nomenclature{BDMA}{Beam-Division Multiple Access}
\nomenclature{BS}{Base Station}

\nomenclature{CCN}{Content-Centric Network}
\nomenclature{CCTV}{Closed Circuit TeleVision}
\nomenclature{CEA}{Cell Estimation Algorithm}
\nomenclature{CH}{Cluster Head}
\nomenclature{CIR}{Channel Impulse Response}
\nomenclature{CoAP}{Constrained Application Protocol}
\nomenclature{CoRE}{Constrained RESTful Environments}
\nomenclature{CR}{Cognitive Radio}
\nomenclature{CSMA}{Carrier Sense Multiple Access}
\nomenclature{CSMA/CA}{Carrier Sense Multiple Access with Collision Avoidance}
\nomenclature{CTD}{Conductivity, Temperature, Depth}

\nomenclature{D2D}{Device-to-Device}
\nomenclature{DAG}{Directed Acyclic Graph}
\nomenclature{DAO}{Destination Advertisement Object}
\nomenclature{DETB}{Dynamic Energy \& Traffic Balance}
\nomenclature{DGAC}{Direction Générale de l'Aviation Civile}
\nomenclature{DL}{DownLink}
\nomenclature{DIO}{Destination Information Object}
\nomenclature{DNS}{Domain Name System}
\nomenclature{DODAG}{Destination Oriented DAG}
\nomenclature{DP}{Dynamic Programming}
\nomenclature{DRL}{Deep Reinforcement Learning}
\nomenclature{DSSS}{Direct Sequence Spread Spectrum}

\nomenclature{E2E}{End-to-End}
\nomenclature{EASA}{European Aviation Safety Agency}
\nomenclature{ENAC}{Ente Nazionale per l'Aviazione Civile}
\nomenclature{EOF}{End Of Frame}
\nomenclature{ePFC}{extended Potential Field Controller}

\nomenclature{FAA}{Federal Aviation Administration}
\nomenclature{FANET}{Flying Ad-hoc NETwork}
\nomenclature{FBS}{Flying Base Station}
\nomenclature{FDM}{Frequency Division Multiplexing}
\nomenclature{FDMA}{Frequency Division Multiple Access}
\nomenclature{FFD}{Fully Functioned Device}
\nomenclature{FIR}{Finite Impulse Response}
\nomenclature{FoV}{Field-of-View}
\nomenclature{FSO}{Free Space Optical}

\nomenclature{G2G}{Ground-to-Ground}
\nomenclature{GA}{Genetic Algorithm}
\nomenclature{GDA}{Gradient Algorithm}
\nomenclature{GIS}{Geographic Information System}
\nomenclature{GPS}{Global Position System}
\nomenclature{GPSR}{Greedy Perimeter Stateless Routing}
\nomenclature{GS}{Ground Station}

\nomenclature{HTTP}{HyperText Transfer Protocol}

\nomenclature{ICI}{Inter-Carrier Interference}
\nomenclature{ICN}{Information-Centric Networking}
\nomenclature{ICT}{Information and Communication Technologies}
\nomenclature{IDFT}{Inverse Discrete Fourier Transform}
\nomenclature{IETF}{Internet Engineering Task Force}
\nomenclature{IIoT}{Industrial Internet of Things}
\nomenclature{IMC}{Internal Model Control}
\nomenclature{IMU}{Inertial Measurement Unit}
\nomenclature{IoD}{Internet of Drones}
\nomenclature{IoT}{Internet of Things}
\nomenclature{IP}{Internet Protocol}
\nomenclature{IPv4}{Internet Protocol version 4}
\nomenclature{IPv6}{Internet Protocol version 6}
\nomenclature{ISM}{Industrial, Scientific and Medical}
\nomenclature{ISP}{Internet Service Provider}
\nomenclature{ITS}{Intelligent Transportation System}

\nomenclature{JSON}{JavaScript Object Notation}

\nomenclature{KPI}{Key Performance Index}
\nomenclature{KPIs}{Key Performance Indices}

\nomenclature{LA}{Learning Algorithm}
\nomenclature{LAR}{Location-Aided Routing}
\nomenclature{LDRA}{Low Delay Routing Algorithm}
\nomenclature{LLN}{Low-power Lossy Network}
\nomenclature{LLSF}{Low-power Lossy Network}
\nomenclature{LoRaWAN}{Long Range Wide Area Network}
\nomenclature{LoS}{Line of Sight}
\nomenclature{LPWAN}{Low Power Wide Area Network}
\nomenclature{LTE}{Long-Term Evolution}
\nomenclature{LTE-A}{Long-Term Evolution-Advanced}
\nomenclature{LTE-BS}{Long-Term Evolution Base Station}
\nomenclature{LTE-U}{Long-Term Evolution - Unlicensed}

\nomenclature{M2M}{Machine-to-Machine}
\nomenclature{MA}{Multiple Access}
\nomenclature{MAC}{Medium Access Control}
\nomenclature{MANET}{Mobile Ad-hoc NETwork}
\nomenclature{MTC}{Machine-Type Communication}
\nomenclature{MTC}{mission critical Machine-Type Communication}
\nomenclature{MCU}{Micro-Controller Unit}
\nomenclature{MEC}{Mobile Edge Computing}
\nomenclature{MINLP}{Mixed Integer Non-Linear Programming}
\nomenclature{MPTCP}{Multipath TCP}
\nomenclature{MSPRT}{Multi Sequential Probability Ratio Test}
\nomenclature{MTU}{Maximum Transmission Unit}

\nomenclature{NC}{Network Coding}
\nomenclature{NFP}{Networked Flying Platform}
\nomenclature{NFV}{Network Function Virtualization}
\nomenclature{NLoS}{Non-Line of Sight}
\nomenclature{NOMA}{Non-Orthogonal Multiple Access}

\nomenclature{O-QPSK}{Offset-Quadrature Phase-Shift Keying}
\nomenclature{OFDM}{Orthogonal Frequency-Division Multiplexing}
\nomenclature{OLSR}{Optimized Link State Routing Protocol}
\nomenclature{OS}{Operating System}
\nomenclature{OTF}{On The Fly}

\nomenclature{P2P}{Peer-to-Peer}
\nomenclature{PCC}{Path Computation Client}
\nomenclature{PCE}{Path Computation Element}
\nomenclature{PCEP}{Path Computation Element Protocol}
\nomenclature{PDR}{Packet Delivery Ratio}
\nomenclature{PHY}{Physical Layer}
\nomenclature{PID}{Proportional-Integral-Derivative}
\nomenclature{PLA}{PolyLactic Acid}
\nomenclature{PLR}{Packet Loss Ratio}
\nomenclature{PSO}{Particle Swarm Optimization}
\nomenclature{PSH}{Problem Specific Heuristic}
\nomenclature{PxIMU}{Pixhawk Inertial Measurement Unit}

\nomenclature{QoE}{Quality of Experience}
\nomenclature{QoL}{Quality of Link}
\nomenclature{QoS}{Quality of Service}

\nomenclature{RFD}{Reduced Functioned Device}
\nomenclature{RMM}{Random Mobility Model}
\nomenclature{RMS-DS}{Root Mean Square Delay Spread}
\nomenclature{ROLL}{Routing Over Low power and Lossy networks}
\nomenclature{ROS}{Robot Operating System}
\nomenclature{ROV}{Remotely Operated underwater Vehicle}
\nomenclature{RPL}{Routing Protocol for Low-power and Lossy networks}
\nomenclature{RTT}{Round Trip Time}
\nomenclature{RSMA}{Rate-Splitting Multiple Access}
\nomenclature{RSSI}{Received Signal Strength Indication}
\nomenclature{RSU}{RoadSide Unit}
\nomenclature{RSS}{Received Signal Strength}

\nomenclature{SAIN}{Space-Air Integrated Network}
\nomenclature{SC-FDM}{Single Carrier - Frequency Division Multiplexing}
\nomenclature{SDMA}{Spatial-Division Multiple Access}
\nomenclature{SDN}{Software Defined Network}
\nomenclature{SDR}{Software Defines Radio}
\nomenclature{SE}{Spectral Efficiency}
\nomenclature{SFD}{Start of Frame Delimiter}
\nomenclature{SF0}{Scheduling Function Zero}
\nomenclature{SINR}{Signal to Interference-plus-Noise Ratio}
\nomenclature{SLAM}{Simultaneous Localization And Mapping}
\nomenclature{SNR}{Signal-to-Noise Ratio}
\nomenclature{SPI}{Serial Peripheral Interface}

\nomenclature{TCP}{Transmission Control Protocol}
\nomenclature{TDMA}{Time Division Multiple Access}
\nomenclature{TI}{Texas Instruments}
\nomenclature{TSCH}{Time Slotted Channel Hopping}

\nomenclature{UAS}{Unmanned Aircraft System}
\nomenclature{UASN}{Underwater Acoustic Sensor Network}
\nomenclature{UAV}{Unmanned Aerial Vehicle}
\nomenclature{UAV-BS}{Unmanned Aerial Vehicle Base Station}
\nomenclature{UDP}{User Datagram Protocol}
\nomenclature{UE}{User Equipment}
\nomenclature{UFP}{Unmanned Flying Platform}
\nomenclature{UGV}{Unmanned Ground Vehicle}
\nomenclature{USB}{Universal Serial Bus}
\nomenclature{USV}{Unmanned Surface Vehicle}
\nomenclature{UTM}{Unmanned Aerial System Traffic Management}
\nomenclature{UUV}{Unmanned Underwater Vehicle}
\nomenclature{UWSN}{Underwater Wireless Sensor Network}
\nomenclature{UL}{UpLink}

\nomenclature{V2I}{Vehicle-to-Infrastructure}
\nomenclature{V2V}{Vehicle-to-Vehicle}
\nomenclature{VANET}{Vehicular Ad-hoc NETwork}
\nomenclature{VLC}{Visible Light Communication}
\nomenclature{VMS}{Variable Message Sign}

\nomenclature{WG}{Working Group}
\nomenclature{WSN}{Wireless Sensor Network}

\printnomenclature

\bibliographystyle{elsarticle-num}
\bibliography{bibliography.bib}

\end{document}

%% file: src/acronyms.tex
\newacronym{6G}{6G}{Sixth-Generation}
\newacronym{6lo}{6lo}{IPv6 over networks of resource-constrained nodes}
\newacronym{6LoWPAN}{6LoWPAN}{IPv6 over Low power Wireless Personal Area Networks}
\newacronym{6top}{6top}{6tisch Operation Sublayer}
\newacronym{6TiSCH}{6TiSCH}{IPv6 over the TSCH mode of IEEE 802.15.4}

\newacronym{A2A}{A2A}{Air-to-Air}
\newacronym{A2G}{A2G}{Air-to-Ground}
\newacronym{ACODS}{ACODS}{Adaptive Computation Offloading Drone System}
\newacronym{AGV}{AGV}{Automated Guided Vehicle}
\newacronym{AI}{AI}{Artificial Intelligence}
\newacronym{AODV}{AODV}{Ad-hoc On-demand Distance Vector}
\newacronym{AP}{AP}{Allocation Policy} 
\newacronym{AQ}{AQ}{Autonomous Quadcopter}
\newacronym{ASAP}{ASAP}{As-Soon-As-Possible Schedule Another Parent}
\newacronym{ASN}{ASN}{Absolute Slot Number}
\newacronym{ASV}{ASV}{Autonomous Surface Vehicle}
\newacronym{AUV}{AUV}{Autonomous Underwater Vehicle}

\newacronym{BDMA}{BDMA}{Beam-Division Multiple Access}
\newacronym{BS}{BS}{Base Station}

\newacronym{CCN}{CCN}{Content-Centric Network}
\newacronym{CCTV}{CCTV}{Closed Circuit TeleVision}
\newacronym{CEA}{CEA}{Cell Estimation Algorithm}
\newacronym{CH}{CH}{Cluster Head}
\newacronym{CIR}{CIR}{Channel Impulse Response}
\newacronym{CoAP}{CoAP}{Constrained Application Protocol}
\newacronym{CoRE}{CoRE}{Constrained RESTful Environments}
\newacronym{CR}{CR}{Cognitive Radio}
\newacronym{CSMA}{CSMA}{Carrier Sense Multiple Access}
\newacronym{CSMA/CA}{CSMA/CA}{Carrier Sense Multiple Access with Collision Avoidance}
\newacronym{CTD}{CTD}{Conductivity, Temperature, Depth}

\newacronym{D2D}{D2D}{Device-to-Device}
\newacronym{DAG}{DAG}{Directed Acyclic Graph}
\newacronym{DAO}{DAO}{Destination Advertisement Object}
\newacronym{DETB}{DETB}{Dynamic Energy \& Traffic Balance}
\newacronym{DGAC}{DGAC}{Direction Générale de l'Aviation Civile}
\newacronym{DL}{DL}{DownLink}
\newacronym{DIO}{DIO}{Destination Information Object}
\newacronym{DNS}{DNS}{Domain Name System}
\newacronym{DODAG}{DODAG}{Destination Oriented DAG}
\newacronym{DP}{DP}{Dynamic Programming}
\newacronym{DRL}{DRL}{Deep Reinforcement Learning}
\newacronym{DSSS}{DSSS}{Direct Sequence Spread Spectrum}

\newacronym{E2E}{E2E}{End-to-End}
\newacronym{EASA}{EASA}{European Aviation Safety Agency}
\newacronym{ENAC}{ENAC}{Ente Nazionale per l'Aviazione Civile}
\newacronym{EOF}{EOF}{End Of Frame}
\newacronym{ePFC}{ePFC}{extended Potential Field Controller}

\newacronym{FAA}{FAA}{Federal Aviation Administration}
\newacronym{FANET}{FANET}{Flying Ad-hoc NETwork}
\newacronym{FBS}{FBS}{Flying Base Station}
\newacronym{FDM}{FDM}{Frequency Division Multiplexing}
\newacronym{FDMA}{FDMA}{Frequency Division Multiple Access}
\newacronym{FFD}{FFD}{Fully Functioned Device}
\newacronym{FIR}{FIR}{Finite Impulse Response}
\newacronym{FoV}{FoV}{Field-of-View}
\newacronym{FSO}{FSO}{Free Space Optical}

\newacronym{G2G}{G2G}{Ground-to-Ground}
\newacronym{GA}{GA}{Genetic Algorithm}
\newacronym{GDA}{GDA}{Gradient Algorithm}
\newacronym{GIS}{GIS}{Geographic Information System}
\newacronym{GPS}{GPS}{Global Position System}
\newacronym{GPSR}{GPSR}{Greedy Perimeter Stateless Routing}
\newacronym{GS}{GS}{Ground Station}

\newacronym{HTTP}{HTTP}{HyperText Transfer Protocol}

\newacronym{ICI}{ICI}{Inter-Carrier Interference}
\newacronym{ICN}{ICN}{Information-Centric Networking}
\newacronym{ICT}{ICT}{Information and Communication Technologies}
\newacronym{IDFT}{IDFT}{Inverse Discrete Fourier Transform}
\newacronym{IETF}{IETF}{Internet Engineering Task Force}
\newacronym{IIoT}{IIoT}{Industrial Internet of Things}
\newacronym{IMC}{IMC}{Internal Model Control}
\newacronym{IMU}{IMU}{Inertial Measurement Unit}
\newacronym{IoD}{IoD}{Internet of Drones}
\newacronym{IoT}{IoT}{Internet of Things}
\newacronym{IP}{IP}{Internet Protocol}
\newacronym{IPv4}{IPv4}{Internet Protocol version 4}
\newacronym{IPv6}{IPv6}{Internet Protocol version 6}
\newacronym{ISM}{ISM}{Industrial, Scientific and Medical}
\newacronym{ISP}{ISP}{Internet Service Provider}
\newacronym{ITS}{ITS}{Intelligent Transportation System}

\newacronym{JSON}{JSON}{JavaScript Object Notation}

\newacronym{KPI}{KPI}{Key Performance Index}
\newacronym{KPIs}{KPIs}{Key Performance Indices}

\newacronym{LA}{LA}{Learning Algorithm}
\newacronym{LAR}{LAR}{Location-Aided Routing}
\newacronym{LDRA}{LDRA}{Low Delay Routing Algorithm}
\newacronym{LLN}{LLN}{Low-power Lossy Network}
\newacronym{LLSF}{LLSF}{Low-power Lossy Network}
\newacronym{LoRaWAN}{LoRaWAN}{Long Range Wide Area Network}
\newacronym{LoS}{LoS}{Line of Sight}
\newacronym{LPWAN}{LPWAN}{Low Power Wide Area Network}
\newacronym{LTE}{LTE}{Long-Term Evolution}
\newacronym{LTE-A}{LTE-A}{Long-Term Evolution-Advanced}
\newacronym{LTE-BS}{LTE-BS}{Long-Term Evolution Base Station}
\newacronym{LTE-U}{LTE-U}{Long-Term Evolution - Unlicensed}

\newacronym{M2M}{M2M}{Machine-to-Machine}
\newacronym{MA}{MA}{Multiple Access}
\newacronym{MAC}{MAC}{Medium Access Control}
\newacronym{MANET}{MANET}{Mobile Ad-hoc NETwork}
\newacronym{MTC}{MTC}{Machine-Type Communication}
\newacronym{mcMTC}{MTC}{mission critical Machine-Type Communication}
\newacronym{MCU}{MCU}{Micro-Controller Unit}
\newacronym{MEC}{MEC}{Mobile Edge Computing}
\newacronym{MINLP}{MINLP}{Mixed Integer Non-Linear Programming}
\newacronym{MPTCP}{MPTCP}{Multipath TCP}
\newacronym{MSPRT}{MSPRT}{Multi Sequential Probability Ratio Test}
\newacronym{MTU}{MTU}{Maximum Transmission Unit}

\newacronym{NC}{NC}{Network Coding}
\newacronym{NFP}{NFP}{Networked Flying Platform}
\newacronym{NFV}{NFV}{Network Function Virtualization}
\newacronym{NLoS}{NLoS}{Non-Line of Sight}
\newacronym{NOMA}{NOMA}{Non-Orthogonal Multiple Access}

\newacronym{O-QPSK}{O-QPSK}{Offset-Quadrature Phase-Shift Keying}
\newacronym{OFDM}{OFDM}{Orthogonal Frequency-Division Multiplexing}
\newacronym{OLSR}{OLSR}{Optimized Link State Routing Protocol}
\newacronym{OS}{OS}{Operating System}
\newacronym{OTF}{OTF}{On The Fly}

\newacronym{P2P}{P2P}{Peer-to-Peer}
\newacronym{PCC}{PCC}{Path Computation Client}
\newacronym{PCE}{PCE}{Path Computation Element}
\newacronym{PCEP}{PCEP}{Path Computation Element Protocol}
\newacronym{PDR}{PDR}{Packet Delivery Ratio}
\newacronym{PHY}{PHY}{Physical Layer}
\newacronym{PID}{PID}{Proportional-Integral-Derivative}
\newacronym{PLA}{PLA}{PolyLactic Acid}
\newacronym{PLR}{PLR}{Packet Loss Ratio}
\newacronym{PSO}{PSO}{Particle Swarm Optimization}
\newacronym{PSH}{PSH}{Problem Specific Heuristic}
\newacronym{PxIMU}{PxIMU}{Pixhawk Inertial Measurement Unit}

\newacronym{QoE}{QoE}{Quality of Experience}
\newacronym{QoL}{QoL}{Quality of Link}
\newacronym{QoS}{QoS}{Quality of Service}

\newacronym{RFD}{RFD}{Reduced Functioned Device}
\newacronym{RMM}{RMM}{Random Mobility Model}
\newacronym{RMS-DS}{RMS-DS}{Root Mean Square Delay Spread}
\newacronym{ROLL}{ROLL}{Routing Over Low power and Lossy networks}
\newacronym{ROS}{ROS}{Robot Operating System}
\newacronym{ROV}{ROV}{Remotely Operated underwater Vehicle}
\newacronym{RPL}{RPL}{Routing Protocol for Low-power and Lossy networks}
\newacronym{RTT}{RTT}{Round Trip Time}
\newacronym{RSMA}{RSMA}{Rate-Splitting Multiple Access}
\newacronym{RSSI}{RSSI}{Received Signal Strength Indication}
\newacronym{RSU}{RSU}{RoadSide Unit}
\newacronym{RSS}{RSS}{Received Signal Strength}

\newacronym{SAIN}{SAIN}{Space-Air Integrated Network}
\newacronym{SC-FDM}{SC-FDM}{Single Carrier - Frequency Division Multiplexing}
\newacronym{SDMA}{SDMA}{Spatial-Division Multiple Access}
\newacronym{SDN}{SDN}{Software Defined Network}
\newacronym{SDR}{SDR}{Software Defines Radio}
\newacronym{SE}{SE}{Spectral Efficiency}
\newacronym{SFD}{SFD}{Start of Frame Delimiter}
\newacronym{SF0}{SF0}{Scheduling Function Zero}
\newacronym{SINR}{SINR}{Signal to Interference-plus-Noise Ratio}
\newacronym{SLAM}{SLAM}{Simultaneous Localization And Mapping}
\newacronym{SNR}{SNR}{Signal-to-Noise Ratio}
\newacronym{SPI}{SPI}{Serial Peripheral Interface}

\newacronym{TCP}{TCP}{Transmission Control Protocol}
\newacronym{TDMA}{TDMA}{Time Division Multiple Access}
\newacronym{TI}{TI}{Texas Instruments}
\newacronym{TSCH}{TSCH}{Time Slotted Channel Hopping}

\newacronym{UAS}{UAS}{Unmanned Aircraft System}
\newacronym{UASN}{UASN}{Underwater Acoustic Sensor Network}
\newacronym{UAV}{UAV}{Unmanned Aerial Vehicle}
\newacronym{UAV-BS}{UAV-BS}{Unmanned Aerial Vehicle Base Station}
\newacronym{UDP}{UDP}{User Datagram Protocol}
\newacronym{UE}{UE}{User Equipment}
\newacronym{UFP}{UFP}{Unmanned Flying Platform}
\newacronym{UGV}{UGV}{Unmanned Ground Vehicle}
\newacronym{USB}{USB}{Universal Serial Bus}
\newacronym{USV}{USV}{Unmanned Surface Vehicle}
\newacronym{UTM}{UTM}{Unmanned Aerial System Traffic Management}
\newacronym{UUV}{UUV}{Unmanned Underwater Vehicle}
\newacronym{UWSN}{UWSN}{Underwater Wireless Sensor Network}
\newacronym{UL}{UL}{UpLink}

\newacronym{V2I}{V2I}{Vehicle-to-Infrastructure}
\newacronym{V2V}{V2V}{Vehicle-to-Vehicle}
\newacronym{VANET}{VANET}{Vehicular Ad-hoc NETwork}
\newacronym{VLC}{VLC}{Visible Light Communication}
\newacronym{VMS}{VMS}{Variable Message Sign}

\newacronym{WG}{WG}{Working Group}
\newacronym{WSN}{WSN}{Wireless Sensor Network}

%% file: src/1-intro.tex
\section{Introduction}\label{sec:intro}
The \ac{IoD} is defined as a network architecture specifically aimed at supporting communications between autonomous vehicles able to fly and a number of network entities deployed on the ground \cite{GBW16, KM21}.
In the \ac{IoD}, drones are conceived as smart objects in charge of flying all over a certain area, to carry out a number of different tasks, spanning from patrolling to sensing environmental data, in order to gather data of interest, even in real-time, for any further use.
In the \ac{IoD} architecture, \acp{UAV} are conceived as networked objects able to communicate among themselves, exchanging data, for example connected with flight coordination capabilities. At the same time, drones are communicating with a reference ground infrastructure that is in charge of storing and elaborating data for enabling services or providing updated information to remote users connected with dedicated application servers. The reference ground infrastructure is also in charge of controlling the airspace, granting the optimal position of drones and monitoring their activities. Since the ground network infrastructure is conceived as a set of logical nodes with high computational capabilities, tasks coordination and mission plan updates can be delegated, without affecting the limited resources onboard of drones.

\ac{IoD} has attracted much interest in the recent literature, due to the flexibility and the adaptability of networks of drones in the most widely differing scenarios, and their ability of enhance the performance of other network architectures. \acp{UAV} are becoming more and more widespread in many fields, due to some technological, tactical and/or practical advantages: (i) high mobility, (ii) easy deployment and re-employability, (iii) real-time monitoring and coordination (strongly dependent on both system architectures and communication technologies), (iv) load transport (depending on the specific cargo application requirements).
Drones can also be employed in places hardly accessible to humans, enhancing the network connectivity, coverage and capacity, especially in combination with other wireless/wired network architectures \cite{GBW16,KM21,SK17, NNB+20}.
Given this wide range of functionalities, a number of different applications for the \ac{IoD} can be enabled. Among them: smart agriculture, search and rescue, surveillance systems, mission-critical services, stock management, sport and training, telecommunications, art and creativity, etc. \cite{SK17}.

Against this, several issues need to be properly investigated in \ac{IoD} scenarios. First, the wireless medium by its own nature is unreliable: it is prone to errors, attenuations, and multipath propagation, thus bringing to data losses and link interruptions.
Even though this is a well-known aspect of wireless communications, this becomes more important when the networked objects are flying by design and moving suddenly from one place to another. This kind of problem has been tackled in literature when referring to \acp{WSN}, \acp{MANET}, and \acp{FANET}. Unfortunately, in the \ac{IoD}, drones are accomplishing several tasks, spanning from coverage enhancement, when they are employed as \acp{FBS}, to pattern identification in crowd situations. In these cases, the intrinsic unreliability of the wireless medium suggests that dedicated design efforts must be devoted. In fact, the communication range is often limited, especially in environments with high signal attenuation \cite{KCZ+18}. The high mobility degree of drones makes this situation even worse, requiring an accurate control and coordination of drones fleets to avoid conflicts, especially in terms of routes and data exchange \cite{FQD+18,SK17}.

Power consumption is another critical point in battery powered devices as drones are.
This problem is not new to those that study \ac{IoT}, since sparing battery and consuming less energy when sensing, communicating, and elaborating data is a mandatory design criterion in that domain. This is motivated by the fact that in the \ac{IoT} domain, network nodes are supposed to be deployed with low or null maintenance. As a consequence, policies of duty cycling and low transmission power are usually involved in this context. What makes the \ac{IoD} a more challenging operating scenario is the fact that energy consumption is not only associated with sensing and communicating but also with movements and flight patterns and mission plans. Further, while \ac{IoT} devices are often communicating low amounts of data, drones are frequently involved in high traffic volumes. This happens for example when drones are acting as \acp{FBS} or supporting video streaming applications. This suggests that onboard memory matters must be faced \cite{IBG20}.
Efficient resource allocation should thus be employed to optimize energy consumption, in terms of both the energy spent for data communication and routes of drones \cite{SK17}.
In some application scenarios, where drones are used as relay points to collect and/or deliver data in multi-hop flying networks, cooperation is a mandatory requirement to increase network connectivity, coverage performance, latency and throughput. Moreover, frequent topology changes make much more difficult to implement effective routing protocols that keep a high level of data reliability and network connectivity \cite{GJV16,NNB+20, LWW21}.
Last but not least, the idea that drones can communicate logically implies some major concerns about security and privacy. Given the constraints in terms of energy consumption and computational capabilities the drones are logically affected by, lightweight cryptography and security solutions are highly recommendable \cite{FQD+18}.
%In this landscape, the rest of the network infrastructure offers several services for information exchange, such as connectivity optimization, service provisioning and data elaboration \cite{GBW16}.
Overall, the \ac{IoD} is proposed to allow the development of networks composed by heterogeneous and interconnected drones with proper resource planning to be integrated within third parties businesses and services \cite{CSG+18}.

The relevance of \ac{IoD}-related topics to the scientific community to solve the issues described above is testified by the very large number of works found in literature, where the \ac{IoD} paradigm is analyzed in all its multifaceted aspects.
%The analysis of the recent literature on \ac{IoD} testifies the strong and always relevant interest on this topic.
%A very large number of works can be found where the \ac{IoD} paradigm is analyzed, in all its multifaceted aspects, and trying to solve or mitigate its inherent issues in a wide set of application scenarios.
%
%The main challenges for the \ac{IoD} are related to: (i) channel modeling, (ii) interference, (iii) deployment, (iv) energy efficiency, (v) path planning/mobility, (vi) resource management, (vii) new/innovative services, and lastly, (viii) security, privacy and safety.
%
%\todo
The motivation for this survey is to analyze communications aspects, technologies, protocols and architectures in the \ac{IoD} scenario, detailing as much as possible the analysis of the state-of-the-art research on the topic, at the same time updating and completing the analysis of the other surveys found in literature \cite{NNB+20,MTA16,GJV16,FQD+18,SK17,CZX14,KCZ+18}.

%categorize them all, by proposing a thorough, varied and original classification of the \ac{IoD} scenario. If compared to the other surveys on the same topic, a significant added value is 

%that this work analyzes, through the surveyed literature, a wide variety of topics, aiming to be complementary to all the other surveys, completing the analysis of each of them, at the same time updating and detailing as much as possible the analysis of the state-of-the-art research on the \ac{IoD} theme.

%Summarizing, this survey provides a more thorough, varied and original classification of the \ac{IoD} scenario, if compared to the other surveys found in literature. A significant value added is that it is not restricted to a single environment. Furthermore, it analyzes, through the surveyed literature, a wider variety of topics, aiming to be complementary to all the aforementioned works, completing the analysis of the missing topics of each of them, at the same time updating and detailing as much as possible the analysis of the state-of-the-art research on this theme.

For all the reasons mentioned above, and for ease of comprehension, the classification proposed in this work follows, at a higher level, the classical protocol stack, starting from the analysis of all the issues related to the physical layer and going on with the data link layer, the network layer, and the transport and application layers. Nevertheless, there are also cross-layer and optimization approaches that cannot be framed in any of the protocol layers. They have thus been analyzed and discussed in a separate section.

%Going into deeper descriptive detail, the contribution of this work is compared with similar contributions of the other surveys on \ac{IoD} found in literature, highlighting the most relevant differences between the proposed classification approach and the other existing ones. Then, the surveyed papers are further categorized and discussed, relating them to the issues specific of the layer under analysis.
The last part of this work is then focused on a discussion on the main research challenges and possible future directions in the \ac{IoD} field.

%Wireless communication with \ac{UAV}:
%\begin{itemize}
%	\item \ac{UAV}-to-\ac{UAV} communication required for coordination, interference mitigation, relaying, routing in the air.
%	\item Satellite and WiFi considered as candidate technologies for providing wireless backhauling => depending on latency bandwidth requirements
%	\item Satellite backhauling brings the advantage of unlimited coverage offering the possibility of connecting the aerial network for any distance. However, the latency introduced by the satellite links (GEO) may affect some real time services such as voice and real time video.
%	\item To avoid satellite delays and the cost, WiFi links can be used albeit reduced coverage and capacity (doubtful \ac{QoS} guarantees..)
%\end{itemize}
%
%\todo
%Presentazione sintetica dello scopo delle varie sezioni e dello scopo della classificazione e tassonomia adottate

Although researchers and research teams worldwide are concentrated on drones' related topics, all the themes that the scientific community is dealing with appear to be fragmented. As a result, all the works that will be described herein, as well as the survey papers, are only focused on specific aspects, leading to non-structured nor systematic studies.
For instance, when discussing the application of drones in real world context, there is a lack of characterization of the operational contexts. Indeed, the social and economical impact is not discussed in detail. At the same time, these aspects are often treated on a regional basis.
Moreover, in the context of drones communications, several solutions have been proposed in terms of modulation techniques. Nevertheless, the theme has never been surveyed before. 
To provide a further example, when optimization approaches are discussed, the proposed frameworks are not as thorough as they could be. For example, it is easy to find papers that optimize both trajectory and energy consumption. Sometimes it is possible to find papers that design the path followed by a drone during a mission without taking into consideration 3D spaces.
Hence, while specific aspects are addressed within the other surveys and overviews on the \ac{IoD}, the purpose of this work is to carry out an organic and thorough analysis, dealing with all the main aspects of the theme.
At the same time, this work investigates the applicability of novel technologies, e.g., mmWave and \ac{VLC}, upcoming networks, i.e., \ac{6G}, and communication paradigms, i.e., \ac{ICN}.
To the best of authors knowledge, this is also the first study on the subject that also considers drone economics, thus discussing the economic impact of the adoption of drones in all the main application fields.

The remainder of this paper is organized as follows: Section \ref{sec:applications} discusses the applicability of the \ac{IoD} paradigm in real use-cases.
Section \ref{sec:related} proposes a detailed study on the related surveys, describing their classification approach. At the same time, it clarifies the differences between the analytical models and approaches proposed so far and the way this work approaches the \ac{IoD}, as a whole.
%In Sections from \ref{sec:phy} to \ref{sec:appl} a number of key problems and research challenges are discussed. The rationale for these Sections can be found in the layered structure on which telecommunication protocols and stacks are based.
%In particular, 
Section \ref{sec:phy} focuses on the physical layer, and provides a detailed characterization on connectivity issues as well as channel modeling approaches. Section \ref{sec:data} focuses on data link layer, discussing the related works on resource allocation and data scheduling.
Section \ref{sec:network} is dedicated to network layer, paying attention mainly to coordination aspects, routing problems and relaying schemes.
Section \ref{sec:appl} is focused on the application layer, which includes \ac{QoE} indicators, computation offloading, task allocation and data collection/distribution.
Section \ref{sec:others} analyzes all the papers proposing cross-layer approaches and optimization strategies in several \ac{IoD}-related aspects, ranging from path planning and collision avoidance to network formation and control, energy efficiency, mobility, network architectures, etc.
Section \ref{sec:miscellaneous} is focused on security aspects and business models, which can be considered as complementary themes when discussing the development of drones as enabling technology toward the \ac{IoD}.
%Even though at the end of each Section, several considerations and remarks are reported,
Section \ref{sec:disc_future} summarizes the main findings and discusses the strengths and weaknesses of the analyzed technological landscape. All the considerations made in the surveyed literature pave the way of future research perspectives which are discussed in detail, with specific focus on the open issues and the most promising technologies.
Finally, Section \ref{sec:conclusions} concludes the work.

%% file: src/3-related.tex
\section{Related Survey and Review Papers}\label{sec:related}
\begin{figure}[htbp]
	\centering
	\includegraphics[width=0.7\linewidth]{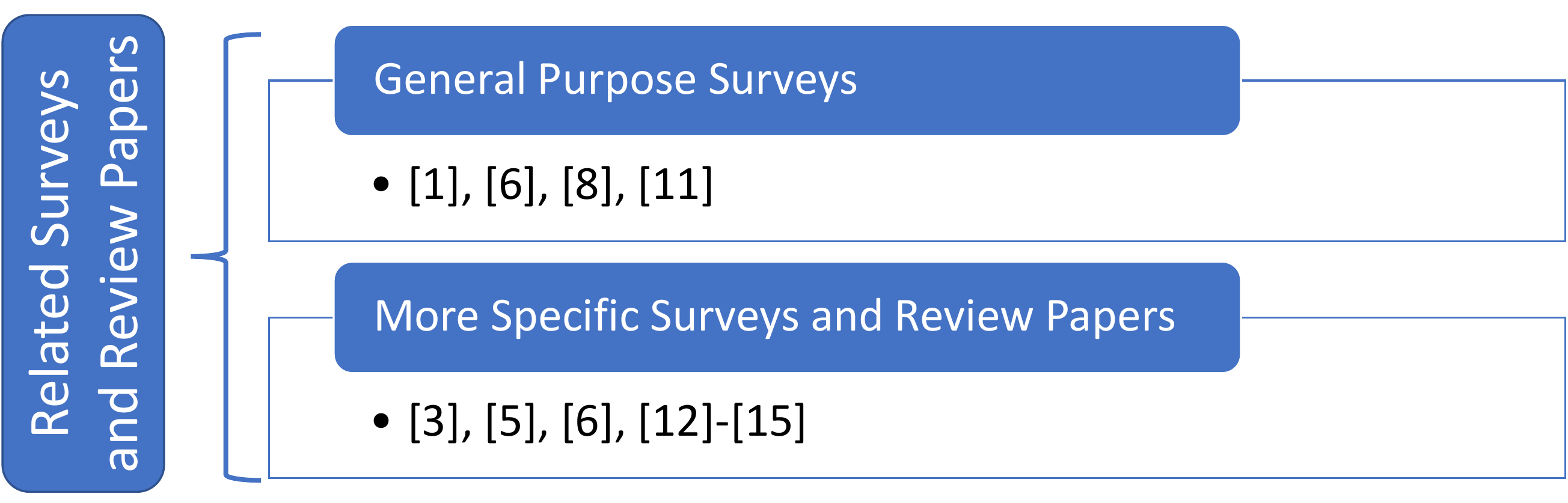}
	\caption{Related works Taxonomy.}
	\label{fig:tax_surveys}
\end{figure}

The main goal of this Section is to analyze all the available survey works related to the \ac{IoD} and to \ac{IoD}-related themes. The aim is to better highlight the key contributions of the present work with respect to what is already available in literature.
The overall organization of this section is reported in Figure \ref{fig:tax_surveys}.

%%%%%%%%%%%%%%%%%%%%%%%%%%%%%%%%%%% GENERAL PURPOSE SURVEYS %%%%%%%%%%%%%%%%%%%%%%%%%%%%%%%%%%%%%%%%%
\subsubsection{General Purpose Surveys}
There are some papers that investigate in detail several aspects related to networks of drones \cite{GBW16}\cite{MTA16}\cite{GJV16}\cite{FQD+18}.\\
The first, and probably most important, work on \ac{IoD} can be found in \cite{GBW16}. It proposes an \ac{IoD} architecture that aims to solve the problem of airspace allocation and management. Other network architectures are first analyzed, to derive valuable lessons on scalability and fault tolerance to be fruitfully exploited in networks of drones, and at the same time highlighting the main differences with the \ac{IoD} scenario. The proposed architecture is then introduced, to provide generic services that can be used by different applications where drones are enabled to carry out tasks. Some challenges for effective \ac{IoD} systems are also explored and discussed, to provide future research directions on the topic.
%It gives a definition of the network infrastructure based on the preliminary study of three distinct large scale network structures: i) air traffic control, (ii) cellular networks, and (iii) the Internet. These networks are different from many points of view, e.g., size, bandwidths, latencies, \ac{QoS} and \ac{QoE} indicators. The proposal is mainly based on a description of the layers of the reference architecture. Each layer describes a systematic approach in defining the features that have to be implemented.
%The main weak point of this work is that it refers only to precise \ac{IoD} network architectures, neglecting all the other related topics (i.e., signal propagation, mobility, relaying and routing, etc.).

The integration of \acp{UAV} in cellular networks is analyzed in \cite{FQD+18}. Several aspects are discussed in detail. First, \acp{UAV} types and main features (flying mechanisms, coverage range and altitude, speed, flight time, and power consumption) are addressed. Several standardization studies on channel modeling and characterization aspects are discussed in detail. The main part of this survey focuses on the different challenges and opportunities for \ac{UAV}-based networks, ranging from the optimization of aerial \acp{BS} location and flying path to the minimization of energy consumption. Different prototyping tests are also surveyed, together with the main aspects of \ac{UAV} regulations and security issues.
%The survey \cite{FQD+18} surely proposes an original classification of the works in literature; it is more focused on legislation and security aspects of \ac{UAV}-assisted cellular networks, rather than on technical issues. Furthermore, the \ac{UAV}-assisted cellular networks is the only scenario analyzed in this work. 

Another interesting contribution is provided by \cite{GJV16}. It proposes a classification of \acp{UAV} based on their employability in the contexts of \acp{VANET} and \acp{MANET}, comparing such use cases with networks composed by \acp{UAV} only (ad-hoc networks). The contribution also highlights some of the main characteristics of \ac{UAV} networks, taking into account single-\ac{UAV} and multi-\ac{UAV} systems. Networks are categorized based on their topology. A large part of this survey is dedicated to routing approaches. The work is completed with a detailed discussion on strategies for energy efficiency and handover.
%Large part of this work is focused on routing and energy efficiency issues in \acp{UAV}-based networks, but neglects several other issues related to \ac{UAV} networks like connectivity, communication technologies, optimization strategies, application scenarios, etc.

In the survey \cite{MTA16} a detailed review of papers on communication technologies is presented for networks of drones in \ac{IoT} scenarios. Literature is analyzed based on the classification of technologies into long-range and short-range. Long-range technologies are discussed in papers focusing on drones control, connectivity increase, relaying, and in emergency situations or when \acp{UAV} are outside the range of direct links. Short range technologies are exploited for different applications like flight control, real-time data acquisition, file transfer, device synchronization, low-power/low-cost communications, and \ac{UAV} localization \cite{MTA16}. Literature is classified based on some application scenarios, i.e., monitoring activities, management of road traffic and disasters, path planning and routing, data collection mechanisms, almost exclusively referred to hardware (sensors, cameras, RFIDs, etc.) on board the drones. This survey also investigates the literature on network architectures and communication technologies, i.e., WiFi, satellite, WiMax, etc.
%The survey \cite{MTA16} discusses exhaustively the most widely used communication technologies in \ac{IoD} scenarios; nevertheless, it completely neglects some newer technologies, like mmWave or \ac{VLC}.

Even if each of the works \cite{MTA16,GJV16,FQD+18} exhaustively analyzes aspects peculiar of \ac{UAV} networks, nevertheless the discussion remains confined to specific topics, such as communication technologies in \ac{IoT} \cite{MTA16}, routing and energy efficiency \cite{GJV16}, or legislation and security aspects in \ac{UAV}-assisted cellular networks \cite{FQD+18}. As a matter of fact, none of these surveys conducts a wide-range analysis of the main challenges present at different layers of the protocol stack.
%%%%%%%%%%%%%%%%%%%%%%%%%%%%%%%%%%%%%%%%%%%%%%%%%%%%%%%%%%%%%%%%%%%%%%%%%%%%%%%%%%%%%%%%%%%%%%

%%%%%%%%%%%%%%%%%%%%%%%%%%%%%%%%%%%%%% MORE SPECIFIC SURVEY/REVIEW PAPERS %%%%%%%%%%%%%%%%%%%%%%%%%%%%%%%%%%%%%%%%%%%%
\subsubsection{More Specific Surveys and Review Papers}
Even more contextualized to specific \ac{IoD}-related topics is the analysis carried out in the survey and review papers \cite{SK17,KCZ+18,FMT+17,CGT+18,CZX14,ZZW+19,FQD+18}.

The survey \cite{SK17} focuses on cooperation models for \ac{UAV}-based networks. Single \ac{UAV} and multi-\ac{UAV} systems are analyzed in detail, with specific reference to relaying and routing strategies for cooperative network formation and coordination. The goal is to provide a detailed insight of the existing models, with their related features, to allow researchers to identify the existing solutions and analyze the performance of both existing and the models proposed. To this end, the available software solutions and simulators that enable cooperation and communications among flying entities forming ad-hoc networks are also compared and discussed, to test the implementation of an \ac{UAV}-oriented network.
%is based on the analysis of the existing literature on \ac{UAV} frameworks aiming at cooperative network formation and coordination issues.
%Even if this survey is detailed and exhaustive in the topic of cooperative \ac{UAV}-based communications, some important challenges like synchronization among nodes, delays, resource reservation, access management and error control, and modified \ac{MAC} protocols are not described in detail.

Physical layer aspects in the aerial/ground environment are analyzed in \cite{KCZ+18}, with an emphasis on channel characterization and modeling. This work focuses mainly on the measurement approaches at the basis of the modeling procedures in \ac{UAV} networks. A classification follows of the different kinds of models, empirical or analytical, through the analysis of the papers on the topic. Empirical studies based on measurement campaigns are described in detail, taking into account several factors like signal frequency, environmental conditions and \acp{UAV} main parameters (coverage, height, and mobility).
%This approach is useful to analyze in detail the possibility of using analytical models to evaluate the performances of different wireless \ac{UAV} communication techniques. Nevertheless, this study does not take into account all the issues related to the upper layers of the \ac{UAV} communications.

Robotic networks for surveillance purposes are discussed in \cite{FMT+17}. A discussion is carried out on the main research directions and challenges of robotics (with a detailed analysis on autonomy issues), with an analysis of the acoustic signal treatment and processing, network protocols and security aspects in underwater communications. This work is interesting because it refers to a very hostile environment for wireless communications, as the underwater scenario is.
%Nevertheless, it refers only to this environment, which presents issues that are totally different from the aerial/ground counterparts. Furthermore, it refers to a specific application scenario (robotic networks for underwater surveillance).

The overview work \cite{CGT+18} deals exclusively with location optimization algorithms. It makes a classification of the algorithms that can be used for optimizing the position occupied by a certain number of \acp{UAV}, especially when they are supposed/asked to act as \ac{BS} (the so-called \acp{UAV-BS}). Different classes of algorithms are analyzed in this survey. They include exact approaches, able to find the global optimum of the optimization problem, as well as heuristic-based approaches, reinforcement learning approaches, or other searching techniques. 
%The discussed \textit{Given heuristic names}, instead, are well-known \ac{DP} or meta-heuristics such as \ac{PSO}, \ac{GA}, or \ac{GDA}.
%An approach that greatly benefit from learning procedures is the \textit{\ac{LA}}, i.e., reinforcement learning.
%\textit{Enumeration}, instead, uses a search technique to find the best solution.
%The latter approach presented is \ac{PSH}, a solution that should be taken into consideration as a tailored one; in general, this solution is required when the model is created according to the peculiarities and specific properties of the problem.
This kind of classification is useful to solve the formulated problems through \ac{MINLP} techniques.
%Once again, this work tackles only a specific aspect, i.e., the optimization of the location of \acp{UAV} in the network, without considering all the others.

The paper \cite{CZX14} overviews area coverage problems in \ac{UAV} networks. The main impacting factors, i.e., coverage capabilities, \acp{UAV} mobility and lifetime, network connectivity, and presence of obstacles in the environment, are discussed with specific reference to the coverage types that are classified based on the \acp{UAV} motion (hover, fly, stall, etc.) and the network deployment. Other types of constraints in the drones utilization, i.e. energy consumption, coordination among \acp{UAV} and path planning problems, are summarily described together with path models adopted to study dynamic coverage problems.
%Also this survey is focused only on the coverage problems of the \ac{UAV} networks, neglecting all the other issues related to this topic.

The mmWave technology is analyzed in \cite{ZZW+19}. It overviews the issues arising from the utilization of mmWave communication in \ac{UAV} networks. The main research challenges are discussed, especially with respect to channel modeling and estimation/acquisition strategies. The adoption of mmWave in cellular networks is also discussed with reference to aerial-to-ground communication issues that include both the \ac{UAV}-to-\ac{BS} and \ac{UAV}-to-\ac{UE} scenarios. Scheduling and sharing of physical (time/frequency) resources among users are taken into account.
%The main drawback of such an approach is to carry out a detailed investigation of a specific aspect of networks of drones. In this case, only a technology is analyzed, neglecting the others, and only some physical layer aspects are discussed.

It is noteworthy that papers \cite{SK17,KCZ+18,FMT+17,CGT+18,CZX14,ZZW+19} analyze a specific aspect of the \ac{IoD} scenario: papers \cite{SK17,KCZ+18} refer to the only network and physical layer respectively; the survey \cite{FMT+17} deals with surveillance in the underwater environment. The works \cite{CGT+18,CZX14} summarily review optimization strategies \cite{CGT+18} and coverage problems \cite{CZX14}. The paper \cite{ZZW+19} is focused on the research perspectives in mmWave communication technologies.

%%%%%%%%%%%%%%%%%%%%%%%%%%%%%%%%%%%%%%%%%%%%%%%%%%%%%%%%%%%%%%%%%%%%%%%%%%%%%%%%%%%%%%%%%%%%%%
\begin{figure}[htbp]
	\centering
	\includegraphics[width=0.5\columnwidth]{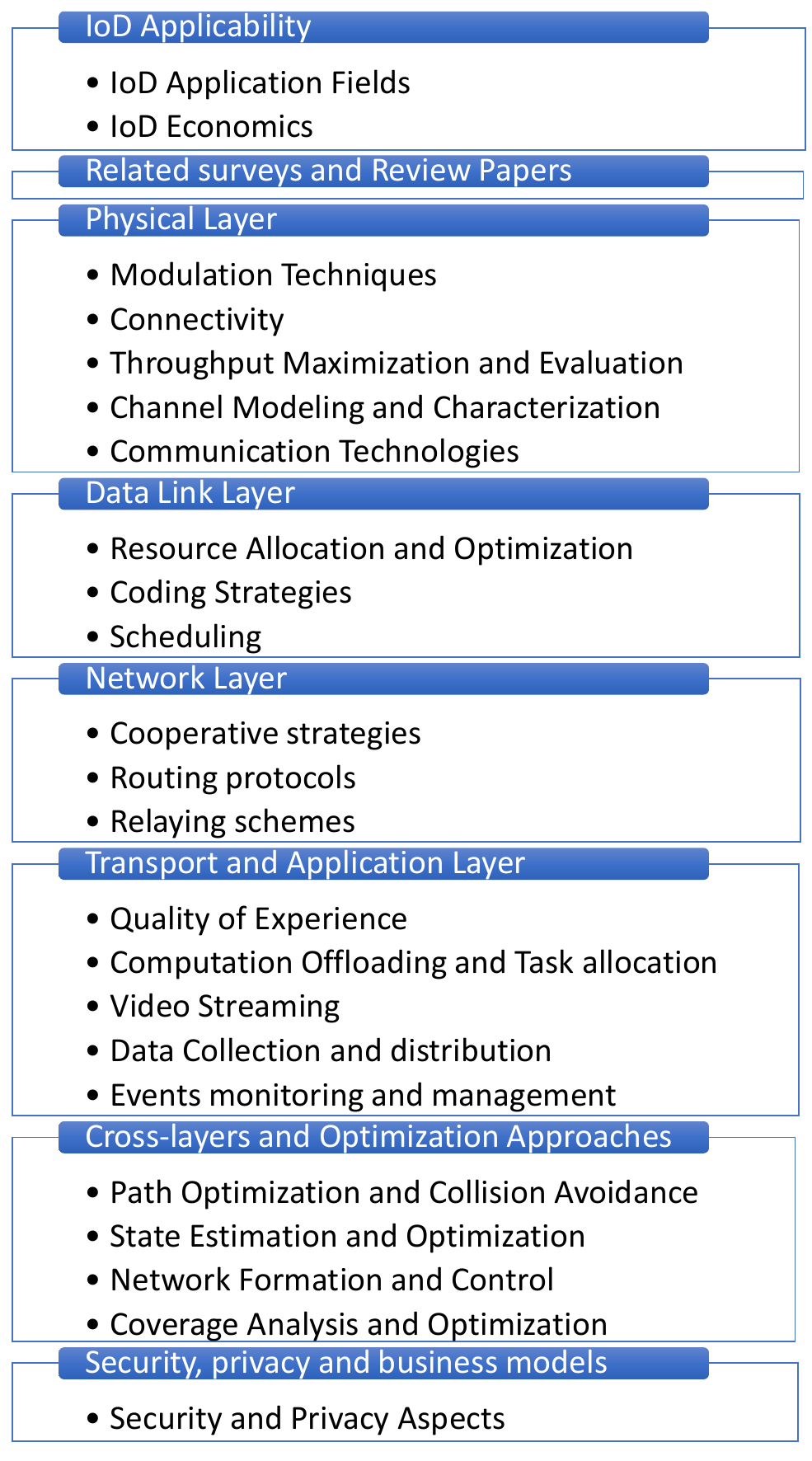}
	\caption{Overall organization of the work.}
	\label{fig:roadmap}
\end{figure}
%\begin{table*}[htbp]
%	\centering
%	\includegraphics[width=\textwidth]{img/surveys}
%	\caption{Comparison among the related surveys.}
%	\label{tab:ref_surveys}
%\end{table*}
\begin{sidewaystable}[htbp]
%	\centering
	\caption{Comparison among the related surveys.}
	\label{tab:ref_surveys}
	\begin{tabular}{l|c|c|c|c|c|c|c|c|c|c|}
		\cline{2-11}
		& \cite{SK17} & \cite{KCZ+18} & \cite{FQD+18} & \cite{GJV16} & \cite{MTA16} & \cite{CZX14} & \cite{FMT+17} & \cite{CGT+18} & \cite{ZZW+19} & \multicolumn{1}{l|}{This survey} \\ \hline
		\multicolumn{1}{|l|}{\ac{IoD} Application Fields}  &   &  &  &  &  X & X &   &   &  & X                                \\ \hline
		\multicolumn{1}{|l|}{Satellite Communications and \ac{6G}}  &   &  &  &  &  & &   &   &  & X                                \\ \hline
		\multicolumn{1}{|l|}{\ac{IoD} economics}                                                                    &   &   &   &   &   &   &   &   &    & X                                \\ \hline
		\multicolumn{1}{|l|}{Modulation Techniques}  &   &  &  &  &  & &   &   &  & X                                \\ \hline		
		\multicolumn{1}{|l|}{Channel Modeling and Characterization}  &   & X & X &   &   & X &   &   & X  & X                                \\ \hline
		\multicolumn{1}{|l|}{Communication Technologies}             &   &   &   &   & X & X &   &   & X  & X                                \\ \hline
		\multicolumn{1}{|l|}{Connectivity}                                                                     &   &   & X & X &   & X & X & X &    & X                                \\ \hline
		\multicolumn{1}{|l|}{Throughput Evaluation and Optimization} &   &   & X &   &   &   &   &   &    & X                                \\ \hline
		\multicolumn{1}{|l|}{Resource Allocation and Optimization}   &   &   & X & X & X &   & X & X &    & X                                \\ \hline
		\multicolumn{1}{|l|}{Scheduling}                                                                       &   &   & X & X &   & X &   &   & X  & X                                \\ \hline
		\multicolumn{1}{|l|}{Cooperative Strategies}                                                           & X &   &   &   &   &   & X &   &    & X                                \\ \hline
		\multicolumn{1}{|l|}{Routing Protocols}                                                                & X &   &   & X & X &   &   &   &    & X                                \\ \hline
		\multicolumn{1}{|l|}{Relaying Schemes}                                                                 & X &   & X &   &   &   &   &   &    & X                                \\ \hline
		\multicolumn{1}{|l|}{\ac{QoS} and \ac{QoE}}                                                                      &   &   &   &   &   &   &   & X &    & X                                \\ \hline
		\multicolumn{1}{|l|}{Task Allocation}                                                                  & X &   &   &   & X &   &   &   &    & X                                \\ \hline
		\multicolumn{1}{|l|}{Video Streaming}                                                                  &   &   & X &   &   &   &   &   &    & X                                \\ \hline
		\multicolumn{1}{|l|}{Data Collection and Distribution}       &   &   & X &   & X &   & X &   &    & X                                \\ \hline
		\multicolumn{1}{|l|}{Novel Communication Technologies and Paradigms}  &   &  &  &  &  & &   &   &  & X                                \\ \hline
		\multicolumn{1}{|l|}{Cross-Layer and Optimization Approaches}  &   &  &  &  &  &  &   &   &  & X                                \\ \hline
		\multicolumn{1}{|l|}{Events Monitoring and Management}       &   &   &   &   & X &   & X &   &    & X                                \\ \hline
		\multicolumn{1}{|l|}{Security and privacy aspects}                                  &   &   & X &   &   &   &   &   &    & X                                \\ \hline
		\multicolumn{1}{|l|}{Drones Identification and Classification}                         &   &   &  &   &   &   &   &   &    & X                                \\ \hline
		\multicolumn{1}{|l|}{Security in Drone Networks}                         &   &   &  &   &   &   &   &   &    & X                                \\ \hline
%		\multicolumn{1}{|l|}{Security Threats and Countermeasures}                         &   &   &  &   &   &   &   &   &    & X                                \\ \hline
%		\multicolumn{1}{|l|}{Security in specific application scenarios}                         &   &   &  &   &   &   &   &   &    & X                                \\ \hline
%%		\multicolumn{1}{|l|}{Authentication Mechanisms}                         &   &   &  &   &   &   &   &   &    & X                                \\ \hline
%		\multicolumn{1}{|l|}{Security Frameworks}                         &   &   &  &   &   &   &   &   &    & X                                \\ \hline
%		\multicolumn{1}{|l|}{Privacy Preservation}                         &   &   &  &   &   &   &   &   &    & X                                \\ \hline
%		\multicolumn{1}{|l|}{Security at Physical Layer}                         &   &   &  &   &   &   &   &   &    & X                                \\ \hline
	\end{tabular}
\end{sidewaystable}

The discussion carried out so far reveals that there is the need to conduct a wider study on the theme. Since the reported works are mainly focused on a subset of the technological aspects of the \ac{IoD}, or specific phenomena, a more cross-cutting classification approach is needed. This is assumed by leveraging a thorough layer-by-layer characterization of mathematical problem formulations, technologies employment, and implementation proposals.
As an outcome, the challenging issues and open research questions will be characterized in networks of drones, as remarked in Section \ref{sec:intro}.
In Figure \ref{fig:roadmap}, the overall organization of the present contribution is shown. Furthermore, for each Section, a detailed representation of the taxonomy will be reported, showing its structure in detail.
In Table \ref{tab:ref_surveys}, the surveyed state of the art is proposed, with a detailed overview of the theme that each work describes and tackles.

%% file: src/2-applications.tex
\section{IoD applicability}\label{sec:applications}
\begin{figure}[htbp]
	\centering
	\includegraphics[width=0.7\columnwidth]{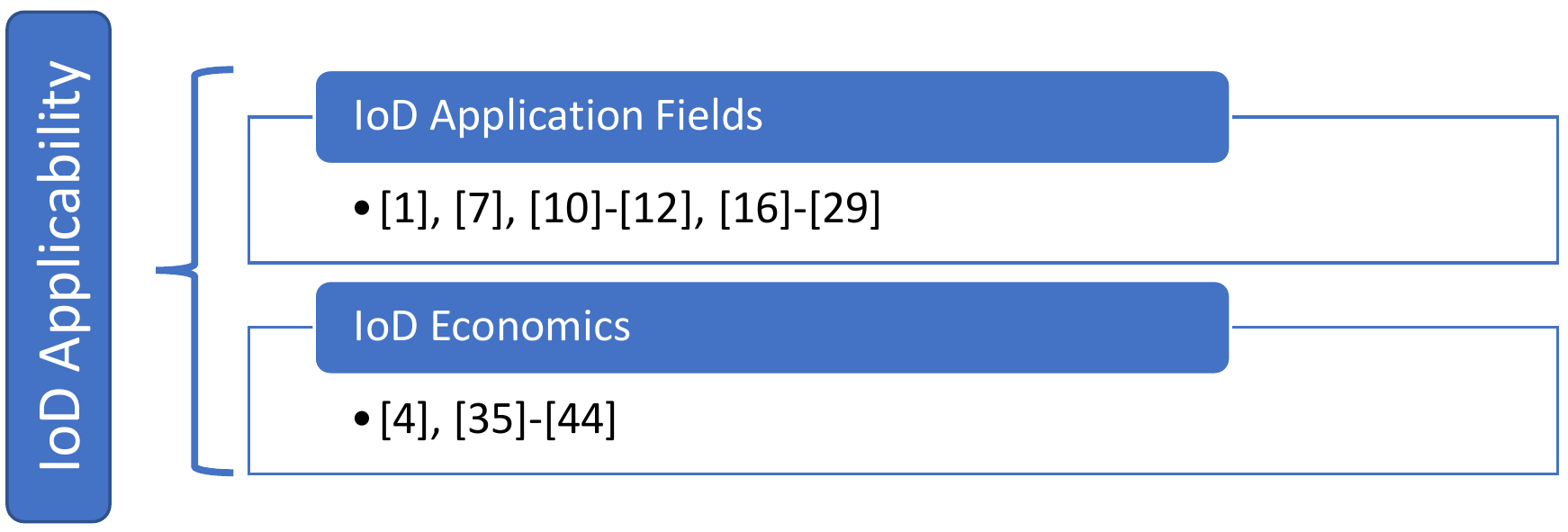}
	\caption{IoD Applicability Taxonomy.}
	\label{fig:tax_applicability}
\end{figure}

This Section discusses the applicability of the \ac{IoD} paradigm with a detailed analysis of all the application fields that have been proposed so far. To dig deeper in the subject, the employment of drones in the many applications that may benefit from their adoption are dealt with from an economical perspective. The overall organization of this section is reported in Figure \ref{fig:tax_applicability}.

\subsection{IoD Application Fields}
As previously anticipated, drones employability in different operational contexts has been widely discussed so far \cite{GBW16, AA17, ASY+18,BA15, CCB+18, CSM17, CSG+18, CZX14, FCC+18, IBG20, KQF+19, MGA+17, MTA16, saad2019vision, VCA+19, VCP17, ZVK+17, WCC17,EBEID201811}.
\begin{sidewaystable}[htbp]
%	\centering
	\caption{Summary of the main applications drones are involved in.}
	\label{tab:drones_applications}
	%	\resizebox{\textwidth}{!}{
	\begin{tabular}{|l|l|l|l|}
		\hline
		\multicolumn{1}{|c|}{\textbf{Application Area}}                                       & \multicolumn{1}{c|}{\textbf{Activity}}                                                                                                    & \multicolumn{1}{c|}{\textbf{Open Challenges}}                                                                                                                   & \textbf{References}                                                                                                                      \\ \hline
		Law Enforcement                                                                       & \begin{tabular}[c]{@{}l@{}}Public Safety\\ Orthographic surveillance\end{tabular}                                                         & \begin{tabular}[c]{@{}l@{}}Multiple sensing units\\ Long lasting missions\\ Seamless connectivity\end{tabular}                                                  & \begin{tabular}{@{}l@{}}\cite{VCA+19, AA17, BA15,VCP17, FCC+18}\\ \cite{MTA16,ZVK+17, ASY+18, KQF+19, EBEID201811}\end{tabular}                               \\ \hline
		Civil Engineering                                                                     & \begin{tabular}[c]{@{}l@{}}Aerial Photogrammetry\\ Area mapping (GIS)\\ Land development/construction\\ Science and research\end{tabular} & \begin{tabular}[c]{@{}l@{}}Multiple sensing units\\ High quality video imaging\end{tabular}                                                                     & \cite{IBG20, MGA+17}                                                                                                                            \\ \hline
		Logistics Tracking                                                                    & \begin{tabular}[c]{@{}l@{}}Unmanned cargo\\ Process enhancement\\ Proactive Maintenance\end{tabular}                                      & \begin{tabular}[c]{@{}l@{}}Multiple sensing units\\ Interaction with environment\end{tabular}                                           & 
		\begin{tabular}{@{}l@{}}\cite{VCA+19, saad2019vision, MGA+17,AA17,BA15}\\ \cite{VCP17,FCC+18, CSM17, KQF+19}\end{tabular}                     \\ \hline
		Military Applications                                                                 & \begin{tabular}[c]{@{}l@{}}Search and rescue\\ Surveillance\\ Border control\end{tabular}                                                 & \begin{tabular}[c]{@{}l@{}}Long lasting missions\\ Multiple sensing units\\ Seamless connectivity\\ Autonomous decision making\end{tabular} & 
		\begin{tabular}{@{}l@{}}\cite{VCA+19,AA17, BA15,VCP17, FCC+18}\\ \cite{MTA16,ZVK+17, EBEID201811}\end{tabular} \\ \hline
		Air Traffic Controlling                                                               & \begin{tabular}[c]{@{}l@{}}Traffic control\\ Security\\ Weather forecast\\ \acp{ITS}\\ Science and research\end{tabular} & \begin{tabular}[c]{@{}l@{}}Multiple sensing units\\ Seamless connectivity\end{tabular}                                                                          & 
		\begin{tabular}{@{}l@{}}\cite{VCA+19, saad2019vision, MGA+17, AA17,VCP17}\\ \cite{FCC+18, MTA16, ASY+18, KQF+19, EBEID201811}\end{tabular} \\ \hline
		Public Safety                                                                         & \begin{tabular}[c]{@{}l@{}}Search and rescue\\ Disaster Management\end{tabular}                                                           & \begin{tabular}[c]{@{}l@{}}Real-time tracking\\ High quality video imaging\end{tabular}                                                                         & 
		
		\begin{tabular}{@{}l@{}}\cite{GBW16, VCA+19, saad2019vision, MGA+17,AA17,VCP17}\\ \cite{FCC+18, MTA16, CZX14,CSM17, CCB+18, KQF+19}\end{tabular} \\ \hline
		Entertainment                                                                         & \begin{tabular}[c]{@{}l@{}}TV series and films\\ Concert and Events live streaming\\ Flight clubs\\ Selfies\end{tabular}                  & \begin{tabular}[c]{@{}l@{}}High quality video imaging\\ Artificial Vision\\ Objects and Pattern Tracking\end{tabular}                                           & \cite{VCA+19, saad2019vision, IBG20,VCP17, FCC+18, CCB+18,WCC17}                                    \\ \hline
		\begin{tabular}[c]{@{}l@{}}Industrial monitoring\\ Processes enhancement\end{tabular} & \begin{tabular}[c]{@{}l@{}}Smart Agriculture and Pharming\\ Power lines/grids\\ Oil and Gas\end{tabular}                                  & \begin{tabular}[c]{@{}l@{}}Multiple sensing units\\ High quality video imaging\\ Interaction with environment\end{tabular}              & 
		\begin{tabular}{@{}l@{}}\cite{GBW16, NNB+20, IBG20, MTA16, AA17,BA15}\\ \cite{CCB+18,ASY+18,FCC+18, ZVK+17,VCP17}\end{tabular} \\ \hline
	\end{tabular}
	%}
\end{sidewaystable}
In many cases, the studies propose a detailed analysis of their capabilities, the possibilities they enable, the persistent configurations and, above all, the involved technologies.
In the telecommunications field, drones are of interest because they can be used both in emergency situations and in the ordinary conditions, i.e., management of data traffic.
As an application example, let be an event with many connected users who want to send data through their 4G connections. In such a situation, the phone cells are typically overloaded. In this scenario drones can act as mobile \acp{BS}, delivering the service with smart and self-adaptive deployment plans to support the load in excess, thus granting connections continuity.
These capabilities can also be extremely useful in those contexts where service interruptions and signal losses are due to network outages. In case of environmental disasters, such as earthquakes and floods, the massive use of fleets of drones may be the only possibility of timely intervention to restore connectivity and in the meantime guarantees the acquisition of data, i.e., video images or environmental parameters.
In non-critical scenarios, several examples of drones employment can be done, such as photogrammetry and aerial survey of both buildings and large wooded areas, as well as in smart agriculture. Here, the detailed analysis of high-resolution images can provide several kinds of information, like for example the quantity of pesticides or irrigation, the effects of abundant rain, the quality of the soil, the presence of diseases, etc. This information allows targeted interventions to save entire crops and improve the quality of production.
\begin{figure}[htbp]
	\centering
	\includegraphics[width=0.8\columnwidth]{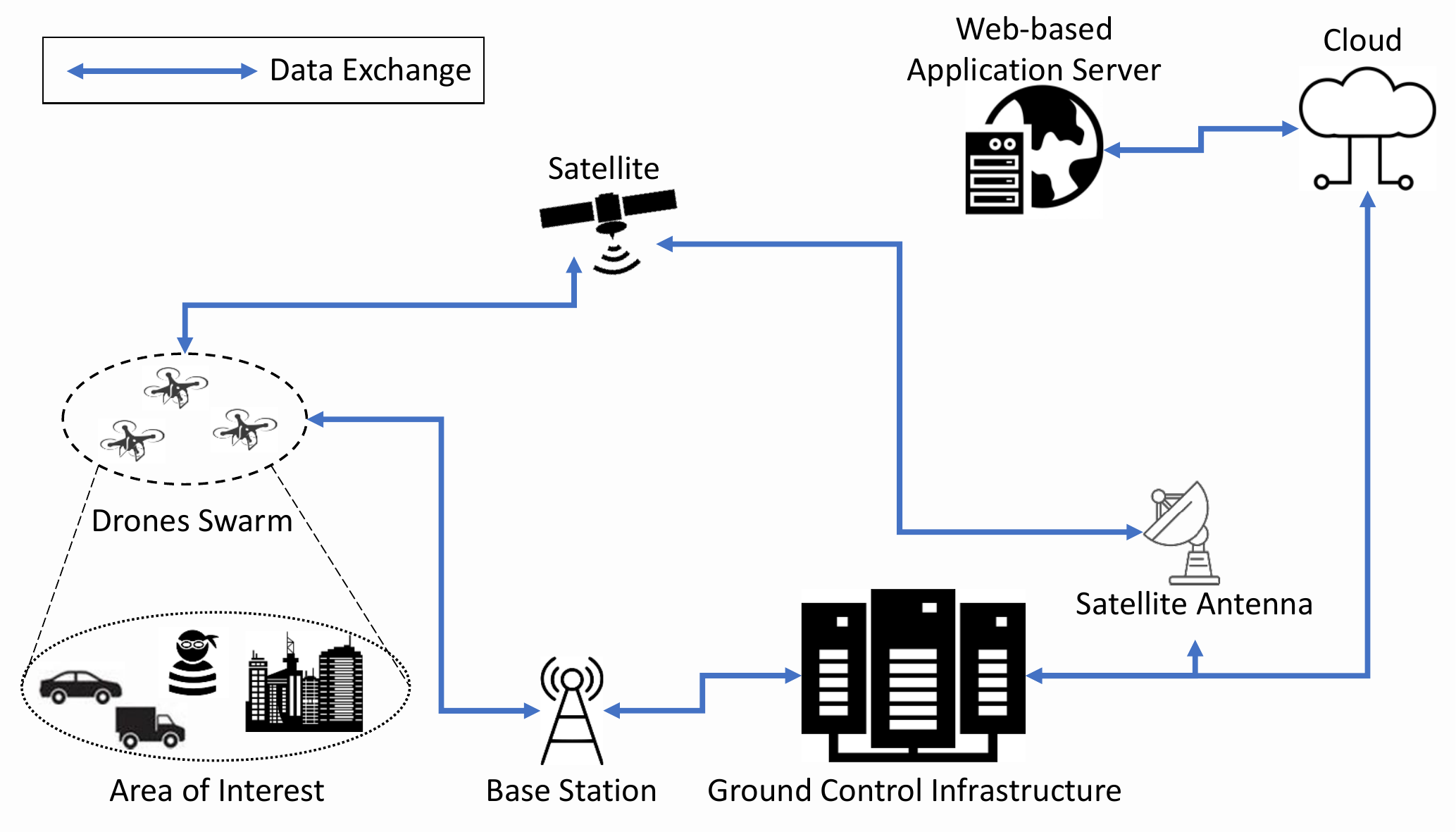}
	\caption{Reference \ac{IoD} architecture.}
	\label{fig:iodtypical}
\end{figure}

The employment of drones has been extended in smart mechanism by integrating the \ac{M2M} communications. This enables drones to interact with sensors and service providers. Such a mechanism can be further evolved towards coordination schemes in between the sensor services with its upper layers, which is, in general, a fog and/or cloud environment. In the latter context, distributed intelligence and cooperation among logical nodes becomes of utmost importance to offload the elaboration provided by \acp{UAV}.

In Figure \ref{fig:iodtypical}, the reference architecture for the \ac{IoD} is depicted. In it, a swarm of \acp{UAV} is flying over an area of interest. While on flight, drones are communicating to the to a Ground Control Infrastructure leveraging a certain number of \acp{BS}.
As suggested by the preliminary studies on \ac{6G} technology, drones are communicating with satellites \cite{CSG+18,Deebak2020,saad2019vision,BMM+20, deebak2020drone, SBC20, YXX+19}. In particular, satellites are mainly communicating to drones to be activated and/or remotely controlled. These control messages are sent by the ground control infrastructure via satellite antennas. The whole set of information regarding the mission plans, the achieved mission goals and, eventually, remote control are handled by web-based applications in a distributed fashion leveraging cloud computing.
In order to enhance the data transmission rate and the system capacity, \ac{6G} network has led a key role in pioneering the introduction of system intelligence using \ac{AI} \cite{deebak2020drone}.

As clearly comes out from this description, the \ac{IoD} architecture must be considered as in a general-purpose fashion. In particular, the main claim of the \ac{IoD} is that the network architecture is extremely versatile and able to meet several requirements at the same time. To clarify this, some of the main references of the present work are reported in Table \ref{tab:drones_applications}, where the application fields are classified together with the roles the drones are covering \cite{CSG+18}.
A detailed discussion of these points will be given in the remainder of the present work.

\subsection{\ac{IoD} Economics}
Beyond the novelty, and possible innovation, that drones may lead in the wide context of civil applications, \acp{UAV} represent a promising perspective in several industrial segments \cite{BII18,PV19, castillo2015projecting, koiwanit2018analysis, NNB+20}.
The economic impacts associated to the wide employment and integration of \acp{UAV} is certainly going to have impacts on job creations. At the same time, industries will be able to cut costs from more effective means of transportation and distribution. The applications may vary for several reasons. First of all, the ever increasing technological readiness of drones is continuously increasing in terms of performances. For instance, drones are able to fly regardless of the payload: in fact, at the time being, \acp{UAV} may be equipped with cameras (e.g., photo or video), sensors, and radars.
As for flight time and autonomy, drones are more and more able to fly for longer periods of time, thanks to enhanced battery and energy supply systems.

The work presented in \cite{PDE+18} discusses the possible economic impact of the introduction of drones in multiple applications. Starting from what's stated in \cite{jenkins2013economic}, and further confirmed in \cite{keaveney2019single,clarke2014regulation,castillo2015projecting}, the employment of drones at scale would bring \$80 billion by 2025. This is motivated by the fact that a plenty of interesting functionalities that drones may expose when employed \cite{andersen2020strategic}, such as improving public support given their intrinsic autonomy. Moreover, security can be enhanced according to the fact that social and economic impacts can be the result of the fact that surveillance or search and rescue activities can be carried out almost on their own by the drones.
Another interesting contribution is \cite{Balafoutis2017}, in which the main focus is on smart agriculture and smart farming technologies. This work discusses the details about precision application technologies, which implies the need for technologies able to work with variable-rate. The rationale for this variability may be application dependent and context dependent.
For example, data acquisition related to crop monitoring when diseases or irrigation problems are detected may lead to sudden intervention , thus saving entire production sectors. At the same time, precision irrigation and weeding has been widely recognized as one of the leading pillar of smart farming. Irrigating on demand and constant monitoring of soil humidity Can be enabled by the analysis of aerial images.	\cite{pricewaterhousecoopers2018skies} provides similar forecasts, but in the UK market. Even though regulations and production sectors are different from other countries, what this study has in common with others is the fact that they share the same opinions in terms of the drones’ role. Defined as game-changer, drones are expected to improve industrial processes at scale by £42 billion by 2030. The study discuss is the role of technology in saving lives in harsh environments and in the case of hard working condition, providing concrete examples in case of utilities and oil and gas plants.
The detailed analysis carried out in this contribution envisions several strategic fields in which the impact of drum is estimated to be of importance. in particular:
\begin{itemize}
		\item Public and Defense, Health, technology.
		\item Technology, media and telecommunications.
		\item Construction and manufacturing
		\item Financial, insurance, professional and administrative services.
		\item Transport and logistics.
		\item Agriculture, mining, gas and electricity.
\end{itemize}

In a nutshell, drones may have direct impact on \cite{PV19}:
\begin{itemize}
	\item Consumer applications, thus including individuals, non-commercial and non-professional. A forecast on drones shipment for this sector indicates a total of 29 million units by 2021.
	\item for enterprise drones, Business Insider Intelligence expects shipments to reach 805,000 in 2021 with a five-year CAGR of 51\% from 102,600 in 2016.
	\item Government drones fall into two categories: military and public safety. Among the two, the former market can be considered as the most mature. The U.S. military has been using drones for combat since 2001, and it rapidly expanded its drone fleet to more than 7,000 by 2012. The Department of Defense budget in 2016 allocated \$2.9 billion for more than 50 new drones for combat and surveillance, according to the Bard Center for the Study of Drones.
\end{itemize}
Industrial players may significantly cut costs and/or improve operations for the enterprises that will introduce or adopt them. As for the industrial applications that will mostly benefit from the introduction of drones, the most significant impacts will be on \cite{PV19}:
\begin{itemize}
	\item Infrastructure (45.2B \$).
	\item Agriculture (32B \$).
	\item Security (10B \$).
	\item Media and Entertainment (8.8B \$).
	\item Telecommunications (6.3B \$).
\end{itemize}

As for delivery, is has been estimated that there could be an increment on transportation speed ($\sim 50\%$), a positive effect on environmental impact (lower by almost $50\%$), increased control over package delivery flow ($\sim 40\%$), and safety ($\sim 30\%$) \cite{koiwanit2018analysis}.
Service provisioning may be related to systems and methods for delivering mail and goods using \acp{UAV}.

To sum up, the number and kind of fields that may benefit from the employment of drones are so numerous that their potential impact in automating and optimizing processes can only be roughly foreseen. At the same time, as happens every time a groundbreaking technology becomes available, massive drones employment may radically change every business not only in terms of time-to-market for products but, even more, in terms of process design and optimization.

%% file: src/4-phy.tex
\section{Physical Layer}\label{sec:phy}
\begin{figure}[htbp]
	\centering
	\includegraphics[width=0.9\columnwidth]{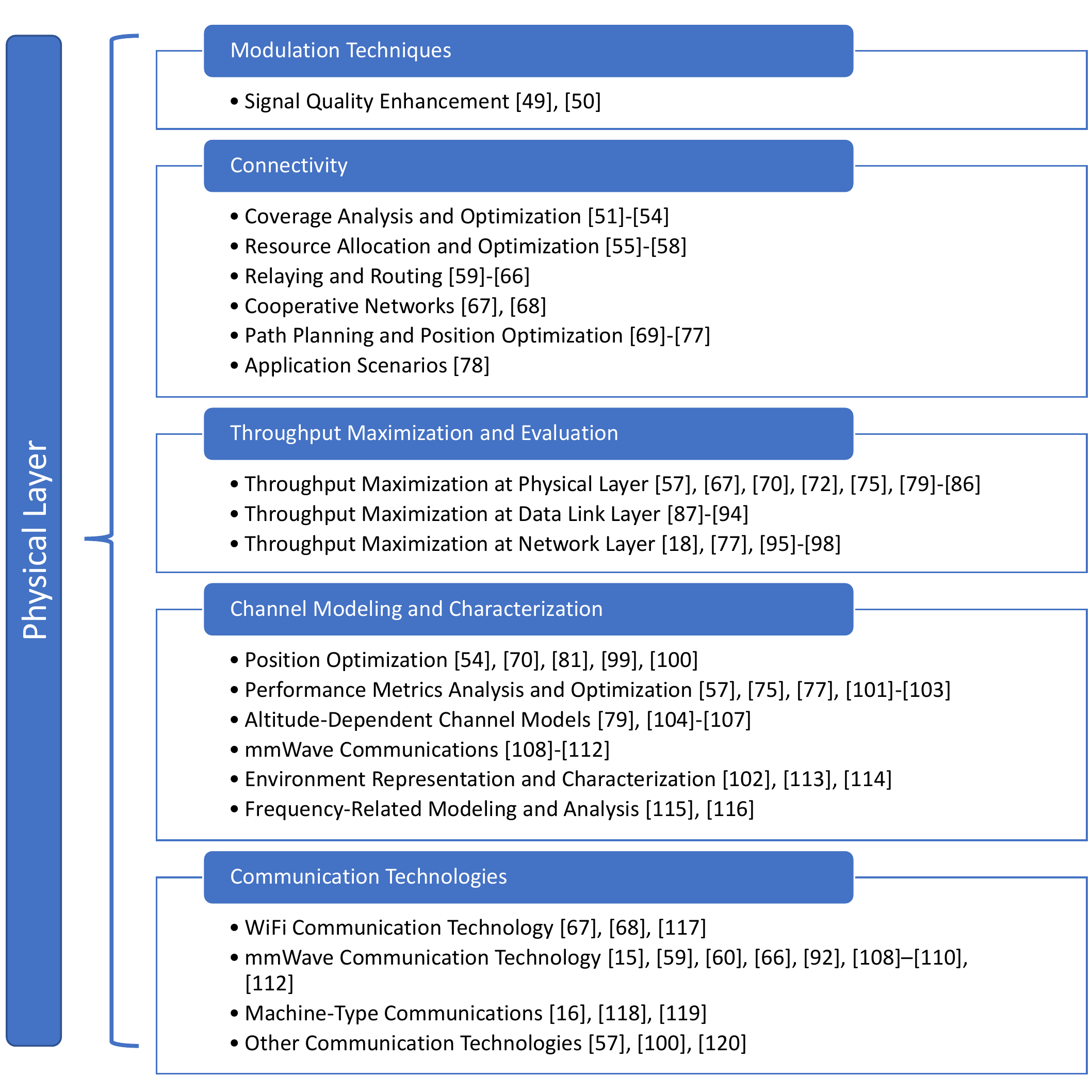}
	\caption{Physical Layer Taxonomy.}
	\label{fig:taxphy}
\end{figure}

This Section discusses all the technological aspects and research challenges connected to the \ac{PHY}. In particular, the main topics of interests are: (i) Modulation techniques, (ii) Connectivity, (iii) Throughput maximization, (iv) Channel modeling and characterization, and (v) Communication technologies. The Section closes up highlighting the lessons learnt on this theme.
The overall organization of this section is reported in Figure \ref{fig:taxphy}.

\subsection{Modulation Techniques}
There are some papers that analyze modulation techniques in the context of \ac{UAV} communications \cite{EEA+19,SAE18,KPR19,AKL+18,JTY20}. They are discussed in the following subsections.

%%%%%%%%%%%%%% SIGNAL QUALITY ENHANCEMENT %%%%%%%%%%%%%%%
\subsubsection{Signal Quality Enhancement}
More strictly related to the analysis of modulation schemes are the contributions \cite{AKL+18,JTY20}.
A modulation technique that  increases the efficiency of the data link between \acp{UAV} is investigated in \cite{AKL+18}. The analyzed modulation scheme is the \ac{SC-FDM}, with the important goal to fulfill long-range communication and long battery life requirements for networks of \acp{UAV}. This scheme is already known in uplink communications for \ac{LTE} systems, but in this study is contextualized to \ac{UAV} communication, by comparing its performance with the classical \ac{OFDM} modulation scheme. Performance analysis is carried out in terms of transmission power optimization, bit error rate and pulse shaping.
An analysis of the \ac{OFDM} modulation is considered in \cite{JTY20}, to propose a noise suppression method for \ac{UAV} communications. The proposed method filters electromagnetic impulse noise at receiving side; then it performs \ac{OFDM} demodulation to suppress noise. The problem is tackled analytically, using models for the various electromagnetic pulses to analyze their influence on the \ac{OFDM}-based communication. The electromagnetic pulse suppression strategy is then presented; it relies on a module that performs the separation of noise from the signal before \ac{OFDM} demodulation, that acts in combination with a dual antenna method to reduce the bit error rate and increase the quality of the received pictures.

%%%%%%%%%%%%%%%%%%%%%%%%%%%%%%%%%%%%%%%%%%%%%%%%%%%%%%%%%%%%%%%
%%%%%%%%%%%%%%%%%%%%%%%%%%%%%%%%%%%%%%%%%%%%%%%%%%%%%%%%%%%%%%%
\subsection{Connectivity}
Connectivity is a topic widely analyzed in literature, as testified by the works \cite{GKD17,RD16,LC17,MSB+15,CSY17,MSB+17,AGS+16,NXL+17,KYW+17,QHG+16,GSY17,PJS+17,OLL+16,OLZ+17,ZJF+18,GPM+18,MKM+15,RDC17,SKS+17,RIG16,AS15,CSB19,LXN+17,MSS16,MSB+16,ARC+16B,CS17_02,CMS+17}. They are mainly focused on: (i) analysis and optimization of coverage and resource allocation, (ii) relaying and routing, (iii) cooperative networks, (iv) path planning and position optimization and (v) connectivity issues in specific application scenarios, as described in what follows.

%%%%%%%%%%%%%% COVERAGE ANALYSIS AND OPTIMIZATION %%%%%%%%%%%%%%%
\subsubsection{Coverage Analysis and Optimization}
\begin{figure}[!h]
	\centering
	\includegraphics[width=0.7\columnwidth]{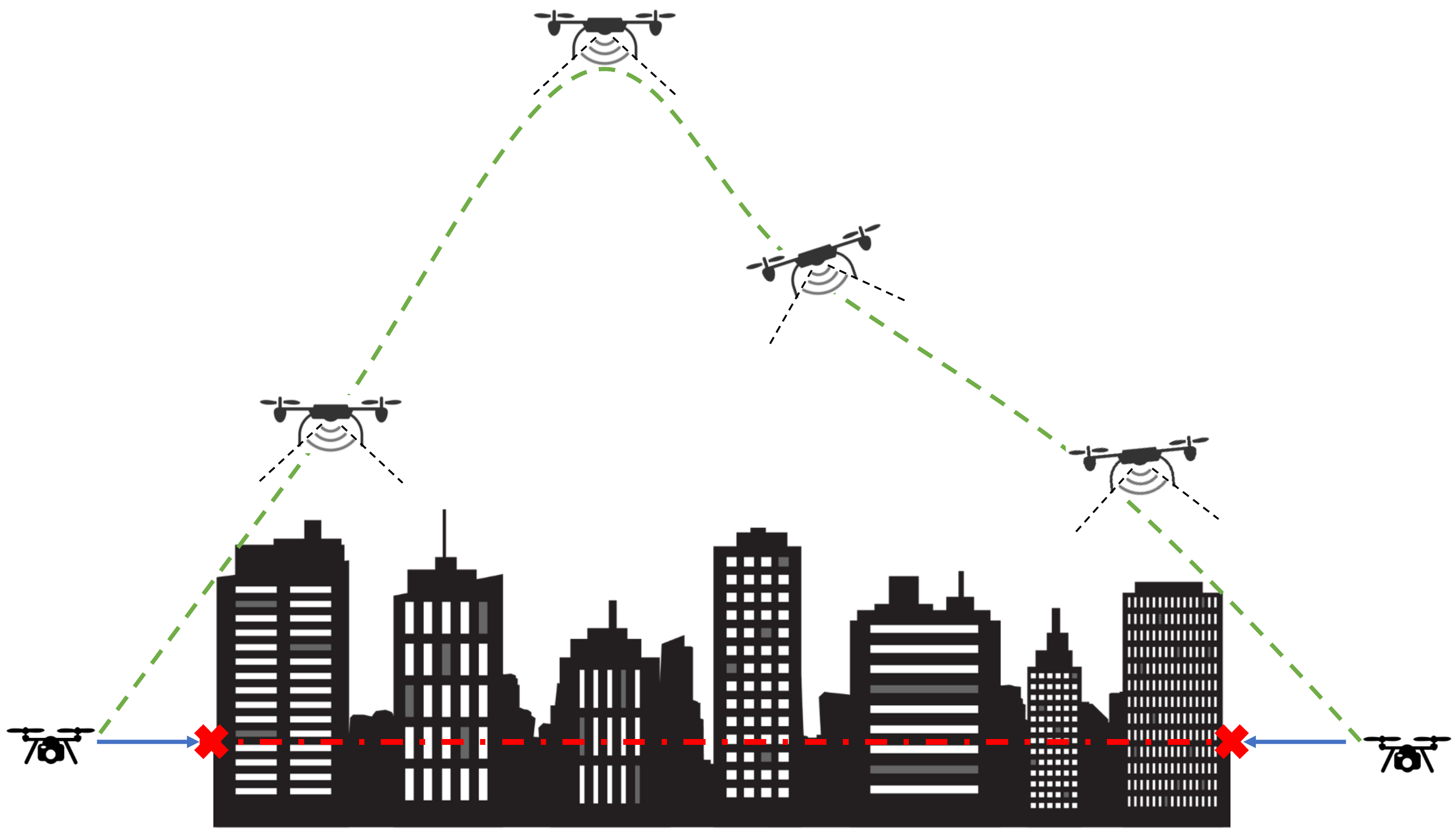}
	\caption{\ac{NLoS} communications enabled by drones.}
	\label{fig:buildings}
\end{figure}

A connectivity analysis can be fruitfully exploited to study coverage performance of networks of drones in different contexts; this is the approach followed by the works \cite{GKD17,RD16,LC17,MSB+15}.
The work \cite{GKD17} proposes a coverage analysis that takes into account connectivity between \acp{UAV} and ground users. In this work, \acp{UAV} at low altitudes provide services to users falling into a coverage area whose shape is influenced by the \acp{UAV} directional antennas. The coverage analysis is carried out statistically as a function of the main \ac{UAV} parameters, among which, the \ac{LoS} connection and \ac{LoS} and \ac{NLoS} (as graphically represented in Figure \ref{fig:buildings}) interference between \acp{UAV} and ground users.
A similar approach is followed in \cite{RD16}, where a coverage probability analysis is carried out in a multi-\ac{UAV} scenario. In particular, here \acp{UAV} provide service to a ground receiver. The analytical expression takes into account both \ac{LoS} and \ac{NLoS} components, and aims at carefully capturing the influence of the dominant interferer on the receiver, with respect to the other interference contributions from the rest of \acp{UAV}.
%, that are instead derived through approximations to simplify the analytical expression of the coverage probability.
Another coverage study in a network that includes \acp{UAV} with relay functionalities is developed in \cite{LC17}, where the connection probability is studied in conjunction with coverage capabilities, for different types of traffic.
%In the work \cite{MSB+16}, connectivity is expressed as the \ac{LoS} probability in a network where \acp{UAV} act in combination with a \ac{D2D} underlaid network. The goal of this work is a coverage and rate performance analysis.
\ac{LoS} and \ac{NLoS} connections are modeled statistically also in \cite{MSB+15}, to study the coverage performance of small cells of drones providing wireless access to ground users. The probability of \ac{LoS} and \ac{NLoS} connections are used to analyze the drones altitudes that maximize the coverage extension and minimize the transmission power.

Even if drones connectivity can be considered as a peculiar aspect of the physical layer, it also impacts on other layers of the protocol stack. For this reason, studies on connectivity are carried out also in several studies proposing strategies for resource allocation and optimization, relaying and routing, cooperative networks, path and position optimization, and analyzing specific application scenarios.

%%%%%%%%%%%%%% RESOURCE ALLOCATION AND OPTIMIZATION %%%%%%%%%%%%%%%
\subsubsection{Resource Allocation and Optimization}
The works \cite{CSY17,MSB+17,AGS+16,NXL+17} take into account connectivity in the framework of resource allocation and optimization strategies. In these works, the probabilities of \ac{LoS} and \ac{NLoS} connections are used to describe theoretically \ac{A2A} and/or \ac{A2G} links. Several factors are considered to model the link connectivity, like the environment, placement and density of buildings, and relative positions of \acp{UAV} and users.
In \cite{CSY17} the model is exploited to develop a resource allocation algorithm in a cache-enabled \ac{UAV}-based network, that optimally allocates the spectrum bands so that the queue stability of users and their bandwidth requirements are satisfied.
In \cite{MSB+17} the link connectivity model is exploited to maximize the average number of bits received by ground users, subject to constraints on the \acp{UAV} maximum flight times and with fair resource allocation guarantees.
In the work \cite{AGS+16} both \ac{LoS} and \ac{NLoS} path loss models are exploited to implement an algorithm that increases the throughput of a public safety network which leverages, at the same time, the unlicensed LTE spectrum of \acp{UAV} and the Wi-Fi spectrum of ground access points. The goal is to balance the load on the basis of the link quality of users.

%%%%%%%%%%%%%% RELAYING AND ROUTING %%%%%%%%%%%%%%%
\subsubsection{Relaying and Routing}
In the works \cite{KYW+17}\cite{QHG+16}\cite{GSY17}\cite{PJS+17}\cite{OLL+16}\cite{OLZ+17}\cite{ZJF+18}\cite{GPM+18} connectivity is analyzed in the context of relaying and routing strategies.
The work \cite{KYW+17} proposes a solution to increase the connectivity among devices adopting mmWave communication, through mobile relaying. To this end, the optimal position of the mobile nodes is found so that the link quality is maximized and the communication range is increased accordingly.
In the work \cite{QHG+16} the loss of connectivity is analyzed to justify the importance of an accurate prediction of node movements for an efficient design of the routing strategy. The proposed routing algorithm includes a feature that predicts the movements of the mobile node, to increase the message delivery ratio and reduce delay in case of short-lived connectivity.
The work \cite{GSY17} models the connectivity between two nodes, in terms of expected connection time, to build a routing protocol that dynamically updates the routing path based on the prediction of the connection times of drones.
In \cite{PJS+17} a route recovery strategy is proposed, that exploits \acp{UAV} as relays interconnecting terrestrial networks in disaster scenarios. The route recovery over damaged networks is based on a topology discovery algorithm that identifies clusters of nodes, and the relative cluster heads, depending on their connectivity capabilities.
In the routing protocol developed in \cite{OLL+16} the most connected path is found to route data between vehicles of a \ac{VANET} and \acp{UAV} assisting them. The path depends on the traffic density and the degree of connectivity on the road.
An extension of the same protocol developed in \cite{OLL+16} is proposed in \cite{OLZ+17}. It considers both routing between \acp{UAV} and vehicles and routing among \acp{UAV}. 
The \acp{UAV} connectivity is a requirement of the layered network architecture presented in \cite{ZJF+18}. The routing algorithm proposed in this work aims at minimizing latency while maximizing the \ac{PDR}. The connectivity information is exploited to improve the routing performance. 
A connectivity model is developed in \cite{GPM+18} to analytically describe the channel conditions of the back-haul links in 5G cellular networks that exploit the mmWave spectrum. The dynamic blockage of back-haul links is modeled in a network with \acp{UAV} acting as relay nodes. This aspect is part of a more complex analytical framework, that captures also the mmWave propagation characteristics and the \acp{UAV} mobility, to dynamically reroute data depending on the variable channel conditions.

%%%%%%%%%%%%%% COOPERATIVE NETWORKS %%%%%%%%%%%%%%%
\subsubsection{Cooperative Networks}
The works \cite{MKM+15,RDC17} analyze connectivity among drones in cooperative networks.
In particular, \cite{MKM+15} proposes a distance control algorithm that finds the optimum distance among drones that cooperate for data transfer tasks, to guarantee the stability of the wireless connection and, consequently, of drones communication.
The continuity of the connectivity is the main focus of the work \cite{RDC17}, that proposes a network infrastructure, composed by drones that exchange data with ground users to manage critical events. The goal is to guarantee the service continuity by keeping the connectivity among \acp{UAV} in case of nodes replacement. To this end, a modification of the \ac{OLSR} protocol, already known in literature, is proposed.

%%%%%%%%%%%%%% PATH PLANNING AND POSITION OPTIMIZATION %%%%%%%%%%%%%%%
\subsubsection{Path Planning and Position Optimization}
There are several works that analyze connectivity to develop path planning and position optimization strategies \cite{SKS+17,RIG16,AS15,CSB19,LXN+17,MSS16,MSB+16,ARC+16B,CS17_02}.
The work \cite{SKS+17} proposes the use of flying back-haul hubs to extend the coverage area in a 5G network. The hubs optimal location is wisely established, by solving an optimization problem, to provide connectivity between the core network and the ground \acp{BS}.
Also the algorithm proposed in \cite{RIG16} aims at finding the optimal position of \acp{UAV} used as \acp{BS} to provide connectivity to ground users. The algorithm makes use of a theoretical channel model that describes the connectivity of the \ac{A2G} links as a function of the propagation paths and the main features of the antenna arrays.
Another path planning algorithm that takes into account connectivity can be found in \cite{AS15}. This work studies the connectivity of a wireless network of balloons with \acp{UAV} acting as relays. The connectivity requirement is expressed as a function of the probability of the link outage among balloons.
%In this context, the connectivity of the network is guaranteed if there is a sufficient number of \acp{UAV} that can reconnect the subparts in which the balloon network has been split by the unreliable links. According to this, the work proposes 
Depending on this, the proposed path planning algorithm finds the optimal \acp{UAV} positions that maximize the network connectivity.
A path planning and resource allocation scheme that models connectivity in cellular-connected \acp{UAV} is found in \cite{CSB19}. In this work, analytical expressions of connectivity are used to model the interference level of the \acp{UAV} on the ground network and compute the optimal paths of \acp{UAV} that minimize it, together with the transmission delay and the time the \acp{UAV} employ to reach their destination.
An approach that takes into account connectivity of \ac{A2A} and \ac{A2G} links is found in \cite{LXN+17}, where connectivity constraints are introduced to study the path planning problem for an aerial sensor network connected to ground \acp{LTE-BS}. The connectivity constraints are expressed in terms of the outage probability of the link, which depends on the relative distance between aerial and ground nodes through \ac{A2A} and \ac{A2G} links. Based on this, the connectivity constraint is derived, which depends on the minimum link capacity.
%Finally, depending on the \acp{UAV} speeds and connectivity constraints, a cooperative motion strategy is proposed that keeps the capacity of the worst link higher than a pre-defined threshold.
The work \cite{MSS16} develops a mobility model of a fleet of \acp{UAV}, where each \ac{UAV} must keep the connectivity with its neighbor, taking also into account residual energy level and area coverage as decision criteria for the path planning procedure.
In \cite{MSB+16} the \ac{LoS} probability is used to derive the coverage probability of an \ac{UAV} providing service to ground users. The approach is similar to what developed in \cite{RD16}, but in this case the coverage probability expression is used to compute the optimal \ac{UAV} position that maximizes the coverage area and coverage lifetime.
The work \cite{ARC+16B} proposes a connectivity model between \acp{UAV} and ground nodes. Connectivity is modeled in terms of the outage probability of the terrestrial-aerial communication link, and the optimal \ac{UAV} height that maximizes the coverage area in presence of fading is found accordingly.
In \cite{CS17_02} the probabilities of \ac{LoS} and \ac{NLoS} connections are used for a theoretical model of \ac{A2A} and/or \ac{A2G} links. This model is used to design a network formation algorithm that guarantees the connectivity between each \ac{UAV} of the aerial network and a gateway node through at most one path.

%%%%%%%%%%%%%% APPLICATION SCENARIOS %%%%%%%%%%%%%%%
\subsubsection{Application Scenarios}
The approach proposed in \cite{CMS+17} differs from all the others described above, since the link connectivity is related to the users \ac{QoE}. Specifically, the \ac{LoS} connectivity, derived in terms of probability of connection loss, is exploited to model the transmission links in a network of \acp{UAV}. The goal is the optimization of the users \ac{QoE}, that in this case depends on the user data rate, delay, and device type.

\subsection{Throughput Maximization and Evaluation}
This section describes all the works proposing strategies for throughput optimization and evaluation, in different \ac{IoD} scenarios and at different layers of the protocol stack (physical, data link and network layer) \cite{MKM+15,WZZ17,LZZ17,RIG16,CG17,CSB19,WDZ19,HNK+16,ZZZ17,AFN+14,YKB11,AGS+16,MSB+16,YCL14,MBM14,SKU16,WGP14,TKN+17,WCL+18,SIL+16,KNK+17,FC12,LNW+16,BA15,LKY+16,CS17_02,QHS+16}.

%%%%%%%%%%%%%%%%%%%% PHYSICAL LAYER %%%%%%%%%%%%%%%%%%%%%
\subsubsection{Throughput Maximization at Physical Layer}
As expected, the majority of the works on throughput maximization and evaluation are focused on the physical layer of the protocol stack \cite{MKM+15,WZZ17,LZZ17,RIG16,CG17,CSB19,WDZ19,HNK+16,ZZZ17,AFN+14,YKB11,AGS+16,MSB+16}.

Some of the most significant contributions of the theme deal with position and path optimization strategies of \acp{UAV} \cite{MKM+15,WZZ17,LZZ17,RIG16,CG17,CSB19,WDZ19,HNK+16}.
More specifically, in \cite{MKM+15} throughput performance is studied to show the effectiveness of a distance control algorithm, with the goal of guaranteeing the stability of the wireless connection in \ac{A2G} links in case of drone replacement in the network. The proposed algorithm dynamically changes the drones position so that a minimum throughput requirement is satisfied. 
The work \cite{WZZ17} proposes a strategy for throughput maximization in a \ac{UAV}-based wireless network where a single \ac{UAV} is used as \ac{BS} to serve ground users. The proposed strategy optimizes jointly the \ac{UAV} path and the scheduling times of users communications. The solution of the optimization problem aims at maximizing the minimum throughput over all users, guaranteeing fairness among them. 
The same network scheme of \ac{UAV}-aided cellular network is proposed in \cite{LZZ17}. In this architecture, the \ac{UAV} offloads the ground \ac{BS} and serves the \acp{UE} at cell edge. The goal of this scheme is the maximization of the minimum throughput of the \acp{UE}, by jointly optimizing the \ac{UAV} trajectory, resource allocation and users partitioning between the \ac{UAV} and the terrestrial \ac{BS}. 
The issue of aggregate data rate maximization in the \ac{DL} channel is tackled in \cite{RIG16}. In this work, an approach is proposed that optimizes the hovering locations of \acp{UAV} acting as flying \acp{BS} to maximize the \ac{DL} aggregate rate, exploiting \acp{UAV} mobility and beam-forming techniques. This goal is obtained through a search algorithm, that finds the hovering locations with a low interference leakage, that maximize the \ac{SNR} of the ground users. 
The work \cite{CG17} considers a wireless network where \acp{UAV} act as relay. In this scenario, an algorithm is proposed that exploits the \ac{LoS} information to find the optimal \ac{UAV} positions that maximize the end-to-end throughput. These positions are found by means of a model that takes into account the signal propagation properties of the \ac{A2G} channel. The computational complexity of the proposed algorithm is shown to be low, if compared to other \ac{UAV} algorithms of position optimization.
The work \cite{CSB19} proposes a path planning algorithm for a network of cellular-connected \acp{UAV}. The problem is modeled as a dynamic game among \acp{UAV}, in which each \ac{UAV} learns the optimal path and the transmission power along the path, that optimizes the latency and the rate per ground user. To this end, the achievable data rate per user is modeled as a function of the channel characteristics, and the \ac{UAV} path is planned so that the interference of \acp{UAV} on the ground network and the transmission latency are both minimized. Allocation of resource blocks assigned to each user is managed accordingly, so that the per-user data rate is increased.
In the work \cite{WDZ19} an algorithm is proposed that optimizes the direction and distance of an \ac{UAV} acting as a \ac{BS}, depending on the amount of randomly moving users in a sector of the served cell. The users throughput is derived as a function of the \ac{SNR}; the optimal \ac{UAV} position is then computed so that the average throughput is maximized in the cell. 
The goal of the work \cite{HNK+16} is the throughput maximization in a multihop \ac{UAV} network. The network model proposed takes into account parameters related to the observation area and signal attenuation due to obstacles. These parameters are exploited to derive an analytical expression of the upper bound of the throughput. The optimal locations of \acp{UAV} that maximize this bound are found accordingly.\\
Some papers analyze throughput for the optimization of spectrum efficiency \cite{ZZZ17,AFN+14}.
More specifically, in \cite{ZZZ17} an optimization algorithm is proposed that takes into account data rate. The algorithm aims at maximizing the spectrum efficiency and energy saving of an \ac{UAV} that relays data between two \acp{GS}, by jointly optimizing the time allocation of data to be received and forwarded, and the \ac{UAV} speed and trajectory. The expression of the maximum instantaneous throughput is derived for the links between the \acp{GS} and the \ac{UAV} and is used to maximize the spectral and energy efficiency.
An approach for energy efficiency and throughput improvement can be found in \cite{AFN+14}. In this work, a study on the adaptive modulation is carried out for networks of \acp{UAS} collecting data from some sensor nodes. Based on the analysis of the impact of the UAS trajectory on the modulation scheme, a problem is formulated that maximizes the throughput per energy of \acp{UAS}, while preserving fairness among transmissions between sensor nodes and \acp{UAS}. The solution proposed exploits an approach based on the game theory.

An analysis on the impact of the link quality on system throughput is found in the works \cite{YKB11,AGS+16}.
In \cite{AGS+16}, a study is carried out on the use of unlicensed spectrum for \acp{UAV-BS} communications in a heterogeneous network with Wi-Fi ground \acp{AP}, to investigate the enhancement of achievable throughput in emergency situations. To this end, a game theoretic approach is proposed, and the related algorithm configures \acp{UAV-BS} transmissions to balance the load between \acp{UAV-BS} and ground \acp{AP}. The solution found takes into account the users link quality and the loads of \acp{UAV-BS} and \acp{AP}, aiming to ensure a satisfactory throughput for all the users.
The work \cite{YKB11} studies the impact of the characteristics of the wireless link on the system throughput in a \ac{UAV}-to-ground link. Several parameters are chosen, i.e., antenna orientation, \ac{UAV} height, yaw and distance between the \ac{UAV} and the ground access point, for experimental evaluation of throughput performance.

The paper \cite{MSB+16} analyzes rate and coverage performance for \ac{UAV}-based communication with \ac{D2D} links. To this end, an analytical model for the system sum-rate is first derived as a function of the \ac{UAV} height and number of users, in a scenario of a single static \ac{UAV} transmitting data to a \ac{D2D} network. Rate performance are derived as a function of \ac{SINR}, \ac{UAV} altitude and density of \ac{D2D} links, also showing the existence of an optimal \ac{UAV} altitude and a users density that maximize the rate.

%%%%%%%%%%%%%%%%%%%% LINK LAYER %%%%%%%%%%%%%%%%%%%%%
\subsubsection{Throughput Maximization at Data Link Layer}
Several works analyze throughput at link layer, even though the discussion of the proposal is carried out at physical layer \cite{YCL14,MBM14,SKU16,WGP14,TKN+17,WCL+18,SIL+16,KNK+17}.

In some papers, the throughput analysis is carried out in the framework of \ac{VLC} systems \cite{YCL14,MBM14,SKU16,WGP14}.
In \cite{YCL14}, an analytical model is developed for the \ac{CSMA/CA} protocol. It considers the relationship between the physical and MAC layers of the protocol stack. In this context, the paper shows that the multi-packet reception capability of the \ac{VLC} system, i.e., the capability to decode multiple packets coded with orthogonal sequences, reduces packets collisions and increases the number of simultaneous transmissions, also increasing throughput.
A throughput evaluation is carried out in \cite{MBM14}. The goal of this paper is to develop a IEEE 802.15.7 simulation module for ns-2, based on an accurate modeling of the physical layer and of the \ac{CSMA/CA} protocol at MAC layer. The mean throughput of a system composed by a coordinator servicing 12 nodes is derived, to validate ns-2 simulation results and compare the module performance with a Markov-based analytical model present in literature. 
A theoretical study of the IEEE 802.15.7 standard that takes into account throughput is found in \cite{SKU16}. The study is conducted ad MAC layer, through theoretical analysis and computer simulations. More specifically, a Markov chain is used to model the MAC layer with the use of the beacon frame and \ac{CSMA/CA}. In this context, closed-form expressions are derived for the main performance metrics, including throughput. The model is then validated through simulation.
Also the work \cite{WGP14} evaluates throughput performance at MAC layer in a \ac{VLC} system where two nodes are in reciprocal \ac{FoV}, and one of them continuously transmits data to the other. The MAC layer throughput is evaluated as a function of the frame payload (at a fixed distance) and the nodes distance (with a fixed frame payload). Throughput evaluations are carried out in two configuration scenarios, one-hop (direct communication between two nodes) and two-hop (three nodes, with one node forwarding packets at network layer between the other two).

Access schemes at MAC layer for throughput improvement are discussed in the works \cite{TKN+17,WCL+18}.
A radio access scheme thought for multi-\ac{UAV} networks with relaying capabilities is proposed in \cite{TKN+17}, to increase throughput. This work proposes a modification of the classical \ac{LTE-A} frame structure, together with a scheduling algorithm that dynamically changes the ratio of \ac{DL} and \ac{UL} subframes, to increase the efficiency of relay communication and, thus, the network throughput, especially in cases of longer distances.
The work \cite{WCL+18} discusses the integration of the mmWave communication in last-generation \ac{UAV}-based cellular networks. This work mainly focuses on the design of a multiple access scheme that, in combination with multiplexing techniques and beam-width selection strategies, improves the system throughput in presence of concurrent transmissions.

Resource allocation schemes for throughput optimization are discussed in \cite{SIL+16,KNK+17}.
In \cite{SIL+16}, a priority-based frame selection scheme is proposed, to optimize throughput and energy saving. This scheme can be exploited to improve the efficiency of data delivery in a sensor network that adopts a single \ac{UAV}. The framework comprises the frame selection scheme, another scheme that adjusts the \ac{CSMA/CA} contention window at MAC layer, and a routing protocol that selectively transmits frames based on their priority and the relative distance between nodes.
A throughput-aware resource allocation algorithm is proposed in \cite{KNK+17}. The model is suitable for data transmission between an \ac{UAV} and \acp{GS}. In this work, the effective throughput is used as requirement to optimally allocate communication resources, expressed as time slots, for communication between the \ac{UAV} and \acp{GS}. Throughput is computed as a function of different parameters, i.e., allocated time slots, modulation scheme and coding rate. The model aims at increasing the minimum throughput, as derived from the solution of the optimization problem.
\begin{figure}[htbp]
	\centering
	\includegraphics[width=0.7\columnwidth]{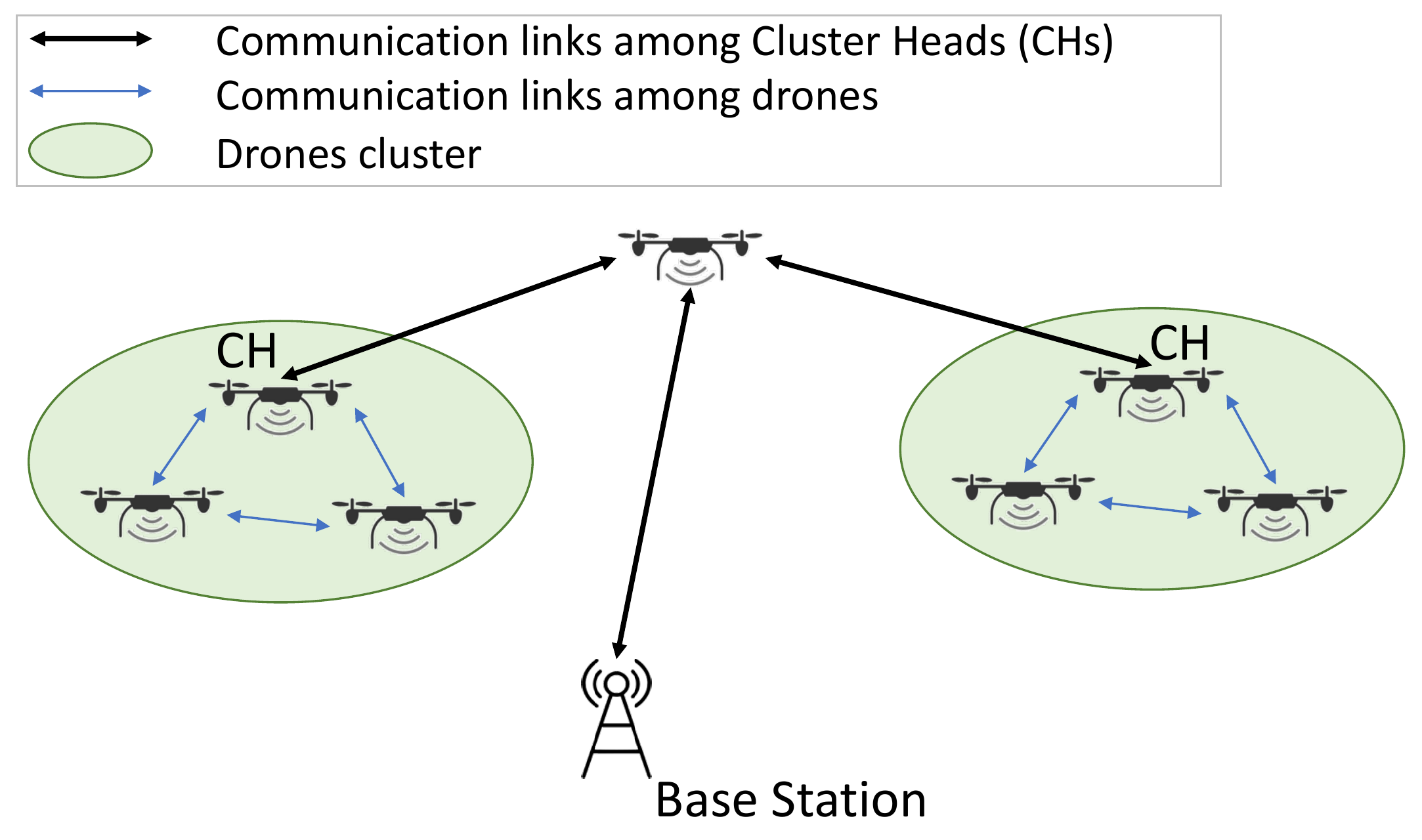}
	\caption{Communications among clusters/drones swarms and Base Station thanks to relay node.}
	\label{fig:cluster}
\end{figure}

%%%%%%%%%%%%%%%%%%%% NETWORK LAYER %%%%%%%%%%%%%%%%%%%%%
\subsubsection{Throughput Maximization at Network Layer}
Throughput is also considered in some papers discussing issues at network layer \cite{FC12,LNW+16,BA15,LKY+16,CS17_02,QHS+16}.

A couple of works focus on relay schemes for throughput improvement \cite{FC12,LNW+16}.
In \cite{FC12}, throughput maximization is discussed for multi-hop networks with clusters of relays, as illustrated in Fig. \ref{fig:cluster}. The goal of this work is to find the optimal deployment of relay nodes, in terms of number of hops and clusters locations, that minimize the outage probability, together with the optimal number of relays that maximizes the throughput.

The work \cite{LNW+16} proposes a relaying scheme that schedules transmission so that the energy consumption is minimized, at the same time maximizing the network throughput in presence of bit errors. To this end, an optimization algorithm is proposed, together with a simplification that provides a sub-optimal scheduling solution in a much faster way, at the same time guaranteeing throughput improvements.

Novel routing schemes that take into account throughput are developed in \cite{BA15,LKY+16}.
A geographical routing (also referred to as geo-routing) scheme is implemented in \cite{BA15}. The solution is known by literature and adopted in \ac{UAV}-based systems. Its goal, if compared to the classical implementation, is to increase throughput and decrease delay. The core of the proposal is a modification of the frame structure at physical layer of the geo-routing scheme with contention-free capabilities, that reduces overhead, thus reducing delay and increasing throughput.
A throughput stability analysis is found in the work \cite{LKY+16}. This work proposes a centralized routing protocol for a fleet of \acp{UAV}, that exploits the information coming from a ground control system to keep connectivity among drones and throughput stability, while shortening the route update times.

Network architectures that take into account throughput are found in \cite{CS17_02,QHS+16}.
The proposal of the work \cite{CS17_02} is a scheme in which \acp{UAV} act as a flying back-haul network connecting ground \acp{BS}. The flying network is dynamically formed by choosing the \ac{A2A} and \ac{A2G} links and the network topology that maximize a given utility function, through a network formation game approach. The chosen utility function considers the achievable data rate, the number of relayed packets, and delay in both \ac{DL} and \ac{UL} directions. 
A hierarchical network structure is proposed in the work \cite{QHS+16}, that provides a stable throughput and extends the network lifetime. It is composed by satellites, airships and \acp{UAV}. Based on an estimation of the weather conditions, a topology control and routing algorithm are proposed, that distribute the traffic load among the network nodes so that bottlenecks of link capacity can be avoided. In this way, an acceptable performance in terms of network lifetime and average throughput is guaranteed, for different weather conditions. 
%%%%%%%%%%%%%%%%%%%%%%%%%%%%%%%%%%%%%%%%%%%%%%%%%%%%%%%%

%%%%%%%%%%%%%%%%%%%%%%%%%%%%%%%%%%%%%%%%%%%%%%%%%%%%%%%%%%%%%%%
%%%%%%%%%%%%%%%%%%%%%%%%%%%%%%%%%%%%%%%%%%%%%%%%%%%%%%%%%%%%%%%
\subsection{Channel Modeling and Characterization}
This section describes all the works related to channel modeling and its analytical description together with the most relevant techniques for characterizing the communication medium in use by drones \cite{LWC+17,MSB+15,GGK+17,RIG16,CG17,MSB+16,MSB+16A,CS17_02,AGS+16,BEY18,LH17,ARC+16A,ARC+16B,GW15,CPA+17,ANM+17,KOG17,XXX16,RYR+19,WTZ+18,ZJ18,CEG+17,BEY18,BSY18,VYS+16,DGI+18}. Different models are presented aiming at mathematically describe several aspects of \ac{IoD} communication scenario, that are described in what follows. 

%in \ac{IoD} scenarios.

%%%%%%%%%%%%%%%%%%%%%% Position optimization %%%%%%%%%%%%%%%%%%%%%%%%%
\subsubsection{Position Optimization}
Some works analyze the communication channel for \ac{UAV} position optimization \cite{LWC+17,MSB+15,GGK+17,RIG16,CG17}.
Specifically, the works \cite{LWC+17,MSB+15} refer to the \ac{A2G} channel. 
An \ac{A2G} channel model is employed in \cite{LWC+17} in a network of \acp{UAV-BS}. This work proposes a position optimization algorithm in the 3D space, that minimizes the \ac{UAV} lifetime. In this scenario, the \ac{A2G} channel model is used to compute the average path loss as a function of the environment, the carrier frequency and the type of the link (\ac{LoS} or \ac{NLoS}). This model is exploited to find the optimal \ac{UAV} hovering altitude that minimizes the transmission power, and that contributes to find the optimal position of \acp{UAV} in the 3-Dimensional space.
A channel model is presented in \cite{MSB+15} to study small cells of \acp{UAV} acting as aerial \acp{BS}. More specifically, the \ac{A2G} channel model is developed, including \ac{LoS} and \ac{NLoS} paths and the related path loss. The probabilities of \ac{LoS} and \ac{NLoS} connections are derived. Further, those values are characterized in terms of average path loss, which depends on the \ac{UAV} height and coverage radius. This channel model is exploited to find the optimal \ac{UAV} altitude that maximizes the ground coverage area, and study the coverage performance of \acp{UAV-BS}.
A channel model is adopted in \cite{GGK+17} to propose a solution that minimizes the energy of micro-\acp{UAV} in cognitive radio systems. The goal of this work is to find the optimal drone position and power level that minimize the \acp{UAV} energy consumption. To this end, the path loss and channel gain are analytically derived, to model the \ac{A2G} and \ac{G2G} channels. These models are then exploited to solve the joint optimization problem of power allocation and drone positioning.
A narrow band channel model is introduced in \cite{RIG16} to study the optimal hovering locations of \acp{UAV-BS} to provide wireless connectivity during time-limited events. The model used for the wireless link is expressed as the sum of different propagation paths with different gains from \acp{UAV} to users. This model is then used to find the optimal \acp{UAV} positions that maximize the user \ac{SNR}. To this end, an antenna array is adopted; this solution is of interest since it introduces the contribution of beam-forming to the \ac{SNR} improvement and locations optimization.
The same approach of \cite{RIG16} is used in \cite{CG17}, where channel models are exploited to find the optimal \ac{UAV} position 
%based on fine-grained \ac{LoS} information. The goal of the algorithm proposed in \cite{CG17} is to find the \ac{UAV} position 
that maximizes the end-to-end throughput, but without knowing the whole radio map of the signal strength. To this end, two different channel models are employed: an \ac{A2G} channel model between the \ac{BS} and the \ac{UAV} that exploits the only \ac{LoS} propagation, and the channel between the \ac{UAV} and the user, that instead models the path loss exploiting both \ac{LoS} and \ac{NLoS} propagation.
%%%%%%%%%%%%%%%%%%%%%%%%%%%%%%%%%%%%%%%%%%%%%%%%%%%%%%%%%%

%%%%%%%%%%%%%%%%%%%%%%%%%%%%% Performance metrics analysis and optimization %%%%%%%%%%%%%%%%%%%%%%%%%%%%%%%
\subsubsection{Performance Metrics Analysis and Optimization}
Channel modeling is tackled for analysis and optimization of other performance metrics in \cite{MSB+16,MSB+16A,CS17_02,AGS+16,BEY18,LH17}.
In \cite{MSB+16,MSB+16A}, the \ac{A2G} channel is modeled to optimize coverage and rate performance in \ac{UAV} networks.
%To this end, the signal power received at users locations from an \ac{UAV}. 
As known by literature, the model used in these works takes into account three different groups of signals: \ac{LoS}, reflected \ac{NLoS}, and multiple reflected signals that cause multipath fading. Similarly to several other works on the topic, \ac{LoS} and \ac{NLoS} links are modeled separately to derive the received signal power at users locations. Based on this procedure, the goal of \cite{MSB+16} is to analyze the coverage and rate performance in a \ac{UAV}-aided network in presence of \ac{D2D} links, while the work \cite{MSB+16A} aims to derive the optimal coverage probability and deploy multiple \acp{UAV-BS} with directional antennas.
Other theoretical approaches that model the \ac{A2G} channel for analysis and optimization of performance metrics are found in \cite{CS17_02,AGS+16,BEY18}.
The work \cite{CS17_02} takes into account both \ac{A2A} and \ac{A2G} channel models to design a multi-hop backhaul network composed by \acp{UAV}. The \ac{A2G} channel is modeled statistically, through \ac{LoS} and \ac{NLoS} links that take into account the higher attenuation due to diffraction and shadowing. The only \ac{LoS} link is instead considered for the \ac{A2A} links. These models are then exploited to evaluate the \ac{SINR} of the \ac{A2G} link and the \ac{SNR} of the \ac{A2A} link, to derive the achievable end-to-end rate and design the \acp{UAV} network formation accordingly.
The same approach to the \ac{A2G} channel modeling adopted in \cite{CS17_02} is found also in \cite{AGS+16}, where the channel is modeled statistically, through \ac{LoS} and \ac{NLoS} path loss, as already done in the past literature. The goal of the work is to balance the load between \acp{UAV} and Wi-Fi ground \acp{AP} during emergency situations, based on the users link quality. The path loss model is exploited to compute the received \ac{SINR} that, together with the load balancing requirement, determines the best association between the user and the \ac{BS}.
Differently from the studies mentioned above, the work \cite{YKB11} focuses on \ac{A2G} channel measurements to characterize path loss exponents in the \ac{UL} and \ac{DL} channel, in two different environmental scenarios. Several parameters are considered in the field experiments: antenna orientation, \ac{UAV} height, yaw and distance. The impact of these metrics on the \ac{RSS} and throughput is evaluated.
A channel model is adopted in \cite{LH17}, that proposes a strategy for optimization of resource allocation so that the transmission delay of packets in an \ac{UAV}-based cellular network is minimized. Part of this study is dedicated to the communication channel model based on non-\ac{LoS} paths with Rayleigh fading. This model is then used to derive the power received at destination node, that is exploited to derive the transmission rate first, and then the mean packet arrival rate and transmission delay, that are optimized by solving an optimization problem.
%%%%%%%%%%%%%%%%%%%%%%%%%%%%%%%%%%%%%%%%%%%%%%%%%%%

%%%%%%%%%%%%%%%%%%%%% Altitude-dependent channel models %%%%%%%%%%%%%%%%%%%%%%
\subsubsection{Altitude-Dependent Channel Models}
Some works develop channel models that take into account the influence of \acp{UAV} altitude \cite{ARC+16A,ARC+16B,GW15,CPA+17,ANM+17}.
The works \cite{ARC+16A,ARC+16B} propose an \ac{A2G} channel model to study the effect of the altitude of an \ac{UAV-BS} on some performance parameters, i.e., power, capacity gain \cite{ARC+16A}, path loss and scattering \cite{ARC+16B}. The optimal \ac{UAV} height is found that brings to the best trade-off between the optimal values of these parameters, through an extension of the \ac{A2G} channel model that takes into account the \ac{UAV} height in path loss and fading. The validity of the proposed model is tested also for low-altitudes \acp{UAV}.
The work \cite{GW15} focuses on the analysis of \ac{A2A} channel features. Its goal is to extend the Rice model, which is usually involved in theoretical characterization of the \ac{A2A} channel. Such studies are usually conducted by varying the \ac{UAV} altitude, to provide a mode accurate description of the multipath effect. Based on real \ac{RSS} measurements, an estimation of the main Rice model parameters that describe the fading and multipath effects is derived, in dependence of the \acp{UAV} altitude. The Rice channel model is extended accordingly.
Modeling of the propagation channel for low-altitude \acp{UAV} is the goal of \cite{CPA+17}. The \ac{UAV} communication channel is studied through narrow-band and broadband measurement campaigns carried out in a suburban area, capturing the main channel characteristics: path loss, fading, power delay profile, multipath components, and root-mean-square delay spread. This work carries out an analysis of the \ac{A2G} channel characteristics for low \acp{UAV} heights.
The work \cite{ANM+17} is entirely focused on channel models. Its primary goal is to model the path loss exponents of the radio link between an \ac{UAV} acting as \ac{UE}, and the cellular network. The proposed model takes into account the \ac{UAV} height, which is exploited to model both the path loss exponent and the shadowing effect. The model is validated in a real scenario, using a \ac{UAV} connected to two different \ac{LTE} networks. 
%%%%%%%%%%%%%%%%%%%%%%%%%%%%%%%%%%%%%%%%%%%%%%%%%%%%%

%%%%%%%%%%%%%%%%% mmWave communication %%%%%%%%%%%%%%%%%%
\subsubsection{mmWave Communications}
Channel modeling and characterization are developed for mmWave communications in \cite{KOG17,XXX16,RYR+19,WTZ+18,ZJ18}.
A characterization of the \ac{A2G} channel is developed in \cite{KOG17} for \ac{UAV} communication in the mmWave band. The goal of this work is to analyze the main characteristics of the \ac{A2G} channel in different scenarios, each one with different kinds of mmWave propagation. Simulations are used to study the channel characteristics between a fixed ground station and a \ac{UAV} that moves along a linear trajectory at a constant velocity, for different \ac{UAV} heights. The most suitable model for the \ac{RSS} behavior is derived accordingly.
In the work \cite{XXX16} channel modeling is discussed for mmWave communication in \ac{UAV} cellular networks. Channel characteristics are illustrated, discussing the most relevant issues of mmWave propagation, in terms of propagation loss, scattering, multipath components, and reflections around the \acp{UAV}. The most suitable channel model is then chosen to derive beam-forming vectors for different users.
mmWave channel for \ac{DL} transmission is studied in \cite{RYR+19}, in a network of \acp{UAV} connected to cellular \acp{BS}. The model takes into account the multipaths and their angle of departure, the height and horizontal distance of the \acp{UAV} from the \acp{BS}, the transmit antenna structure at the \ac{BS} (composed by an antenna array), and the path loss of the \ac{DL} channel. The \ac{LoS} component of the paths is considered as predominant in this model, that is not suitable for low \acp{UAV} altitudes. The goal is the evaluation of energy efficiency in multiple access schemes for mmWave \ac{DL} communications.
The work \cite{WTZ+18} analyzes the coverage performance of cellular networks assisted by \acp{UAV} in mmWave communications. To this end, a channel model is introduced. It is designed as to exploit the only LoS component of the links between \acp{UAV} and mobile terminals. Due to the adoption of mmWave technology, the model takes into account the possibility that the \ac{LoS} signal is blocked by obstacles. Also small scale fading and shadow fading effects are considered, through probability distributions. This model is exploited to define the strategy of user association and a cooperative clustering scheme for \acp{UAV} that maximize the coverage performance of the network.
The characteristics of a mmWave channel in \ac{UAV} communications with an antenna array are investigated in \cite{ZJ18}. The channel is characterized through the \ac{IDFT} to obtain an analytical description of the channel behavior in the discrete time domain. The main parameters of this description are estimated to provide a channel tracking method that takes into account angle information, Doppler shift and \ac{UL}/\ac{DL} channel gains. 

%Research directions on channel modeling and characterization are discussed in \cite{ZZW+19} for \ac{UAV} mmWave communications. The main issues of channel modeling are tackled, with respect to their characteristics (propagation, Doppler effect due to \acp{UAV} mobility, environment, etc.). Issues for the estimation and prediction of channel parameters are also addressed. Finally, \ac{UAV} mmWave communication is discussed for \ac{UAV}-aided cellular networks.
%%%%%%%%%%%%%%%%%%%%%%%%%%%%%%%%%%%%%

%%%%%%%%%%%%%%%%%%%%%%%%%%%%% Environment representation and characterization %%%%%%%%%%%%%%%%%%%%%%%%%%%%%%%
\subsubsection{Environment Representation and Characterization}
Some papers exploit channel models in specific environmental scenarios \cite{CEG+17,BEY18,BSY18}.
An algorithm to learn the channel characteristics between a \ac{UAV} and a ground user is proposed in \cite{CEG+17}. To this end, the radio channel is characterized through a propagation model based on a representation of obstacles whose configuration
%that approximates the \ac{A2G} propagation by storing the configuration of obstacles 
is stored into a map, together with other channel parameters. These data are used to estimate and reconstruct the channel propagation characteristics based on a reduced set of measurements, with relatively small prediction errors in the radio map reconstruction.
In \cite{BEY18} an \ac{A2G} channel model is described and used for \acp{UAV-BS} providing wireless service to ground users. As discussed in other works \cite{CS17_02,AGS+16}, the probability of \ac{LoS} link between the \ac{UAV} and the user is evaluated, as a function of the environment and the horizontal and vertical distances between the \ac{UAV} and the user. Based on this, the path loss is modeled and used to evaluate the conditions in which the ground user can be served by the \ac{UAV}.
A type of \ac{A2G} channel model that takes into account detailed information (i.e., shapes, intersections and heights of buildings, reflection, diffraction, propagation mechanisms, etc.) is presented in \cite{BSY18}. Its main goal is to accurately characterize the path loss characteristics of the links between \acp{UAV-BS} and users, to derive a model suitable for altitudes both lower and higher that the buildings heights. The model parameters are optimized, based on the \ac{UAV} altitude, so that the root-mean-squared-error is minimized.
%%%%%%%%%%%%%%%%%%%%%%%%%%%%%%%%%%%%%%%%%%%%%%%%%%%%%%%%%%%%%%%%%%%%%%%%%%

%%%%%%%%%%%%%%%%%%%%%%%%%% Frequency-related modeling and analysis %%%%%%%%%%%%%%%%%%%%%%%%%%%
\subsubsection{Frequency-Related Modeling and Analysis}
A couple of works deals with the specific problem of characterizing the channel in the frequency domain \cite{VYS+16,DGI+18}.
The work \cite{VYS+16} proposes a technique to estimate and compensate the \ac{ICI} for high speed \acp{UAV}. Based on the analytical expression of \ac{OFDM} symbols, the channel matrix estimation is performed in the frequency domain, to accurately estimate the Doppler effect and the amplitude of all the paths. A classification framework that takes into account both spatial and spectral information is then adopted, to correctly evaluate the received data and the channel estimation and equalization method.
A theoretical channel characterization is proposed in \cite{DGI+18} to equalize channels that present selectivity in both time and frequency domains (the so-called double-selective channels). In this work, doubly-selective channel equalization is studied for the transmission of continuous phase modulated signals, that usually present a high computational complexity. The goal is to present new equalization techniques that simplify the equalizers, even at a cost of approximations. To this end, the channel impulse response is characterized through a causal \ac{FIR} system.
%%%%%%%%%%%%%%%%%%%%%%%%%%%%%%%%%%%%%%%%%%%%%%%%%%%%%%%%%%%%%%%%%%

% Cosa vuol dire? Quali sono i problemi?
%Cosa fa \cite{KCZ+18}
%Cosa fa \cite{ZZW+19}

\subsection{Communication Technologies}

The right choice of the wireless technology used for drones communication (\ac{LTE}, Wi-Fi, \ac{VLC}, WiMax, etc.) is of great importance to optimize the communication performance of networks of drones. It depends on many factors, including the type of the \ac{UAV} tasks, their duration, the environment, drones mobility, communication range, limitations on transmitted data or energy consumption:
%Accordingly, the chosen communication technology should take into account the specific scenario to maximize the advantages, and minimize the disadvantages, of drones communication. Specifically, 
technologies like cellular networks, WiMax and satellite communications guarantee a wide coverage area, high transmission bandwidth, high throughput and reliable connectivity, but are energy-consuming and often introduce high delays.
Counterwise, technologies like Bluetooth and Zigbee have a low power consumption, data degradation, implementation costs and delays, but also a low transmission bandwidth, throughput, and are more subject to interferences caused by obstacles.
All these aspects are covered in the works \cite{BDL13,MKM+15,RDC17,KYW+17,RYR+19,WCL+18,QHG+16,GPM+18,KOG17,XXX16,ZJ18,ZZW+19,AA17,OOF+17,ASA17,GGK+17,AGS+16,FAA18}.

%A \ac{D2D} communication scheme is introduced in \cite{LGY+15}, in public safety network scenarios. The goal of this work is to extend the coverage of the network in zones where relaying by means of ground stations is very difficult. In this work, \acp{UAV} act as relay nodes, providing connection via the \ac{D2D} technology. Multi-hop \ac{D2D} communication is exploited to find the optimal position of drones that maximize the data rate between a \ac{BS} and a terminal device.

%%%%%%%%%%%%%%%%%%%%%%%%%%%%% WiFi communication technology %%%%%%%%%%%%%%%%%%%%%%%%%%%%%%%
\subsubsection{WiFi Communication Technology}
\begin{figure}
	\centering
	\includegraphics[width=0.6\columnwidth]{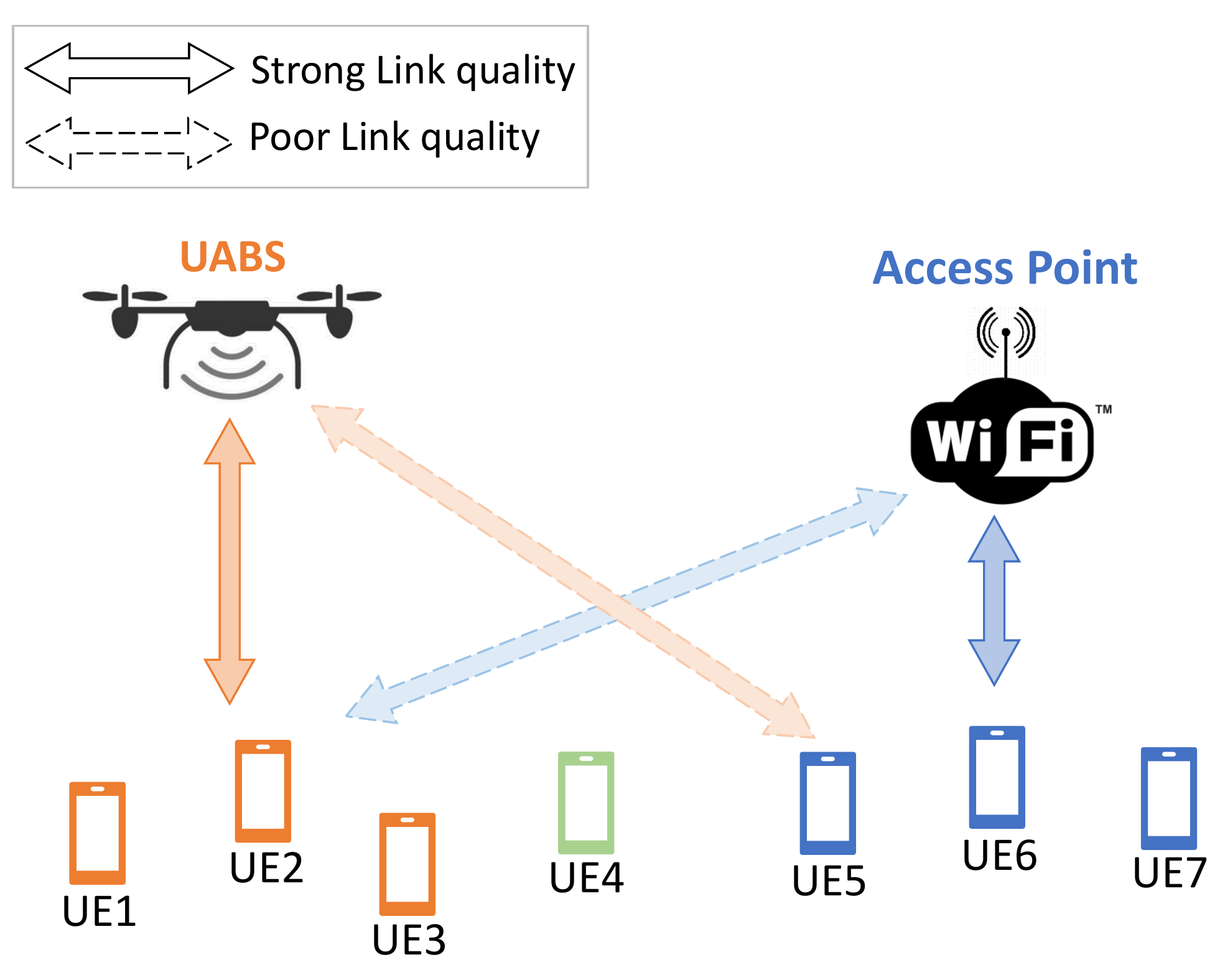}
	\caption{Cellular technology aided by an Unmanned Aerial Base Station (UABS) combined with Wi-Fi to provide connectivity to User Equipments (i.e., users).}
	\label{fig:wifi-cellular}
\end{figure}

WiFi technology is explicitly addressed in the papers \cite{BDL13,MKM+15,RDC17}. The work \cite{BDL13} mentions the Wi-Fi technology to build an ad-hoc network and implement a communication protocol for high-quality video transmission between drones sensing an area and mobile devices. The proposed protocol introduces an ad-hoc routing protocol that allows three drones to work together in the same network and expand the sensing area.
Wi-Fi is used also in \cite{MKM+15}, to implement a distance control algorithm, that finds the optimal distance between two \acp{UAV} that maximize the stability of the wireless communication, and minimize the number of \acp{UAV} utilized to cover a given area.
In \cite{RDC17} Wi-Fi is chosen as the most suitable technology for an air-to-ground communication network composed by \acp{UAV} and ground users, as illustrated in Fig. \ref{fig:wifi-cellular}. The goal of this study is to keep a service continuity when a \ac{UAV} is replaced because of limited battery lifetime. The choice of Wi-Fi is motivated with its ease of deployment, the use of unlicensed spectrum, and the need to support applications with different vales for throughput in the same network.
%%%%%%%%%%%%%%%%%%%%%%%%%%%%%%%%%%%%%%%%%%%%%%%%%%%%%%%%%%%%%%%%%%%%%%%%%%%%%%%%

%%%%%%%%%%%%%%%%%%%%%%%%%%%%% mmWave communication technology %%%%%%%%%%%%%%%%%%%%%%%%%%%%%%%
\subsubsection{mmWave Communication Technology}
Several works analyze the mmWave technology for \ac{UAV}-aided networks \cite{KYW+17,RYR+19,WCL+18,QHG+16,GPM+18,KOG17,XXX16,ZJ18,ZZW+19}.

The mmWave technology is introduced in the work \cite{KYW+17} to propose a strategy for an \ac{UAV} to find automatically and accurately its optimal relay position, so that data can be forwarded in short time. mmWave communication advantages (very high frequency and increased bandwidth, short wavelength that allows beam-forming) and disadvantages (attenuation in free space propagation, blocking effect due to obstacles and objects) are described in detail, to justify the need of large use of relaying in mmWave communications, and introduce the proposed strategy.

The adoption of mmWave in \ac{UAV}-aided cellular networks is investigated in \cite{RYR+19,WCL+18,QHG+16}. In \cite{RYR+19,WCL+18} multiple access schemes are discussed for mmWave communications. The mmWave channel model is introduced in \cite{RYR+19} to evaluate the energy efficiency of different multiple access schemes in a cellular-connected network that uses \acp{UAV}. The paper \cite{WCL+18} proposes a multiple access scheme design for 5G mmWave UAV networks acting at both the physical and MAC layers, where concurrent transmissions are multiplexed into a single beam so that interference in the single beam is transformed into concurrent transmissions, and interference among adjacent beams is reduced. In \cite{QHG+16}, different aspects related to the mmWave technology are investigated: the channel propagation characteristics, the implementation of beam-forming techniques that account for channel variations, the impact of the Doppler effect due to \acp{UAV} mobility, and the adoption of spatial-division multiple access for mmWave to increase the network capacity. Some algorithms for optimization of \acp{UAV} mobility are also considered, to avoid signal blockage. Finally, the relationship between \ac{UAV} positioning and user discovery is discussed, to analyze the reciprocal impact of one over the other.

The mmWave technology is exploited in \cite{GPM+18} to model backhaul links in \ac{UAV}-assisted networks in urban environments, with \acp{UAV} acting as relay nodes. This work develops a mathematical model that takes into account the multipath propagation and the blockage effect of the mmWave backhaul link, and the mobility of both signal blockers (humans, objects, etc.) and \acp{UAV}. These phenomena are modeled accurately by using the New Radio technology, which is a backhaul architecture with reconfiguration capabilities that reroutes dynamically data to alternative paths, to increase connectivity and data reliability. The study of mmWave backhaul performance is conducted in both spatial and temporal domains.

The main goal of \cite{KOG17} is to characterize the mmWave \ac{A2G} channel used by \acp{UAV} communications through simulations. Two mmWave bands are chosen for simulation, analyzing the \ac{RSS} and \ac{RMS-DS} of multipath components (MPCs) for different \ac{UAV} heights and environments (urban, suburban, rural, and over sea). Simulation results are also exploited to build a channel sounder.

Training and tracking strategies for mmWave \ac{UAV} communications are presented in \cite{XXX16,ZJ18}. The design of beam-forming codebooks with different beam widths is proposed in \cite{XXX16}, that starts from the analysis of the channel propagation characteristics, to fasten the beam-forming training and tracking phases. The impact of the Doppler effect due to \acp{UAV} mobility is investigated, also carrying out an analysis of spatial-division multiple access schemes and signal blockage events. A channel tracking strategy is presented in \cite{ZJ18}. Based on an analysis of the mmWave \ac{UAV} channel with a beam squint effect, the proposed channel tracking method exploits the information on angle and Doppler reciprocity to obtain the channel state information exploiting only one pilot symbol, and reducing the feedback overhead.
%%%%%%%%%%%%%%%%%%%%%%%%%%%%%%%%%%%%%%%%%%%%%%%%%%%%%%%%%%%%%%%%%%%%%%%%%%%%%%%%

%In the paper \cite{ZZW+19} \ac{UAV} mmWave communication is analyzed, by first illustrating the main issues concerning the channel characteristics and modeling, and then discussing the research challenges and the most promising solutions for the \ac{UAV} mmWave cellular networks. 
%

%%%%%%%%%%%%%%%%%%%%%%%%%%%%% Machine-type communications %%%%%%%%%%%%%%%%%%%%%%%%%%%%%%%
\subsubsection{Machine-Type Communications}
The works \cite{AA17,OOF+17,ASA17} analyze different technologies for \ac{MTC} in \ac{UAV} networks. In \cite{AA17} they are mentioned to discuss \ac{D2D} communications in public safety applications scenarios. This work is mostly a review paper on the recent advances in \ac{D2D} technology, and tackles the issues of device discovery, \ac{D2D} clustering and relaying, and Vehicle-to-Vehicle (V2V) communications. In this context, together with the \ac{D2D} technology as standardized by 3GPP rel.13, to increase coverage capabilities in critical situations, several other technologies are discussed, that allow \ac{D2D} communication in unlicensed spectrum: Wi-Fi, Bluetooth, ZigBee, etc. \cite{AA17}. All these technologies have also the advantages of high data rates, low latencies and implementation costs, and a wide adoption in mobile phones and devices. The paper \cite{OOF+17} focuses on \ac{mcMTC} and investigates some connectivity options like \ac{D2D} links and drone-assisted networks to satisfy the \ac{mcMTC} requirements. Specifically, \ac{LTE} technology is combined with Wi-Fi for \ac{D2D} communications between mobile devices and \acp{UAV}, to increase connectivity and data reliability. In addition, the mmWave technology is studied in the scenario of small cells of drones that act as \acp{AP} to increase the network coverage where the performance of the \ac{LTE} network is poor. The technologies introduced in \cite{ASA17} are used to describe the main specifications of localization techniques in a \ac{M2M} network composed, among others, by drones. The main goal of this paper is to propose a mathematical model that describes the structure of the \ac{M2M} network that adopts agents for efficient localization of nodes.
%%%%%%%%%%%%%%%%%%%%%%%%%%%%%%%%%%%%%%%%%%%%%%%%%%%%%%%%%%%%%%%%%%%%%%%%%%%%%%%%

%%%%%%%%%%%%%%%%%%%%%%%%%%%%% Other communication technologies %%%%%%%%%%%%%%%%%%%%%%%%%%%%%%%
\subsubsection{Other Communication Technologies}
Other communication technologies can be found in \cite{GGK+17,AGS+16,FAA18}.

The \ac{CR} technology is analyzed in \cite{GGK+17} as a promising solution to mitigate the main issues in spectrum utilization of the other wireless technologies (Wi-Fi, Bluetooth, or cellular communications) for \ac{UAV}-based applications. The \ac{CR} technology allows to access the spectrum in an opportunistic way, by sensing the spectrum through advanced radio techniques, and to transmit data over the spectrum bands that are not utilized by other transmissions. It also allows utilizing simultaneously the same spectrum to serve different users, but without exceeding an interference threshold. This paper proposes an energy-efficient optimization strategy, that finds that optimal \ac{UAV} location and transmit power level so that a \ac{UAV} can transmit data using the \ac{CR} technology, while respecting the data rate threshold of the spectrum owner.

The work \cite{AGS+16} investigates the possibility of using the \ac{LTE-U} technology for communication among \acp{UAV-BS}, to increase throughput by exploiting the unlicensed spectrum. In critical scenarios, \acp{UAV-BS} make use of \ac{LTE-U} to fill coverage gaps due to the damaged infrastructure that adopts Wi-Fi. The goal of this hybrid infrastructure is to quickly deploy an on-the-fly cellular network able to fill the gap of throughput requirements of the damaged Wi-Fi ground network, through a game theoretic approach that selects the radio access technology technique that achieves load balancing among LTE-U \acp{UAV-BS} and Wi-Fi \acp{AP}.

Optical wireless communication is studied in \cite{FAA18} for so-called \ac{FSO} communication systems, that exploit the optical signal for high data rate communications over relatively long distances. This paper focuses on \ac{FSO} systems where \acp{UAV} act as relays. The study is conducted by assuming that relays are buffer-aided, and they can move in the space (i.e., they are not stationary), and evaluating the impact of these two assumptions on the FSO system performance. Two application scenarios are also discussed and validated through simulation by taking into account the outage probability as performance metric.
%%%%%%%%%%%%%%%%%%%%%%%%%%%%%%%%%%%%%%%%%%%%%%%%%%%%%%%%%%%%%%%%%%%%%%%%%%%%%%%%
%In this Section ...
%In the works \cite{MTA16,ZZW+19} ... 
%
%%Cosa vuol dire? Quali sono i problemi?
%Cosa fa \cite{MTA16}
%Cosa fa \cite{ZZW+19}

\subsection{Comparison with Other Surveys}
Several topics related to the physical layer are discussed in the surveys \cite{CZX14,SK17,KCZ+18,MTA16,GJV16,ZZW+19}. 
Connectivity is discussed in \cite{CZX14,SK17}. Nevertheless, in both these works its analysis is only related to coverage strategies and in some specific application scenarios or network architectures.
Some works on throughput maximization and evaluation are mentioned in \cite{SK17,KCZ+18,MTA16}. The related analysis mostly refers to routing strategies, even if some insights are provided also for hybrid network architectures \cite{KCZ+18,MTA16}. A throughput analysis is also provided in \cite{GJV16,KCZ+18} but without analyzing novel strategies for its optimization.
Channel modeling and characterization is partially discussed in \cite{KCZ+18,ZZW+19}. Nevertheless, in the survey \cite{KCZ+18} the channel analysis and modeling is tightly bound to measurement systems for the \ac{A2G} links of low-altitude \ac{UAV} networks and platforms. Similarly, in the review work \cite{ZZW+19} channel characterization is studied with the only reference to the mmWave technology, pointing out the most significant challenges and opportunities related to this topic.
Together with the work \cite{ZZW+19}, also the survey \cite{MTA16} analyzes and classifies in detail the most widely used communication technologies for \ac{IoD} scenarios. The drawback of the approach followed in \cite{MTA16} is that this survey neglects other newer technologies, such as mmWave or \ac{VLC}.

If compared to the surveys mentioned above, the value added of this survey is to enhance and complete the analysis on the topics already covered by them, also adding topics that have not been discussed. More in detail, in the present survey the connectivity issues are also related to proposals for resource allocation and optimization, relaying and routing, cooperative networks and path planning and optimization strategies. A more thorough throughput analysis is carried out at network layer, more focusing on optimization strategies and covering also other aspects at physical and link layers. Channel modeling and characterization is described with reference to position (including the \ac{UAV} height) optimization strategies, adding the discussion on frequency-related modes, and collecting and describing more papers that propose an analytical representation and characterization of the environment. Communication technologies are broadly discussed, including mmWave, \ac{CR}, \ac{LTE}, \ac{FSO} and \ac{MTC} systems. Finally, papers on modulation techniques, not present in all the other surveys, are here introduced and described in detail.

\subsection{Lessons Learnt}
Important lessons emerge from the analysis of the physical layer aspects covered in the \ac{IoD} papers.

The modulation techniques analyzed for \ac{UAV}-based networks are not novel by themselves. The main novelty lies in their application in the \ac{FANET} context. As known in fact, drone detection and classification is a very important issue for \acp{UAV}, especially to face security issues, and this task can be accomplished by detecting the modulated signal at physical layer as testified by some of the surveyed papers. Efficient modulation techniques borrowed from other wireless network systems like \ac{LTE} also help to increase the efficiency of the transmitted signal, with particular respect to noise reduction and transmission power optimization.

The analysis of the drones connectivity and channel modeling are tightly bound together, and are both important to increase coverage performance, to optimally allocate resources at different layers of the protocol stack, and to increase the path reliability in routing, path planning and position optimization strategies. Nevertheless, connectivity and channel models and schemes tend to be not so accurate because they cannot take into account jointly several aspects (the presence of obstacles in urban environments, signal dispersion and interferences, height and mobility of drones, etc.), thus being valid only under simplifying hypotheses. Furthermore, the optimization problems for resource allocation, path planning and routing algorithms bring to solutions that are computationally expensive, or suboptimal.

Analysis and optimization of throughput allows to increase the system performance in data exchange and to optimize different metrics in \ac{IoD} systems (mutual distance, positions and paths of drones, spectrum and energy efficiency, network topologies). Also in this case, the throughput optimization in most cases translates into solving an optimization problem, whose solution is computationally expensive, or found approximately, or only in simplified scenarios. Furthermore, throughput maximization is counterbalanced by a higher energy consumption, which can become a serious drawback for battery-powered drones.

Performance of networks of drones can be increased also by choosing the most appropriate communication technology, depending on several factors like the specific environmental scenario, type and duration of tasks, communication range and energy consumption of drones, etc. Several works on this topic are mainly review papers; the most interesting aspects raised by papers on this topic lie in the exploitation of peculiarities of the chosen technology to optimize specific transmission aspects like spectrum utilization, data rate, and distance of nodes.

%
%To summarize, the future research directions related to the physical layer are:
%\begin{itemize}
%	\item Air-to-air channel models
%	\item Shadowing for large sized or small sized aircraft, building shadowing at low altitude, terrain shadowing beyond \ac{LoS} conditions.
%	\item Integration of multiple antennas, together with an evaluation of the most adequate number of elements and their location.
%	\item Channel stationarity and non stationary models.
%\end{itemize}

%% file: src/5-data.tex
\section{Data Link Layer}\label{sec:data}
\begin{figure}[htbp]
	\centering
	\includegraphics[width=0.6\linewidth]{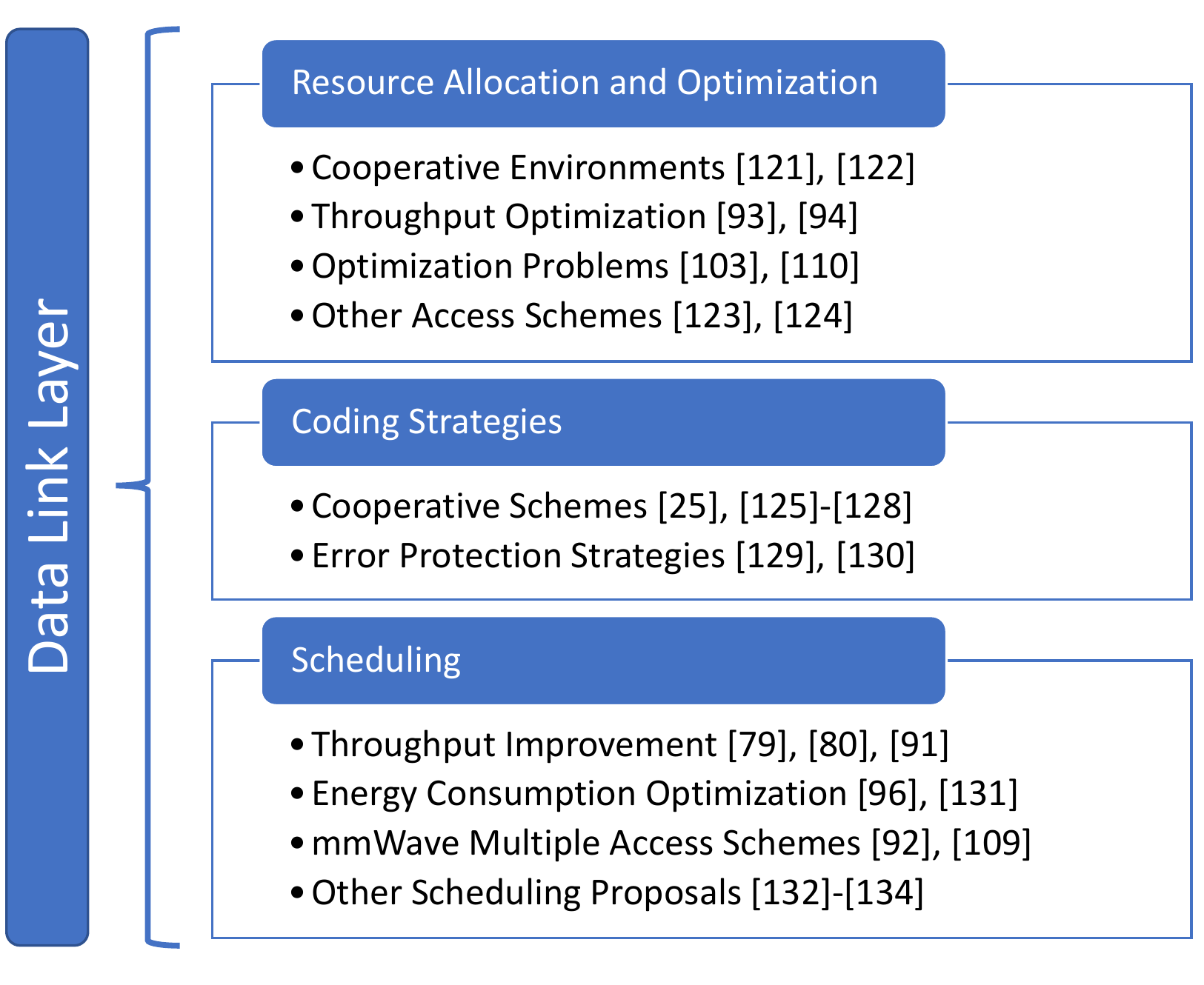}
	\caption{Data Link Layer Taxonomy.}
	\label{fig:taxdata}
\end{figure}

In this Section, some relevant issues at link layer are discussed. They are mainly focused on two different aspects: (i) the allocation and optimization of resources, and (ii) data scheduling. The Section closes up highlighting the lessons learnt on the theme.
The overall organization of this section is reported in Figure \ref{fig:taxdata}.

\subsection{Resource Allocation and Optimization}
\label{datalink_resource_allocation}
Papers on resource allocation at link layer are mostly focused on the optimization of different performance metrics, such as \ac{PDR}, packet error probability, bandwidth, delay, throughput, and energy efficiency.

%%%%%%%%%%%%%%%%%%%%%%%%%%%%%%%% Cooperative environments %%%%%%%%%%%%%%%%%%%%%%%%%%%%%%%%%%%%
\subsubsection{Cooperative Environments}
Schemes at \ac{MAC} layer in cooperative environments are proposed in \cite{MNP+17,M16}.
A packet loss tolerant algorithm is proposed in \cite{MNP+17} to mitigate the communication disruption and help improving cooperation in networks of robots. The proposed algorithm aims at accomplishing a specific task despite link losses, guaranteeing cooperation and coordination among wireless networked robots. The network of robots is composed by one leader and several followers, and the goal of the algorithm is to set position, velocity and control of the followers so that they are always able to follow the leader. Depending on the amount of packet loss, the control input is periodically corrected to reduce the error, or bias, that each robot accumulates in the link with the other robots.
A \ac{MAC} protocol is proposed in \cite{M16} to reduce the packet error probability in \acp{WSN} where a \ac{UAV} collects data from sensor nodes. The proposed protocol is part of a data collection scheme that exploits cooperation between a \ac{UAV} and different sensor nodes. In this protocol, sensor nodes are allowed to transmit data based on a polling signal that gives them the chance to transmit data, which can belong to the single sensor or come from cooperative sensor nodes that collect data from other nodes. The \ac{MAC} protocol manages also the construction of the cooperative sensor node list and establishes which data must be retransmitted through the chosen cooperative sensor node.
%%%%%%%%%%%%%%%%%%%%%%%%%%%%%%%%%%%%%%%%%%%%%%%%%%%%%%%%%%%%%%%%%%%%%%%%%%%%%%%%%%

%%%%%%%%%%%%%%%%%%%%%%%%%%%%%%%% Throughput    optimization %%%%%%%%%%%%%%%%%%%%%%%%%%%%%%%%%%%%
\subsubsection{Throughput Optimization}
Schemes for throughput optimization are presented in \cite{SIL+16,KNK+17}.
A data acquisition framework is proposed in \cite{SIL+16} to increase the efficiency of data gathering in \ac{UAV} networks. The main goal is the optimization of the system throughput through a scheme that assigns different priorities to \acp{UAV} in different areas. To this end, a modification of the contention window mechanism at MAC layer is proposed, that lowers the range of the contention window sizes to areas with high priority, and increases it for lower priority areas. This approach allows reducing packet collisions and losses, especially when the \ac{UAV} moves in the forward direction.
A resource allocation method is presented in \cite{KNK+17} for a wireless network composed by \acp{UAV} and \acp{GS}. Exploiting the \ac{TDMA} scheme, resources are allocated in the communication channel between the \ac{UAV} and the \ac{GS} by taking into account mission-specific and environmental requirements. The beforehand knowledge of the \acp{UAV} throughput, that depends on the \ac{UAV} specific mission, is exploited in the \ac{TDMA}-based allocation scheme, to derive the optimal allocation of the different time slots and respect the throughput requirements of each \ac{UAV}. Two different priority degrees of the \acp{UAV} are also considered in the time slots assignment.
%%%%%%%%%%%%%%%%%%%%%%%%%%%%%%%%%%%%%%%%%%%%%%%%%%%%%%%%%%%%%%%%%%%%%%%%%%%%%%%%%%

%%%%%%%%%%%%%%%%%%%%%%%%%%% OPTIMIZATION PROBLEMS %%%%%%%%%%%%%%%%%%%%%%%%%%%%%%
\subsubsection{Optimization Problems}
The mechanisms proposed in the works \cite{RYR+19,LH17} bring to constrained optimization problems that can be solved through standard methods and/or approximations with reduced computational complexity.
Several \ac{MA} schemes are analyzed in \cite{RYR+19} for mmWave \ac{UAV} cellular-connected networks, with respect to energy efficiency. Two specific MA schemes, i.e., \ac{RSMA} and \ac{NOMA}, are analyzed in detail and optimized to maximize the energy efficiency for mmWave \ac{DL} transmissions, taking into account real propagation patterns from antenna elements in last-generation wireless networks. The energy efficiency optimization problems for both the MA schemes are computationally very expensive, so an approximate algorithm is proposed to provide a solution.
The work \cite{LH17} proposes a resource allocation mechanism aiming to minimize mean packet transmission delay in a cellular network with \acp{UAV} placed in a 3D multi-layered structure. Delay is first derived analytically through a queuing system analysis that models the mean packet arrival rate. Then, it is minimized by solving an optimization problem that allocates spectrum and power for each layer in a two-layer \ac{UAV} network, through standard procedures.
%%%%%%%%%%%%%%%%%%%%%%%%%%%%%%%%%%%%%%%%%%%%%%%%%%%%%%%%%%%%%%%%%%%%%%%%

%%%%%%%%%%%%%%%%%%%%%%%%%%%%%%%%% OTHERS    %%%%%%%%%%%%%%%%%%%%%%%%%%%%%%%%%
\subsubsection{Other Access Schemes}
Other access schemes for resource optimization are developed in \cite{WDZ+17,ZSW18}.
The goal of the access scheme proposed in \cite{WDZ+17} is the optimization of the accuracy in the medium access procedure. The proposed scheme helps an \ac{UAV} joining an ad-hoc \ac{UAV} network by correctly identifying the \ac{MAC} protocol without demodulating the property field \ac{MAC} header, which can be very difficult in particular situations (i.e., when channel attributes strongly change due to interferences). The identification of the \ac{MAC} protocol is performed adaptively through machine learning methods that include a classifier able to identify \ac{MAC} protocols according to features extracted from the channel attributes, improving the access accuracy.

The paper \cite{ZSW18} proposes a \ac{MAC} layer protocol that optimizes the data transmission by sharing the information on the reciprocal position among \acp{UAV}. This information is useful to reduce the packet overhead at \ac{MAC} layer, and to exchange control packets between neighboring nodes with the goal of increasing the effectiveness of mutual communication between drones and pointing directional antenna of transmitting \ac{UAV} towards the \ac{UAV} receiving data, so that the link can be established quickly, the delay in data reception is lowered and the throughput increases. 
%%%%%%%%%%%%%%%%%%%%%%%%%%%%%%%%%%%%%%%%%%%%%%%%%%%%%%%%%%%%%%%%%%%%%%%%

\subsection{Coding Strategies} \label{datalink_coding_strategies}

Some papers analyze and propose coding strategies at link layer for \ac{UAV}-based networks \cite{KSD+19B,KSD+19,VCA+19,ZHM+17,ZHM+18,CF19,CJK+19}.
Most of the surveyed papers on this topic analyze \ac{NC} \cite{KSD+19B,KSD+19,VCA+19,ZHM+17,ZHM+18}. \ac{NC} is a technique that aims at saving time and bandwidth resources when trying to recover information from packets transmitted at link layer. It relies upon the principle that a transmitted packet includes kind of ``repair information'' for previously sent packets. The destination node uses this received repair packet in combination with previously received packets already stored in the receiving buffer, to recover the original information contained in the previously received packets. This concept can also be extended to multi-hop environments, since the information in a repair packet can be properly encoded so that it can be exploited by multiple nodes. The advantage of \ac{NC} techniques is that the transmission time of a repair packet is usually much smaller than the time needed for an independent transmission of several packets, thus saving time and bandwidth, and increasing spectral efficiency \cite{VCA+19}.

\subsubsection{Cooperative Schemes}
In the context of \ac{UAV}-assisted networks, \ac{NC} is usually exploited in combination with cooperative techniques in multi-hop scenarios \cite{VCA+19,KSD+19B,KSD+19,ZHM+17,ZHM+18}. Cooperation is a topic peculiar of the network layer (as pointed out in Section \ref{network_cooperative_strategies}); nevertheless, the \ac{NC} techniques analyzed in these contributions are developed at link layer.
%\subsubsection{Network coded Scenario}
Network coded scenarios are tackled in \cite{VCA+19}. This work proposes a strategy to improve the \ac{QoS} in \ac{UAV}-aided networks that transmit high-quality video streams. \ac{QoS} is evaluated in terms of reduced delay and increased throughput. in this context, \ac{NC} acts in combination with an application layer ARQ protocol. In this paper, \ac{NC} is applied at link layer, and relies on a cooperative technique where several nodes adopt a broadcast transmission of packets. Each node stores multiple packets, and performs an XOR operation among them to build a single coded packet, and forwards it to a destination node which, in turn, decodes the packer XORing it with the packets already stored, thus increasing the spectral efficiency. This coding technique is used to code packets coming from the same flow in a video stream, and is implemented in combination with a selective-repeat ARQ protocol at application layer, to increase the overall network \ac{QoS} in dynamic \ac{UAV}-based topologies.
The works \cite{KSD+19B,KSD+19} focus on a drone acting as a relay node in Network Coded Cooperation (NCC) for wireless networks. The adoption of an \ac{UAV} as relay node gives a higher degree of freedom in terms of mobility. As a relay node, the \ac{UAV} receives packets from different sources, in different time slots, performs \ac{NC} over all the received signals, and transmits the coded packet to all the destinations. Based on the statistical distribution of the fading channels linking the sources, the \ac{UAV} and the destinations, the  contributions of these paper is to treat this problem analytically, by deriving expressions for the network coded noise and its variance, the overall \ac{SNR} at destination node and the rate \cite{KSD+19B}, as well as a closed-form expression for the outage probability and the impact of drone height on the outage probability \cite{KSD+19}.
Cooperative multi-hop networks of drones are analyzed in \cite{ZHM+17,ZHM+18} for data exchange by means of \ac{NC}. An Instantly Decodable Network Coding (IDNC) technique is proposed in these works, whose goal is to reduce the energy consumption and delay due to the \ac{NC} operations in a cooperative multi-hop network. This problem is modeled through a cooperative game theoretical approach \cite{ZHM+17,ZHM+18}. The \ac{NC} metrics, such as decoding delay, completion time, delivery time and rate, are considered in the optimization problem and contribute to partition the network into coalition groups that take into account energy efficiency, together with delay. The main contribution in \cite{ZHM+17} is to define a scalable, low-complexity approach for the cooperation in the IDNC game that takes into account energy consumption. This approach is further enhanced in \cite{ZHM+18} by decentralizing the dynamic choice of the  transmission power by each of the network nodes; furthermore, the stability of the cooperative game is studied, demonstrating its convergence to a stable coalition structure.

\subsubsection{Error Protection Strategies}
Other coding strategies are useful to protect information from errors and noise, as testified in papers \cite{CF19,CJK+19}.

A packet coding strategy is adopted in \cite{CF19} for delay constrained applications transmitting video over fading channels. The proposed scheme makes use of the so-called cross-packet coding strategy, that is based on several transmissions of the same information message occurring from the moment of message availability to the message decoding deadline. During this period, the subsequent retransmissions of the message are coded with different code rates, so that additional protection bits in subsequent retransmissions can contribute to increase the probability of successful decoding of the message. Differently from the conventional cross-packet coding strategies, the proposed scheme does not consider any feedback and does not remove protection bits relative to the previous transmissions, so to reduce the retransmission delay and keep a low frame error rate.

A convolutional denoising encoder is exploited in \cite{CJK+19} to reduce noise due to the drone flight and the wind when transmitting recorded audio signals in \ac{UAV} sensor networks. It makes use of convolutional neural networks that are trained with datasets obtained by mixing drone flying and wind noises with speech signal. The time-frequency features of the speech signal are extracted and used to eliminate the noise contribution from the target speech.

%Cosa vuol dire? Quali sono i problemi?
%Cosa fa 
%Lo facciamo noi

\subsection{Scheduling}

Scheduling of information exchanged among drones is another challenge usually tackled at link layer. Papers on this topic schedule the data transmission so that specific metrics, i.e., successful data rate and throughput, energy consumption and interference among drones, can be optimized.

%%%%%%%%%%%%%%%%%%%%%%%%%%%%%%%%% THROUGHPUT    %%%%%%%%%%%%%%%%%%%%%%%%%%%%%%%%%
\subsubsection{Throughput Improvement}
Scheduling at link layer can be useful to optimize communication throughput, as testified by \cite{TKN+17,WZZ17,LZZ17}.
A radio access scheme thought for improving relaying capabilities in a multi-hop \ac{UAV}-based network is proposed in \cite{TKN+17}. In this scheme, the number of \ac{DL} and \ac{UL} subframes scheme is dynamically changed by first changing the structure of the frame, and then scheduling the \ac{UL} and/or \ac{DL} subframes allocation in the frame to each node, so that the throughput of the communication links is improved also for relatively long distances between the communicating nodes.
A \ac{UAV}-aided network is studied in \cite{WZZ17,LZZ17} where a \ac{UAV} is used as a flying \ac{BS} to serve a group of ground users. A scheme is proposed in \cite{WZZ17} that maximizes the minimum average throughput among all users by jointly optimizing the scheduling of data sent by users and the \ac{UAV} trajectory, given a periodic/cyclical \ac{TDMA} scheme of fixed period. The optimization problem is solved through an iterative algorithm that alternatively optimizes the user scheduling and \ac{UAV} trajectory at each iteration, and whose convergence to the optimal solution is proven to be guaranteed. The same network scenario is presented also in \cite{LZZ17}, but with the addition of a ground \ac{BS} to the \ac{UAV} and the ground users. The performance metrics optimized in \cite{LZZ17} are the same of \cite{WZZ17} (average throughput and \ac{UAV} trajectory), with, in addition, the users partitioning between the \ac{UAV} and the ground \ac{BS}.
%%%%%%%%%%%%%%%%%%%%%%%%%%%%%%%%%%%%%%%%%%%%%%%%%%%%%%%%%%%%%%%%%%%%%%%%%%%

%%%%%%%%%%%%%%%%%%%%%%%%%%%%%%%%% ENERGY CONSUMPTION    %%%%%%%%%%%%%%%%%%%%%%%%%%%%%%%%%
\subsubsection{Energy Consumption Optimization}
Also the energy consumption can be reduced through effective scheduling, as illustrated in \cite{LNW+16,LNW+15}.
Another scheduling algorithm is proposed in \cite{LNW+16} to optimize the energy consumption in cooperative networks of drones with relaying capabilities, at the same time guaranteeing a target rate of successfully received packets. The goal of the algorithm is to schedule the transmitted packets so that the energy consumption is minimized, at the same time guaranteeing a target bit error rate. The problem is solved by means of integer programming techniques, even if its solution is NP-hard. A suboptimal algorithm is then proposed to reduce the computational overhead due to the problem complexity.
A packet scheduling strategy is proposed in \cite{LNW+15} for \ac{UAV}-aided \acp{WSN}, where multiple \acp{UAV} relay data from a \ac{WSN} sensor to a ground \ac{BS}. The proposed scheme aims at saving energy in presence of lossy channels, and at increasing the \acp{UAV} lifetime. The scheduling mechanism relies on the channel quality indications reported by the \acp{UAV} to the \ac{BS}, which in turn establishes the optimal subset of packets that each \ac{UAV} must send to the \ac{BS}, so that a target packet success rate is guaranteed, also minimizing the maximum energy consumption of all the \acp{UAV}. Since the optimization problem is NP-Hard and very expensive computationally, a suboptimal solution is proposed that achieves a faster convergence.
%%%%%%%%%%%%%%%%%%%%%%%%%%%%%%%%%%%%%%%%%%%%%%%%%%%%%%%%%%%%%%%%%%%%%%%%%%%%%%%%%

%%%%%%%%%%%%%%%%%%%%%%%%%%% mmWave multiple access schemes %%%%%%%%%%%%%%%%%%%%%%%%%%%%%%
\subsubsection{mmWave Multiple Access Schemes}
The works \cite{XXX16,WCL+18} study users scheduling in multiple access schemes for \ac{UAV} communications adopting the mmWave technology.
A study on \ac{SDMA}, or \ac{BDMA}, in mmWave communications is carried out in \cite{XXX16}.
The scheme is based on a \ac{BS} equipped with multiple transceivers, and each \ac{UE} equipped with a single transceiver, so that multiple users can transmit over a single beam and the achievable rate can be boosted. The main challenges of such a scheme are described, i.e., how to wisely schedule different users into a group, and allow different groups to access simultaneously the \ac{BS} while minimizing the reciprocal interference.
%The same issue of user scheduling in \ac{UAV} mmWave communications is discussed in \cite{ZZW+19}, where the user scheduling issue is discussed in BDMA schemes.
Multiple access for mmWave communication in \ac{UAV}-aided 5G networks is analyzed in \cite{WCL+18}. The proposed scheme takes into account the transmission of multiple highly directional beams peculiar of mmWave communication, allowing access to different users for each beam, both for the \ac{UL} and \ac{DL} links.
%by introducing a flexible and link-adaptive constellation.
The goal of the scheme is to adapt the beamwidth to the channel quality indication and transmission requirements of users, to establish the best user grouping strategy that selects the users that have to concurrently transmit data in a single beam. The algorithm proposed aims at improving the overall system throughput.
%%%%%%%%%%%%%%%%%%%%%%%%%%%%%%%%%%%%%%%%%%%%%%%%%%%%%%%%%%%%%%%%%%%%%%%%

\subsubsection{Other Scheduling Proposals}
Other scheduling proposals at link layer are found in \cite{PC20,MKD16,NSA10}.
\cite{PC20} proposes a scheduling scheme that takes into account bandwidth requirements during both upload and download phases as a function of services requests to provide dedicated tradeoffs. The simulated solution is a priority-based service scheduling scheme that able to increase service quality while granting fresh data service and lowering the average request serving latency.

The paper \cite{MKD16} proposes different data collection algorithms in a wireless sensor network assisted by an \ac{UAV}. The goal of the algorithms is to increase the efficiency of data collection by taking into account transmission data rate and contact duration time between the sensors and the \ac{UAV}. Based on synchronization and join messages between the sensors and the \ac{UAV}, contact duration time, data rate and available time slots are derived. This information is then exploited to schedule the assignment of the time slots available to the single sensor to transmit data. The effectiveness of the algorithms is evaluated with respect to number of collected packets and fairness in data transmission among sensors.

In \cite{NSA10} a fault detection and scheduling mechanism is studied for a drone that exchanges information from different sensors through a communication network. The focus of this study is the transmission of information on fault detection in drone motors. To this end, the scheduling mechanism proposed in this work aims at preserving as possible the medium access, also enhancing the accuracy and rapidity of information exchange among the different parts of the system. Communication constraints and the possibility of packet drops are also taken into account.

%Cosa vuol dire? Quali sono i problemi?
%Cosa fa \cite{ZZW+19}

\subsection{Comparison with Other Surveys}
The surveys discussing link layer aspects for drones communications are found in \cite{GJV16,SK17,FMT+17}.
In \cite{GJV16} the surveyed papers on data link layer are analyzed with the only focus on energy consumption aspects. The classification of the related approaches is based on the way energy is saved, which nevertheless is not the only issue of \ac{UAV} communication at link layer.
%A first approach is to schedule the sleeping activity with a single radio used for both signaling and data traffic. The second approach is very similar to the first one, but utilizes two different radios, one for signaling and the other for data traffic. The goal of the third approach is to adjust the transmission power levels to regulate the connectivity among neighbors nodes and change the network topology by deciding the nodes to keep active. The fourth approach to save energy divides nodes into clusters, and selects a \ac{CH} as a coordinator for communication among clusters and data aggregation \cite{GJV16}. The survey \cite{GJV16} conducts a detailed analysis at link layer, but only related to the energy consumption issue.
Link layer approaches are mentioned in \cite{SK17} for the formation of \ac{UAV} ad-hoc networks. In this work, the main challenges, i.e., synchronization among nodes, delays, resource reservation, access management and error control, and modified \ac{MAC} protocols are summarized, citing only one paper that overviews them. So, unfortunately, no any detailed discussion is carried out in \cite{SK17} on this topic.
In \cite{FMT+17} some link layer issues are addressed with respect to the cooperative robotic networks. They are mainly related to the severe limitations of signal propagation, that require the design of new \ac{MAC} protocols. Some works discussing the most promising approaches are discussed, ranging from the well known \ac{TDMA}, \ac{FDMA} or \ac{CSMA} schemes, together with Medium Access Collision Avoidance (MACA) and hybrid schemes, with an additional analysis on the amount of control information for channel reservation. The main drawback of this classification is that it refers to the only cooperative robotic networks operating in underwater environment, and the surveyed works mentioned in \cite{FMT+17} do not basically propose any novel technique at \ac{MAC} layer.

If compared to the papers mentioned above, the main goal of this section has been to integrate and further develop their analysis, presenting all the most recent approaches discussed for \ac{UAV} networks in a broader range of link layer related topics.

\subsection{Lessons Learnt}
The analysis of the proposals discussed in this section allows to acquire important knowledge on the pros and cons of resource optimization and scheduling strategies at link layer, as well as coding strategies.
Algorithms proposed to reduce packet losses and increase throughput are important to improve the performance of communication links especially in critical scenarios. Nevertheless, significant modifications to \ac{MAC} layer protocols are needed (i.e., the variation of some protocols parameters, additional prediction/estimation algorithms or metrics, etc.), that can create interoperability problems with the \ac{UAV}-based networks that adopt the classical link layer protocols. Also cooperative schemes are useful to improve communication efficiency and reduce packet losses at \ac{MAC} layer, but they introduce a data overhead needed to share information in the cooperative nodes. This complicates the implementation of scheduling mechanisms needed to decide which node must transmit the information, when, and at what frequency. Tight synchronization mechanisms are another critical point that raises in this scenario.
Another consideration is that the optimality of resource allocation and scheduling often translates into finding the solution of complex optimization problems, whose convergence si not guaranteed, and that can be solved only at the cost of simplifications, that nevertheless bring to suboptimal solutions.
Coding strategies are helpful to increase the efficiency of the communication link, in terms of time and bandwidth saving, and increased spectral efficiency. They are always associated to cooperation strategies, to increase \ac{QoS} metrics in cooperative networks. The drawback of this strategy is the additional time needed to decode packets, and the increased computational complexity for encoding and decoding operations, if compared to the simple packet retransmission. This explains why the research efforts are actually focused on low-complexity techniques, that can scale well in networks with a large number of \acp{UAV}.

%% file: src/6-network.tex
\section{Network Layer}\label{sec:network}
\begin{figure}[htbp]
	\centering
	\includegraphics[width=0.7\columnwidth]{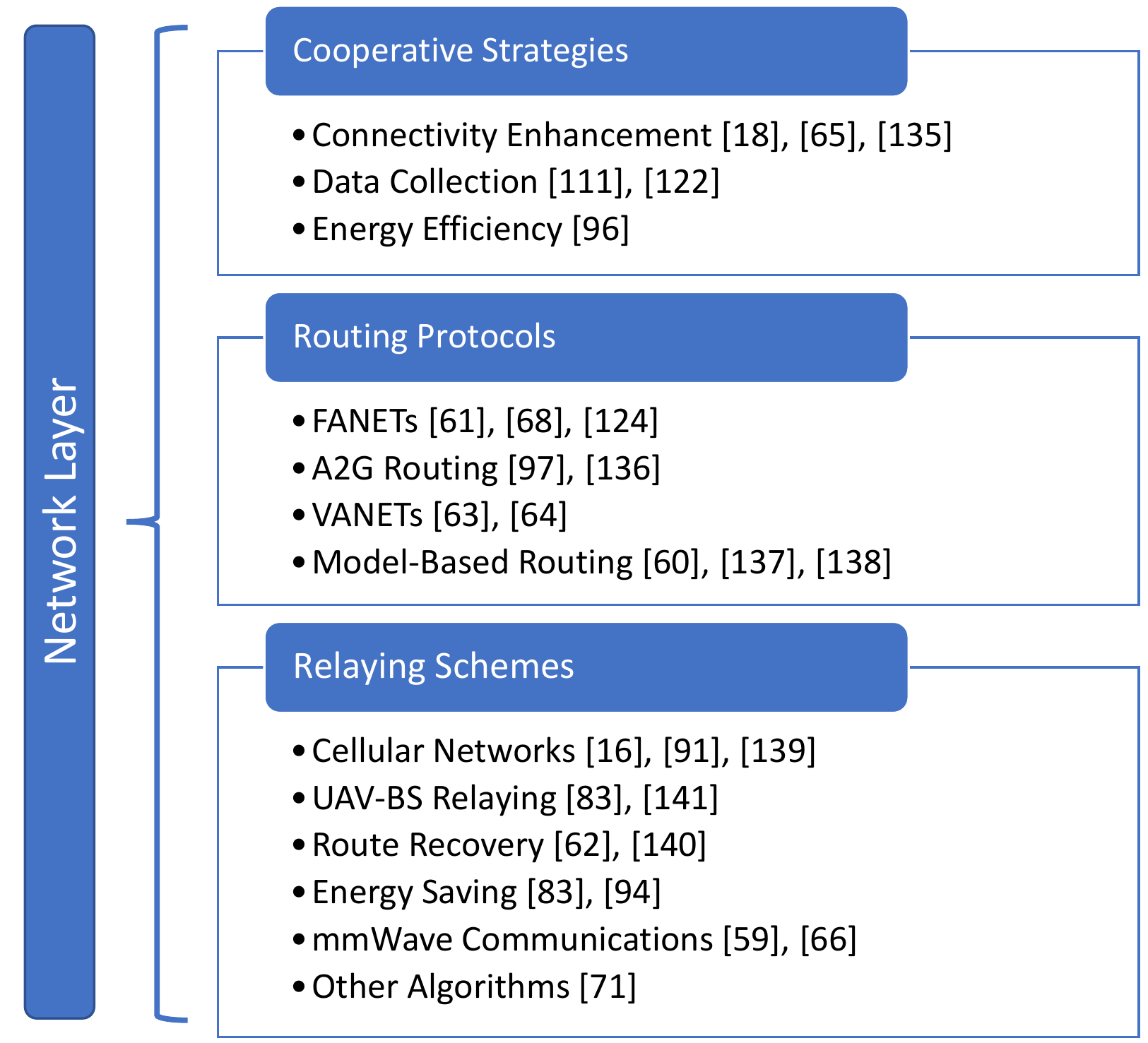}
	\caption{Network Layer Taxonomy.}
	\label{fig:taxnetwork}
\end{figure}

The present Section focuses on: (i) cooperative strategies, (ii) routing protocols, and (iii) relaying schemes.
The organization of this section is reported in Figure \ref{fig:taxnetwork}.
In particular, this Section aims at discussing all the main aspects and issues related to the network layer of the \ac{IoD} i.e., all the proposals and strategies in the multi-hop flying network scenarios.

\subsection{Cooperative Strategies}
\label{network_cooperative_strategies}
Cooperative strategies are discussed in \cite{KAM16,ZJF+18,BA15,M16,WTZ+18,LNW+16} to optimize different metrics, i.e., connectivity \cite{KAM16,ZJF+18}, energy efficiency \cite{ZJF+18}, data collection \cite{M16}, coverage performance \cite{WTZ+18}, and throughput \cite{BA15}.

%%%%%%%%%%%%%%%%%%%%% Connectivity %%%%%%%%%%%%%%%%%%%%%%%%%
\subsubsection{Connectivity Enhancement}
An architecture for \acp{ICN} that uses moving routers is proposed in \cite{KAM16}. An \ac{UAV} that acts as a \ac{CCN} router is used to deliver contents between physically disjoint networks. The goal of this proposal is to exploit the flexible capabilities of CCN routing, i.e., the utilization of content caches and content names, to improve communication performance at network layer, at the same time providing connectivity to disjoint networks thanks to the capability of the router to move autonomously.
The scheme is also extended by proposing a communication architecture which uses different \ac{UAV}-based routers to cooperatively process routing among disjoint networks. 

A low latency routing algorithm is presented in \cite{ZJF+18}, for a \ac{UAV}-based cooperative network providing \ac{IoT} services. The proposed algorithm is suitable for large scale networks of \acp{UAV}, where the latency of the classical routing algorithms developed in literature increases, and the connectivity decreases because of high interference and delay, together with the rapid topology changes due to the high mobility of \acp{UAV}. The \ac{LDRA} proposed in this paper exploits the cooperation among \acp{UAV} that have relaying capabilities,. It uses only partial information on \acp{UAV} location and connectivity, distributing the different data flows by optimally choosing the relay \acp{UAV}, so that latency is minimized. 

A contention-free geo-routing scheme is proposed in \cite{BA15} to decrease latency and increase throughput in \ac{UAV} systems. The scheme is based on a cooperative relaying strategy, where multiple nodes concurrently transmit the same frame towards the destinations. In the first hop of the routing scheme, a source transmits a frame into the network, and nodes closer to the destination relay concurrently the frame. Thanks to the OFDM transmission, the multiple copies of the same frame more easily bring to a successful frame decoding. In the second hop, each receiver decides whether to forward the frame or not, based on both the position information provided by transmitters of the previous hop and the positive progress towards the sink. Due to the concurrent transmission of multiple nodes, the communication range and throughput increase, and latency decreases.

%%%%%%%%%%%%%%%%%%%%%%%%%%%%%%%%%%%%%%%%%%%%%%%%%%%%%

%%%%%%%%%%%%%%%%%%%%% Others %%%%%%%%%%%%%%%%%%%%%%%%%
\subsubsection{Data Collection}
A cooperative scheme is studied in \cite{M16} for efficient data collection in \acp{WSN} that use \acp{UAV} as relay nodes. Even if the main focus of this work is the proposal of a \ac{MAC} protocol as described in Section \ref{datalink_resource_allocation}, a cooperative architecture is also proposed for the data collection system. The scheme is based on a model where an \ac{UAV} collects data from sensors along its flight path and provides them to cloud servers. The servers have complete knowledge of sensor nodes positions and data, and generate the flight paths of \acp{UAV}, sharing them with the \acp{UAV}. In the proposed scheme, the \ac{UAV} requests the retransmission of missing data from cooperative sensors, which are collected in a list.

The cooperative scheme proposed in \cite{WTZ+18} aims at increasing the coverage performance of \ac{UAV}-aided cellular networks where the \acp{UAV} act as \acp{BS} caching the most popular contents and providing them to mobile terminals, so that the ground \acp{BS} are offloaded. \acp{UAV} are grouped into cooperative clusters, to provide cached contents to a group of mobile terminals. Transmission of the cached content by the \ac{UAV} in the cluster is possible only if some requirements, i.e., the amount of energy to feed communication modules, the presence of the requested content in the \ac{UAV} cache and the maximum connection capacity of the \ac{UAV}, are respected.

%%%%%%%%%%%%%%%%%%%%%%%%%%%%%%%%%%%%%%%%%%%%%%%%%%

%%%%%%%%%%%%%%%%%%%%% Energy Efficiency %%%%%%%%%%%%%%%%%%%%%%%%%
\subsubsection{Energy Efficiency}
Energy efficiency is the main focus of the cooperative relaying scheme proposed in \cite{LNW+16}. The proposal is based on the adoption of a swarm of \acp{UAV} that relay data coming from remote sensors to a ground \ac{BS}, to overcome the presence of lossy channels, save energy and extend the \acp{UAV} lifetime. The scheme is based on a relaying cooperative protocol that exploits the information on the reception quality reported by the \acp{UAV} to the \ac{BS}. Accordingly, the \ac{BS} schedules the packet load to be transmitted and the Adaptive Modulation and Coding (AMC) scheme for each \ac{UAV}, to find the best trade-off between the success rate of packet transmission and the energy consumption. The optimization problem is NP-hard and mathematically intractable; thus an approximated solution is proposed that decouples the processes of energy optimization and AMC selection.
%%%%%%%%%%%%%%%%%%%%%%%%%%%%%%%%%%%%%%%%%%%%%%%%%%%%%%%

\subsection{Routing Protocols}
There are many works that propose routing strategies in \ac{UAV}-based networks \cite{ZSW18,GSY17,RDC17,LKY+16,MCK+17,OLL+16,OLZ+17,SBI+16,QHG+16,ZZJ+17}, for different application scenarios and network architectures.

%%%%%%%%%%%%%%%%%%%%%% FANETS %%%%%%%%%%%%%%%%%%%%%
\subsubsection{FANETs}
Papers \cite{ZSW18,GSY17,RDC17} develop routing strategies for \acp{FANET}.
Different communication protocols are proposed in \cite{ZSW18} in the \ac{FANET} scenario. Among them, a routing protocol is presented, which exploits a reinforcement learning approach. Specifically, it takes into account the \acp{UAV} positions in the routing policies, updating them by means of a reward function that depends on the network utility, so that the optimal paths can be chosen that minimize the delay in data delivery. The main advantages of the protocol are that a global knowledge of the network is not needed, and that the protocol has capabilities to continually evolve for self-optimization purposes.

A predictive routing protocol is proposed in \cite{GSY17} to tackle the main challenges in \acp{FANET}, i.e., \ac{UAV} speed, connection loss, changing network topology, etc. The protocol is based on a predictive 3D estimation of the expected connection time between two adjacent intermediate nodes with directional transmission, with a continuous update of position and speed information to improve prediction accuracy. When the estimated connection time is close to expire, an alternative path is chosen during data transmission, to guarantee service continuity. The adoption of omnidirectional transmission and beam-forming is also allowed; the former to increase the connection among nodes, and the latter to increase the transmission distance and reduce the packet collisions and the set-up time of the routing paths.

A modified routing mechanism to evaluate the impact of drone replacement in \acp{FANET} is proposed in \cite{RDC17}. The main goal of this work is to study how to replace dynamically drones in a \ac{FANET} without degrading the overall network performance, and taking into account the network main characteristics, i.e., its dynamic variation in time, its support for heterogeneous traffic, and its limited lifetime. In the last part of this paper, after a description of the most used routing protocols in \acp{FANET} and their classification into reactive and proactive protocols, a modification of the proactive routing protocol chosen for the analysis is proposed, that adapts a protocol attribute to the battery charge of the node, so that the connection during handover between \acp{UAV} (because of battery depletion) can be maintained. 
%%%%%%%%%%%%%%%%%%%%%%%%%%%%%%%%%%%%%%%%%%%%%%%%

%%%%%%%%%%%%%%%%%%%%%%%%%% AIR-TO-GROUND ROUTING %%%%%%%%%%%%%%%%%%%%%%%%%%%%
\subsubsection{A2G Routing}
In \cite{LKY+16,MCK+17} routing strategies are proposed in in \acf{A2G} environments.
The paper \cite{LKY+16} proposes a centralized routing protocol for networks of \acp{UAV} controlled by a Ground Control System (GCS). In the proposed scheme, all the drones are managed by the GCS, which is the component demanded to collect and manage control-related information from drones, including their routing tables. Such an approach avoids the continuous exchange of periodic information on the link cost values among \acp{UAV} (the GCS can extract this information from the geographic information of \acp{UAV}). Furthermore, the disconnection time of the network due to the \acp{UAV} mobility can be minimized because the GCS knows the schedules of drones mobility and can predict the topology changes. The proposed protocol is implemented using real \acp{UAV} and GCS.

Two different opportunistic routing protocols are developed in\cite{MCK+17} in  a scenario where a \ac{UAV} moving at a predefined height and speed provides connectivity to different mobile sensors, that move in the same direction with different speeds. Opportunistic routing is suitable for scenarios with a changing network topology, because of its capability to increase throughput. In both the proposed protocols nodes that want to transmit data but are not in the \ac{UAV} communication range, choose the neighbor nodes in the \ac{UAV} range that can act as packet forwarders, with the goal to increase the \ac{PDR}. In the first proposed protocol, a source node transmits packets to all its neighbors in its range, while in the second it transmits packets to the forwarder with the highest velocity. The \ac{PDR} increase comes at a cost of an increased delay and routing overhead, if compared to the direct connection between source nodes and the \ac{UAV}.

%%%%%%%%%%%%%%%%%%%%%%%%%%%%%%%%%%%%%%%%%%%%%%%%%%%%%%%%%%%%%%%%%%%%

%%%%%%%%%%%%%%%%%%%%%% VANETS %%%%%%%%%%%%%%%%%%%%%
\subsubsection{VANETs}
\begin{figure}[htbp]
	\centering
	\includegraphics[width=0.8\columnwidth]{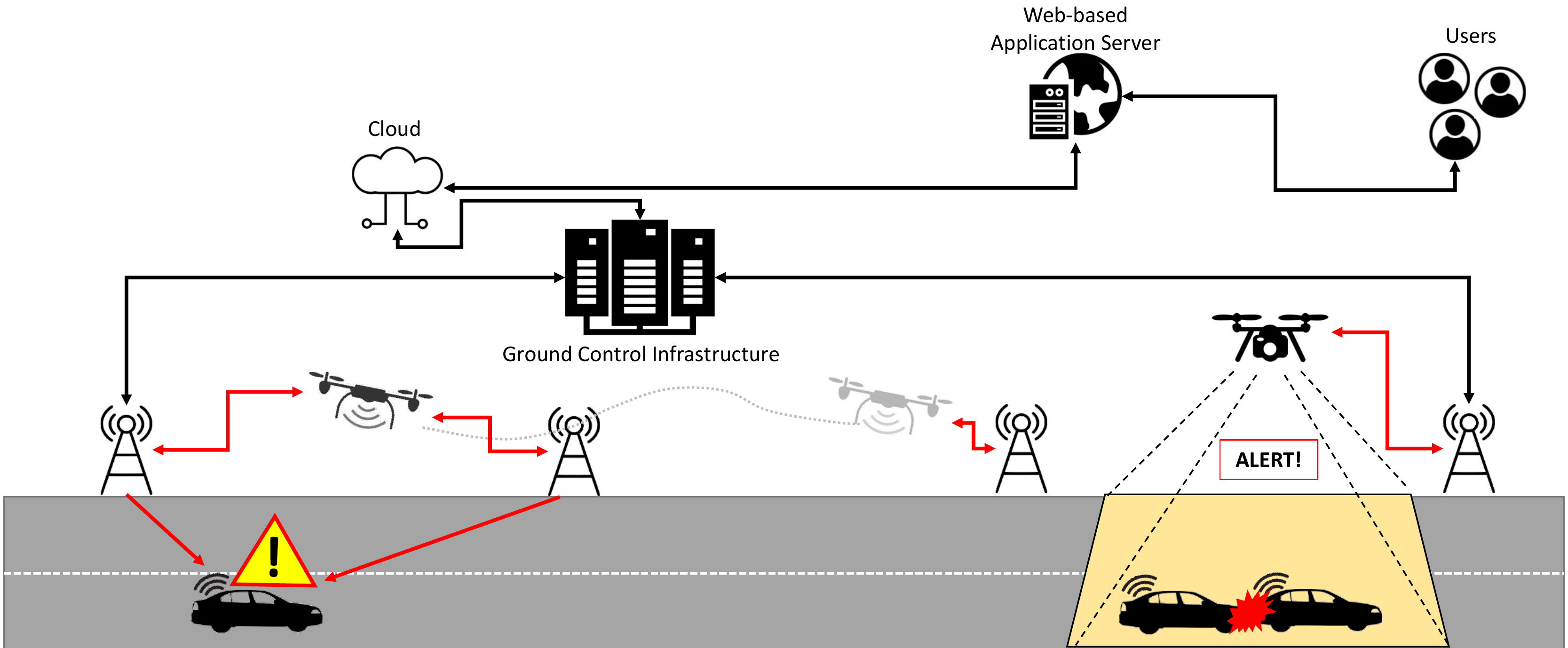}
	\caption{Drones in a \acp{ITS} scenario with connected vehicles.}
	\label{fig:iod_vanets}
\end{figure}

Routing protocols for \acp{VANET} are developed in \cite{OLL+16,OLZ+17}.
For sake of better clarity, Fig. \ref{fig:iod_vanets} provides an illustrative example of a \ac{UAV}-assisted \ac{VANET} where routing operations and protocols are applied. In a nutshell, drones are herein assumed to monitor a road as part of an \ac{ITS}. Once a situation of interest is detected, i.e., an accident, the ground control infrastructure is in charge of providing updated information to the all the vehicles on the road. Thanks to the employment of \ac{VMS} systems, and leveraging \ac{V2I}, all the road users can be promptly notified about the dangerous situation and, at the same time, the closest available \ac{UAV} can be relocated, when needed. As will be discussed later on in this section, \acp{ITS} are also using \ac{V2V} communications. In fact, connected cars and direct links between the drones and the vehicles may be used to simplify message exchanges, thus lowering latencies and improving communication effectiveness.
%\begin{figure}[htbp]
%	\centering
%	\includegraphics[width=\columnwidth]{img/vanets}
%	\caption{Drones in a \acp{VANET} scenario.}
%	\label{fig:vanets}
%\end{figure}

%\begin{figure}[htbp]
%	\centering
%	\subfigure[Drones in a \acp{VANET} scenario.]{\label{fig:vanets}\includegraphics[width=\columnwidth]{img/vanets}}	\\
%	\subfigure[Drones in a \ac{IoD} scenario.]{\label{fig:iod_vanets}\includegraphics[width=\columnwidth]{img/iod_vanets}}
%	\label{fig:vanet_vs_iod}
%	\caption{The role of drones in a \ac{VANET} with respect to an \ac{IoD} architecture.}
%\end{figure}

The routing protocol developed in \cite{OLL+16} aims at increase the reliability of data delivery for \ac{UAV}-assisted \acp{VANET} in urban environments. The proposed protocol is based on the information exchanged between \acp{UAV} and vehicles on the traffic density and the connectivity of the vehicles. Based on this, the \acp{UAV} can decide their location to relay data in the points of the network where connectivity among ground vehicles is missing because of the presence of obstacles which are present in typical urban environments. Accordingly, vehicles can choose to use \acp{UAV} as forwarders if there are no available paths in direct vehicle-to-vehicle communications.

A routing protocol that finds the shortest end-to-end connected path in \ac{VANET} environments is proposed in \cite{OLZ+17}. It takes into account the high mobility of vehicles and the frequent changes in the network topology due to unpredictable movements of vehicles. The proposed protocol exploits heterogeneous communications through the cooperative interaction between an ad-hoc network of \ac{UAV} and a \ac{VANET}. The goal of the \ac{UAV} network is to restore communication links that fall down because of the presence of obstacles in the urban environment. To this end, two different protocols are used: the first, more delay-tolerant, exploits the global knowledge of the \acp{UAV} about the connectivity status of the road segments to route data between \acp{UAV} and ground vehicles; the second, more reactive, routes data among \acp{UAV} only when needed.
%%%%%%%%%%%%%%%%%%%%%%%%%%%%%%%%%%%%%%%%%%%%%%%%

%%%%%%%%%%%%%%%%% MODEL-BASED ROUTING %%%%%%%%%%%%%%%%%%%%%
\subsubsection{Model-Based Routing}
The routing protocols proposed in \cite{SBI+16,QHG+16,ZZJ+17} are based on analytical models. 
A routing protocol for networks of \acp{UAV} is proposed in \cite{SBI+16}. It exploits an application layer functionality that predicts future trajectories of \acp{UAV} and that is integrated in the a routing protocol already known by literature, to improve connectivity in the \ac{UAV}-based network and reduce packet losses. Different mobility models and exploration algorithms are used for comparison between the proposed protocol and other known routing protocols.

A routing framework for hybrid space/air networks is proposed in \cite{QHG+16}, to mitigate the high bit error rates and long delays that can occur because of the mobility of nodes and the varying network topology. The framework exploits a routing algorithm based on a hybrid time-space graph, that is described by two subgraphs. The first subgraph describes deterministically the space network. The second subgraph aims at predicting the contact time and contact probability among the \acp{UAV} of the air network, and for this reason it is based on a semi-Markov prediction model. The graph is transformed into a state-space graph to establish the optimal next hop, based on a forwarding rule that can exploit \acp{UAV} in the air network or satellites in the space network as relays. The goal of the proposal is to improve the message delivery ratio, the end-to-end-delay and the power consumption.

The routing algorithm proposed in \cite{ZZJ+17} exploits 3D cubes as partitioning regions to forward data among \acp{UAV}. More specifically, data are relayed by choosing only one \ac{UAV} as relay node per region. The optimal path among different cubes is chosen based on the maximum successful transmission probability derived from an analytical model. Then, the most appropriate \ac{UAV} chosen to relay data in each cube is selected by jointly considering again the successful transmission probability and the \acp{UAV} mobility. The goal of this routing algorithm is to improve the end-to-end delay, jitter, and \ac{PDR}.
%%%%%%%%%%%%%%%%%%%%%%%%%%%%%%%%%%%%%%%%%%%%%%%%%%%

\subsection{Relaying Schemes}

%{\color{red}Qui potrebbe andare una figura sul relay. Più droni che sono connessi per inoltrare i dati a due stazioni a terra, oppure un drone che fa da ponte tra due stazioni di terra (una trasmittente e una ricevente).}

\ac{UAV}-based relaying schemes are presented in \cite{AA17,LGY+15,TKN+17,LXX+17,PJS+17}.

%%%%%%%%%%%%%%%%%%%%%%% CELLULAR NETWORKS %%%%%%%%%%%%%%%%%%%%%%%%%%
\subsubsection{Cellular Networks}
The papers \cite{AA17,LGY+15,TKN+17} exploit drones as relays in cellular networks.
The review paper \cite{AA17} analyzes \ac{D2D} communications in different application scenarios. Part of this work is focused on the utilization of drones as \ac{D2D} relays, to extend cellular network coverage especially in extraordinary conditions when the ground \acp{BS} can be damaged or switched off in critical scenarios (like catastrophes, network damage, etc.). In such situations, drones transfer the signal from far \acp{BS} to mobile devices, filling coverage holes. This paper does not go into deeper details on the implementation of \ac{D2D} relaying functionalities. The same topic is described also in \cite{LGY+15}, that tackles the wireless coverage issue in case of disasters or incidents in public safety wireless networks. This work proposes a multi-hop \ac{D2D} scheme that extends network coverage wherever ground relay is not possible. The proposed scheme jointly finds the optimal \ac{UAV} position and the resource allocation, in time or frequency domain, so that the data rate of each hop in the path from the \ac{BS} to the mobile device is maximized.

A radio access scheme is proposed in \cite{TKN+17} for muti-hop relay networks composed by \acp{UAV}. This scheme increases the efficiency of relay communications between an observation \ac{UAV} transferring data to a ground \ac{BS} exploiting \acp{UAV} as relay nodes. This study is based on the assumption that the time-division relaying technique exploited in \ac{LTE-A} can be suitable to increase the efficiency of relay communications among \acp{UAV}, in terms of both reduction of mutual interference among \acp{UAV} and communication order by the relay node to transfer data from the observation \ac{UAV} to the \ac{BS}, especially for long distances. Since the existing \ac{LTE-A} scheme is not suitable for \ac{UAV}-based relaying because of some limitations in the frame structure, this work proposes an \ac{LTE-A} access scheme that increases the relay communication efficiency over long distances, taking also into account the \acp{UAV} height.

Also the solution presented in \cite{LGY+15} exploits a relay \ac{UAV}, called Floating Relay (FR), to dynamically and adaptively provide additional coverage in a macrocell served by a ground \ac{BS} to tackle the issue of an increasing traffic volume in the macrocell. The discussion focuses on some important aspects like frequency reuse, the interference both between FR cells and the macrocell and among FR cells, bandwidth allocation for the backhaul network, and coverage capabilities.
%%%%%%%%%%%%%%%%%%%%%%%%%%%%%%%%%%%%%%%%%%%%%%%%%%%%%%%%%%%%%%

%%%%%%%%%%%%%%%%% UAV-BS RELAY SCHEMES %%%%%%%%%%%%%%%%%%%%
\subsubsection{UAV-BS Relaying}
Other relay schemes exploiting \acp{UAV} for data transmission to a ground \ac{BS} can be found in \cite{SDZ17,HNK+16}.
The relaying scheme proposed in \cite{SDZ17} is conceived for image transmission in an \ac{UAV}-based network. In this work, some survey \acp{UAV} transmit sensed images to a \ac{BS} through a relay \ac{UAV}. The scheme is based on a planning model that computes the best set of relay points where the relay \ac{UAV} can meet the survey \acp{UAV}, that minimize the average traveling distance of the relay \ac{UAV}. Then, a reinforcement learning approach is used to find the optimal time to visit each point. This technique exploits a reward function based on the rate of image acquisition. The proposed solution avoids the need of any collaboration between survey \acp{UAV} and the relay \ac{UAV}, and reacts to the varying network traffic to improve the quality in end-to-end delay and frame delivery ratio.

Similarly to \cite{SDZ17}, the paper \cite{HNK+16} focuses on a relay scheme where a relay \ac{UAV} forwards data collected from multiple observation \acp{UAV} to a ground \ac{BS}. The observation \acp{UAV} cover a wide area. Depending on the main parameters that influence the network performance, i.e., signal attenuation and obstacles, a formula for the upper bound of the throughput in the network is derived, together with the optimal location of the relay \ac{UAV} that maximizes the upper bound of the throughput.
%%%%%%%%%%%%%%%%%%%%%%%%%%%%%%%%%%%%%%%%%%%%%%%%%t%%%

%%%%%%%%%%%%%%%%%%%%%%% ROUTE RECOVERY SCHEMES %%%%%%%%%%%%%%%%%%%%%%%%%%
\subsubsection{Route Recovery}
Relaying is adopted for route recovery schemes in \cite{LXX+17,PJS+17}.
A relay scheme is proposed in \cite{LXX+17} that aims to solve the problem of jamming attacks in \acp{VANET}. The scheme exploits an \ac{UAV} that relays data from vehicles to \acp{RSU} placed at fixed locations. If the \ac{RSU} serving the vehicle is in a heavily jammed area, the \ac{UAV} becomes in charge of relaying data to another \ac{RSU} with better channel conditions. The \ac{UAV} decides whether or not to relay data from the vehicle to another \ac{RSU} which is far from the jammer, depending on the channel quality in the direct link between the vehicle and the serving \ac{RSU}. The interaction between the \ac{UAV} and the jammer is studied through an anti-jamming relay game approach, derived from the game theory. Reinforcement learning techniques are also used to derive the optimal relaying strategy of the \ac{UAV} without knowing the jamming model.

A route recovery scheme is proposed in \cite{PJS+17} in ad-hoc networks where \acp{UAV} act as relays. The scheme is based on probe packets sent by the \acp{UAV} to discover the route topology and stitch partial paths and avoid network holes in damaged networks. Based on the captured topology, an algorithm for optimal \acp{UAV} deployment is adopted, which minimizes the \acp{UAV} traveling time and distance and avoids duplicate coverage. It is exploited to decide how to replace he network holes through \acp{UAV}. To this end, an algorithm is adopted that dispatches a reduced number of \acp{UAV}, allowing an improvement of routing performance both on local and global scales.
%%%%%%%%%%%%%%%%%%%%%%% %%%%%%%%%%%%%%%%%%%%%%%%%%%%%%%%%%%%%%%%%%%

%%%%%%%%%%%%%%%%% ENERGY SAVING SCHEMES %%%%%%%%%%%%%%%%%%%%
\subsubsection{Energy Saving}
The relay schemes proposed in \cite{SIL+16,ZZZ17} are focused on energy saving.
A relaying strategy is proposed in \cite{SIL+16} as part of a routing algorithm aiming to save the energy consumption in a \ac{UAV}-aided \ac{WSN} for data acquisition. The routing algorithm is based on a priority-based scheme, that takes into account the \ac{UAV} mobility. More specifically, sensors are classified into frames, each one with an assigned transmission priority, so that sensors in more urgent areas transmit their packets with a higher priority. Sensors of a frame are further grouped into clusters, choosing a \ac{CH} that transmits gathered data to the \ac{UAV}. The proposed routing algorithm aims to deliver data from sensors to the \ac{CH} by choosing the optimal relay node with the better channel quality and the shorter distance from the \ac{CH}, so that the energy consumption from source to destination can be saved. 

The relaying scheme proposed in \cite{ZZZ17} makes use of a single \ac{UAV} that helps two ground stations without any direct connection to reach each other by following a circular trajectory. The goal of the proposed scheme is to maximize the spectrum and energy efficiency by jointly optimizing the time allocation of the \ac{UAV} relaying, its speed, and its trajectory. The solution of this optimization problem reveals a trade-off between the energy consumption for the propulsion of the relay \ac{UAV} and the maximization of energy and spectrum efficiency of the whole system.
%%%%%%%%%%%%%%%%%%%%%%%%%%%%%%%%%%%%%%%%%%%%%%%%%%%%%

%%%%%%%%%%%%%%%%% mmWave COMMUNICATIONS %%%%%%%%%%%%%%%%%%%%
\subsubsection{mmWave Communications}
The works \cite{KYW+17,GPM+18} propose routing schemes for mmWave communications.
Mobile relays in mmWave communications are studied in \cite{KYW+17}. A relay method suitable for mmWave communications is proposed, to bypass obstacles and/or increase the communication range. In this scenario, an \ac{UAV} acting as relay adjusts dynamically its path based on real-time measurements of the link qualities of different mmWave beams. In this way, the \ac{UAV} is able to dynamically choose its optimal position accurately and in a short time, despite the unpredictability and varying nature of the wireless link. The proposed scheme is also enhanced to multiple relays that can also use directional beams to extend the communication range.

A methodology for dynamic rerouting in networks exploiting mmWave technology is presented in \cite{GPM+18}. The proposed methodology is applied to mmWave backhaul links that dynamically reroute depending on the channel conditions, to improve the flexibility and reliability of the backhaul solution. Dynamic routing is possible thanks to \acp{UAV} that work as aerial relay nodes, so that the negative effect of occlusions and signal blockage due to obstacles can be reduced. The methodology presented in this work considers the influence of some components, like the signal propagation model and the blockage probability of the mmWave links due to obstacles typical of urban environments, and the \acp{UAV} mobility.
%%%%%%%%%%%%%%%%%%%%%%%%%%%%%%%%%%%%%%%%%%%%%%%%%%%%%

%%%%%%%%%%%%%%%%% OTHERS %%%%%%%%%%%%%%%%%%%%
\subsubsection{Other Algorithms}
The relaying algorithm proposed in \cite{AS15} is applied in a wireless backhaul network composed by balloons. Some \acp{UAV} are used as relays to increase the network reliability, which fluctuates because of the limited control on the balloons mobility that can bring to temporary link failures. The proposed algorithm aims to plan the \acp{UAV} paths and schedule them all so that the network reliability is maximized. The path planning algorithm is developed for the single \ac{UAV} and also for multiple \acp{UAV}, supposing to know the availability prediction of the links among balloons and the traffic matrix of the network.
%%%%%%%%%%%%%%%%%%%%%%%%%%%%%%%%%%%%%%%%%%

\subsection{Comparison with Other Surveys}
The network layer has been widely studied in the past literature, and related papers mainly discuss on cooperation, routing and relaying aspects at network layer of \ac{UAV}-based networks.
The survey papers that cover this topic are \cite{SK17,MTA16,GJV16}.
The work \cite{SK17} analyzes different aspects of the network layer, ranging from cooperative networks, routing and relaying strategies. It carries out an exhaustive analysis of frameworks, models and approaches applied to multi-\ac{UAV} cooperative systems, data routing and relaying strategies.
Routing strategies are analyzed in \cite{MTA16} for networks of \acp{UAV}. In this work only data routing is analyzed in detail, presenting the most significant protocols in \acp{MANET} and \acp{VANET}.
The survey \cite{GJV16} discusses the applicability of the most important types of routing protocols, whether static, proactive, reactive, hybrid or geographic, to the \ac{MANET} scenarios. Networks prone to delays and disruptions are also analyzed, discussing the suitability of other kinds of routing protocols in this more challenging context.
With this well-articulated background in mind, the main goal of the network layer analysis carried out in this section has been to update the discussion of network layer strategies in \cite{SK17,MTA16,GJV16} through a detailed survey of the most recent literature on this topic.

\subsection{Lessons Learnt}
Multi-hop strategies at network layer have been proposed in many papers on networks of \acp{UAV}. The surveyed works discussing cooperation among drones testify that cooperation is a good solution to increase connectivity among separate networks, to increase the effectiveness of routing algorithms and the energy efficiency, and to reduce latencies in data exchange. Except for the papers proposing simplified scenarios, these improvements translate into optimization problems, whose optimal solution comes at a cost of high computational effort. 

As expected, routing strategies are one of the main issues at network layer. This holds true also for networks of drones, where many challenges peculiar of \acp{FANET} (i.e., \acp{UAV} mobility, fast changes in networks topology, error-prone communication channels, etc.) must be tackled. To this end, the proposed approaches make use of strategies ranging from proactive and predictive techniques, to centralized architectures, analytical models, and information on \acp{UAV} position, velocity, etc. The key-aspect of such proposals is the optimal choice of the \acp{UAV} as data forwarders, so that the above-mentioned challenges can be effectively managed without degrading the network performance (in terms of \ac{PDR}, delay, and routing overhead). Lessons learnt from the analyzed works demonstrate that the joint optimization of all these parameters is extremely difficult, and the most suitable routing strategy is strongly context-dependent.

Another important lesson is learnt from the analysis of the relaying schemes proposed. They are adopted for different purposes, i.e., extension of coverage and communication range, interferences reduction in \ac{G2G} multi-hop communications, energy efficiency, throughput optimization, and defense against jamming attacks. Studies in this direction show that relaying can be effective especially in environments subject to interferences from external sources (physical obstacles, jammers, etc.) or intrinsic to the specific short-range communication technology, as mmWave is. Most of the problems presented in the surveyed works are tackled by finding the optimal number and positions of relay nodes; but, once again, the optimization problems suffer from the same drawbacks highlighted for cooperation strategies.

%% file: src/7-apps.tex
\section{Transport and Application Layer}\label{sec:appl}
The Transport and Application Layers have been conceived as independent. It is worth noting that, in most cases, transport layer is taken for granted. This is motivated by the fact that many applications are designed focusing on the expected performance level and, hence, \ac{QoS} and \ac{QoE} Indicators are used. Accordingly, no survey paper so far investigated application layer in detail.
Indeed, several references start from the concepts of \ac{NFV}, \ac{SDN} and/or distributed computing schemes to propose innovative schemes that demonstrate offloading possibilities in specific applications.
At times, applications are conceived as solutions released on top of consolidated telecommunications technologies. In those cases, the protocol stack is not discussed in detail, nor it is modified. This happens frequently when \ac{LTE}, 4G, or 5G communications or, more in general, consolidated networking solutions are involved.
Overall, many scientific contributions rely on the Internet model, which involves \ac{TCP} and \ac{UDP}. Therefore the transport layer is not among the major concerns and is assumed to be \ac{TCP}- or \ac{UDP}-based.
For all the scientific contributions that will be discussed in this section, a focus on the transport layer will be provided whenever the solution is different from the two cited above.

\begin{figure}[htbp]
		\centering
		\includegraphics[width=0.7\columnwidth]{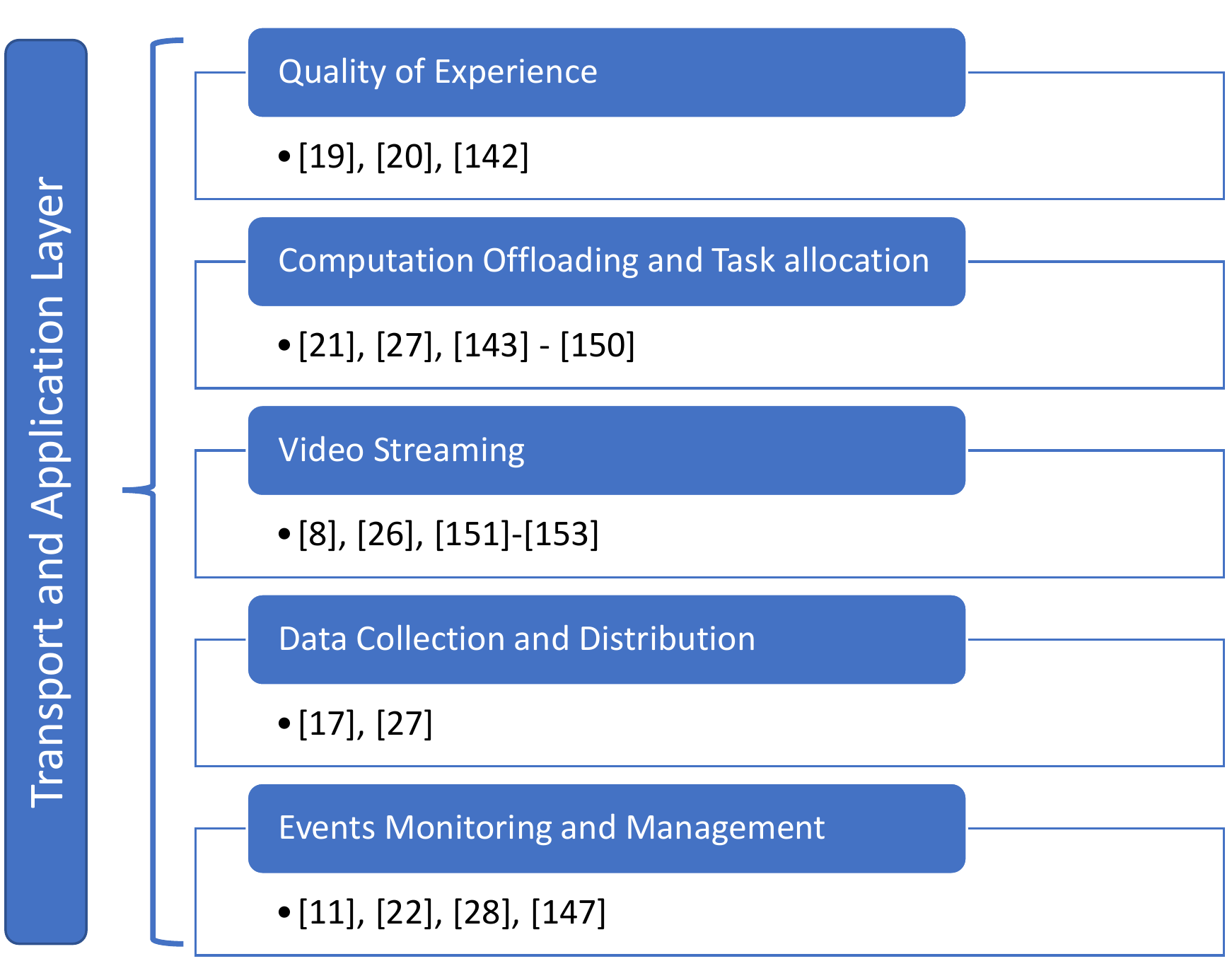}
		\caption{Transport and Application Layer Taxonomy.}
		\label{fig:taxapps}
	\end{figure}

The present Section proposes application layer-related contributions, mainly connected to: (i) \ac{QoE}, (ii) Computation offloading, (iii) Video Streaming, (iv) data collection and distribution, (v) events monitoring and management, and (vi) task allocation. The Section closes up highlighting the lessons learnt on the theme.
In Figure \ref{fig:taxapps}, the overall organization of the present Section is summarized.

\subsection{\acrlong{QoE}}
In the \ac{IoD} context, the network is specifically designed to transmit images and video recordings. For some applications, a very low latency may be highly required (i.e., real-time monitoring). In some other cases, instead, ultra high quality is needed. This heterogeneity leads to the definition of \ac{QoE} indexes, which may enable several possibilities.

One of the most thrilling perspective in research for mobile telecommunication networks is represented by the possibility to employ \acp{UAV} as mobile \acp{BS}. In fact, with the possibility to mount a base station on top of an aerial platform, \ac{BS} planning and deployment can be sensibly simplified.
At the same time, this possibility may significantly improve methodologies for coverage area optimization. Further, mobile \acp{BS} could be helpful in solving several problems connected to base station capacity and inter-cell interference, which are among the major stakeholders when studying cellular networks.
Those problems are tackled by a number of contribution proposed so far and, in particular \cite{CSM17} demonstrates the strengths and weaknesses of the employment of a deterministic approach to analyze these problems. In particular, this work analyzes the aforementioned parameters with a variable threshold for the received power.
Accordingly, an optimal altitude and power consumption model is derived for aerial \acp{BS} in different environments which mainly refer to sub-urban, urban and highly populated urban environments. Among the main outcomes, a channel model for \ac{A2G} communications is obtained.

Granting \ac{QoE} does not specifically mean to improve a specific application and/or functionality. In fact, \ac{QoE} analysis can also be used to design advanced coordination functionalities.
In \cite{CCB+18}, \ac{QoE} is considered in the context of heterogeneous networks, where mobile \acp{BS} are mounted on top of \acp{UAV}. In this work the flight plan is the output of a dedicated \ac{QoE}-aware algorithm, based on the Q-learning approach. Simulation results obtained in this contribution testify that the flight plan diversification improves the overall \ac{QoE} of the users.
In the work \cite{KSJ+19} drones are configured in \acp{FANET} covering area where no fixed (i.e., ground) network infrastructures are deployed, or are hard to reach. The proposal tackles the lack of standardized routing protocols for \ac{FANET} applications to allow efficient communication between devices thanks to an adaptive routing protocol based on fuzzy logic and demonstrates its added value in terms of both \ac{QoS} and \ac{QoE}.

\subsection{Computation Offloading and Task Allocation}
%The increasing popularity of drones is not only suggesting new methodologies and possibilities because derived from their employability. In fact, t
The tasks that drones can carry out may involve more than one drone at a time. This means that drones may be organized in swarms or clusters.
Several contributions so far investigate the tasks that drones can handle suggesting that swarms may be a preferable choice when the application (or reference scenario) may benefit from the increased coverage capability a drone can offer.
Even though a swarm of drones may represent an interesting opportunity, each of them is still limited in terms of computational power and energy resources.
Such limitation suggests an optimization of the task list of the droned, for example by balancing the load of tasks among the swarm components.
%that the task list may not be accomplished on schedule without dedicated optimization, for example offloading a single drone of some of the task it is in charge of and demanding the discarded operation to another component of the swarm.
In this fashion, \cite{RWSK+18} proposes an opportunistic solution to provide computational offloading for a swarm. The scheme leverages the IEEE 802.11ac communication technology and is based on an artificial neural network with a prediction module specifically designed to provide updated decisions on whether it is preferable to keep the task list of a drone, or offload it.
The method aims to be 
%demonstrates its scalability through numerical results and it has been proved to be 
time-effective in deciding whether to offload tasks to other clusters or not.

The same problem is discussed in \cite{VJK18}, where different swarms are supposed to have different task assignment lists. In particular, whenever one of the clusters is more heavily charged with respect to another, the proposed scheme opportunistically decides to carry out the offloading to another cluster, with respect to a number of different parameters and costs. The effectiveness of the scheme lies into an increased drone lifetime and a shorter response time.

The paper \cite{FCC+18} deals with the problem of delay mitigation for video streaming applications, which is of great relevance int the context of cellular networks, especially in high traffic conditions. The proposal of this work is the design of algorithms and criteria for offload selection and drone positioning. This goal is reached by adopting \acp{UAV} that carry lightweight commercial micro-cells with small form factor, so to better handle congestion for macro-cells. 

The problems of complex and time-consuming calculations to be carried out during a mission, together with the limited resources typically available on-board the \acp{UAV}, are discussed in \cite{MAA+17, CWC+20}. Here, computation offloading strategies are, once again, proposed as feasible solutions to mitigate the issues related to over-exploitation of constrained resources. The approach proposed in \cite{MAA+17} is referred to as \ac{MEC} and is based on the game theory, with a focus on a sequential game involving drones, \acp{BS} and edge servers acting as players with computation tasks. The contribution demonstrates the existence of a Nash Equilibrium and design an offloading algorithm to derive possible tradeoffs between energy consumption and achievable delay. In \cite{CWC+20}, the problems related to temporary overloading of the network are solved relying on \ac{DRL} techniques.

In the context of surveillance systems, \cite{JYK+17} discusses the theme of safety-related applications in hazardous locations. 
The proposal leverages the concept of computational offloading by proposing an \ac{ACODS} solution, that is based on a response time prediction module for providing task offloading decisions. Task offloading management is achieved via a \ac{MPTCP} algorithm, a solution already known in literature, that is based on uses multiple communication interfaces to communicate and transfers the packets through one or multiple interfaces. The adoption of the \ac{MPTCP} performance of the network is improved in terms of reliability, throughput, and delay.

Drones are employable in a number of industrial-grade applications, as argued in \cite{ZVK+17}. In this case, the discussion proposes an interesting point of view on the \ac{IoT} scenario, by proposing an application that provides predictive maintenance to energy distribution systems through the employment of \acp{UAV} with visual capabilities combined with a 5G network infrastructure. In particular, thanks to the 5G communication network, massive capacity, zero delay, elasticity and optimal deployment can be reached in in scenarios with a massive deployment of \ac{IoT} devices for broadband and mission critical services.
%Drones are here conceived as a promising aid to surveillance systems in important and demanding Preventive Maintenance in Service Critical Infrastructures applications, i.e., electricity and gas transmission and distribution networks.

Drones employability has been discussed also in fog computing and \ac{MEC}. In the former case, drones are conceived as enablers for distributed offloading intensive computation tasks to an edge/cloud server \cite{MSH+17, YUA+21}. For this reason, the paper \cite{MSH+17} proposes a game theoretical approach in which drones are the players and the cost function to be minimized is a combination of energy overhead and delay. Thanks to this approach, drones can detect, identify and classify objects and/or situations dealing with intensive tasks such as pattern recognition and video preprocessing.
%Given their constrained nature, drones may not be able to afford such highly intensive computation tasks. 
%During missions where 
%, and with respect to the differences between the applications, the drones need to detect, identify and classify objects and/or situations. Thus, they are brought to deal with intensive tasks such as pattern recognition and video preprocessing. Given their constrained nature, drones may not be able to afford such highly intensive computation tasks. 

Due to the limited computing resources of drones swarm, it is usually difficult to handle computation-intensive tasks locally, hence the cloud based computation offloading can be adopted a sa solution. This become of greater relevance as long as low latency and high reliability are required. The proposal in \cite{HRC+19} is based on fog computing for swarm of drones. Here, latency and reliability are taken as constraints of an optimization problem where the energy consumption is the optimization target function to be minimized.

%On the same theme, \cite{SK17} proposes some solutions. In fact, as previously discussed, this work proposes a thorough analysis of cooperative frameworks for \acp{FANET}.

\subsection{Video Streaming}
\begin{figure}[htbp]
	\centering
	\includegraphics[width=0.7\columnwidth]{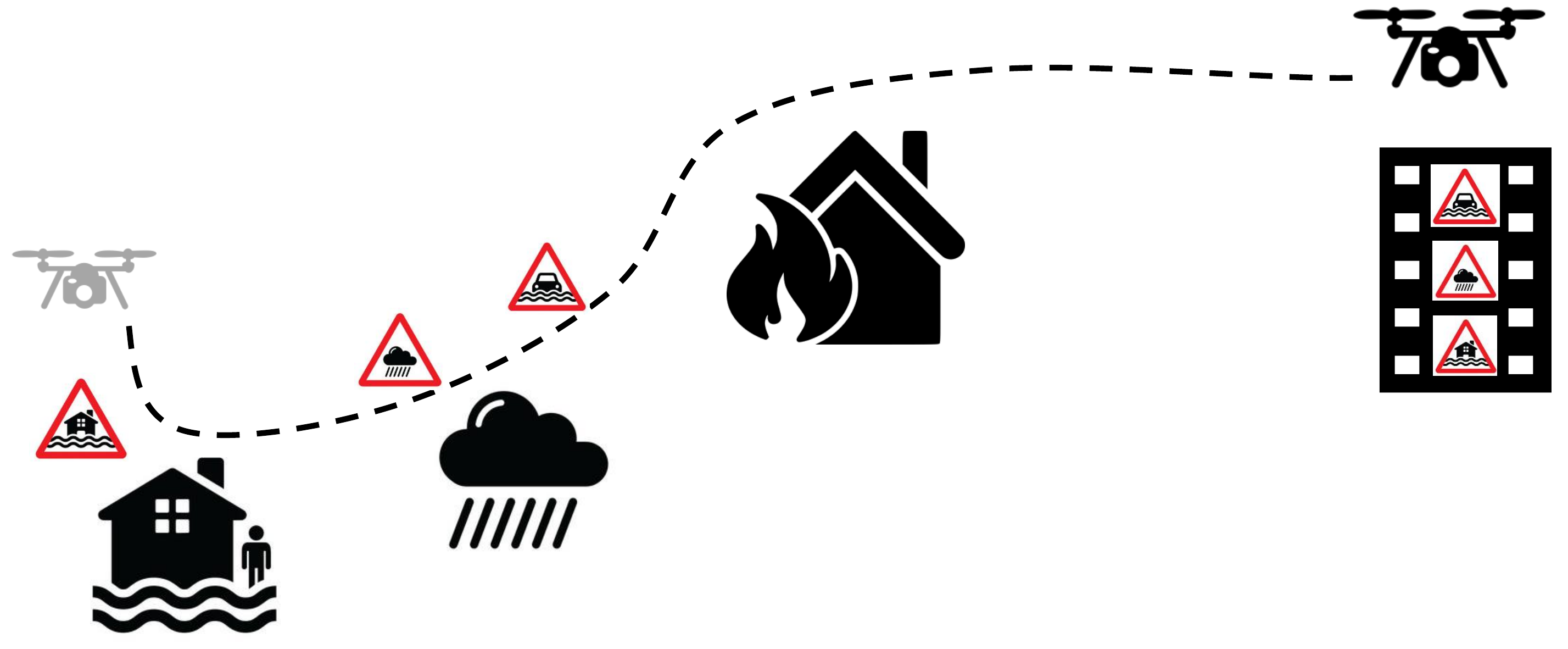}
	\caption{Drone gathering video streaming signals in environmental relevant scenarios.}
	\label{fig:videos}
\end{figure}

Drones employability in both civil and military applications include several possibilities such as search and rescue, coverage, and aerial imaging in different environmental scenarios, as illustrated in Figure \ref{fig:videos}. Once the drone is flying, gathering images and videos may not be as useful as it could be in video streaming contexts.
This aspect may become more and more challenging when referring to a swarm of drones. In this case, coordination among the swarm is not simply beneficial, but becomes a need.

Several contributions so far describe the existing communication and routing protocols, including \ac{AODV}, \ac{LAR}, and \ac{GPSR} protocols \cite{GJV16}. These solutions have been deeply investigated in order to identify their limitations when applied to networks that include drones or swarm of drones.

According to the specific application scenario, the main requirements and characteristics may vary. At the same time, it might happen that drones are asked to fly over multiple areas of interest, which leads to the need for designing optimized trajectories to coordinate the flight of the swarm in a continuous way.
Video streaming based on drones employability may be sensibly enhanced by leveraging multicast wireless video streaming to transmit data. In particular, video multicast streaming using IEEE 802.11 poses challenges of reliability, performance, and fairness under tight delay bounds \cite{VCP17}. The latter assumption in strengthened by the fact that ``ready-to-fly'' consumer drones use a fix-bitrate technique to encode video, for example, 1080p or 4K resolution, which provides poor performance in high mobility conditions where the communication channel may be subject to sudden and relevant fluctuations.
In this work, a dedicated section is dedicated to multi-homing. Since a multi-homed network is defined as a network is connected to multiple \acp{ISP}, the proposed approach helps avoiding connection failures. To reach this goal, it is necessary to integrate in different layers of both the network and the transport layers in both \ac{IPv4} and \ac{IPv6} networks.

Video streaming may also be conceived not only as application enabler but as a method for handling and controlling network traffic and design. As a matter of fact, such a perspective enables interesting perspectives in terms of variable traffic demands, application heterogeneity, and application requirements. 
The contribution presented in \cite{WLS+17} proposes dynamic cloud service placement to handle real-time video streaming and control commands for a drone that is controlled by a remote user. In this case, the main aim is to first grant, and then improve, the \ac{QoE} for streaming services provided by drones using \ac{UDP}.
A similar problem is faced in \cite{LWH+18}, where the drone is remotely controlled and the high demands in both the \ac{UL} and \ac{DL} directions are thoroughly characterized in terms of \ac{QoS}.

Another situation of relevance for video streaming and real-time surveying is data gathering from disaster-struck regions, as discussed in \cite{NHP+18}. In fact, in all cases where the ground network infrastructures are damaged, drones may offer rapid deployment for data source recovery. Without loss of generality, the simple deployment may be ineffective without an adequate energy management strategy. The case study discussed in this work envisages a scenario in which drones operate alongside the wireless network infrastructure to establish a \ac{LoS} link for communication while investigating a power allocation strategy for the \acp{BS}.
%In this contribution, the operating frequency is in the 28 GHz band.
As a result, the work highlights that it is of outermost importance to incorporate the multi-tier heterogeneous network to extend the network coverage in such challenging scenarios.

\subsection{Data Collection and Distribution}
Thanks to the high mobility of drones, their employability has been discussed in a number of applications, such as service delivery, pollution mitigation, farming, and rescue operations. On the theme, \cite{ZVK+17} proposes something similar to what discussed in \cite{MTA16}.
In \cite{ZVK+17}, in fact, \acp{UAV} are proposed as a value-adding utility to \ac{IoT} devices and networks in the context of \acp{MTC}. The contribution represents an interesting survey of all the \ac{UAV}-based architectures that can enhance smart sensors, cameras, actuators, and, more in general, \ac{IoT} devices since they enable a brand-new perspective: an eye in the sky.

Data management is not only a matter of efficient upload (like happens in data gathering applications). In fact, \cite{ASY+18} argues that a high rate, but also cost-efficient, easy-to-deploy, and scalable, backhaul/fronthaul framework, is of outmost importance in the context of 4G/LTE/5G communications and wireless networks.
In this work mobile \acp{BS} are conceptualized as \acp{UFP}. This definition is motivated by the fact that \acp{UAV} are only one of the kinds of flying platforms that can be used in this context. In particular, flying platforms may be drones or balloons.
%(often referred to as \acp{NFP}).
To provide advanced communications services in \ac{FSO}, the proposal in \cite{ASY+18} investigates the feasibility of a framework that involves both backhaul/fronthaul allowing traffic flows between the access and core networks via point-to-point links.

\subsection{Events Monitoring and Management}
In \cite{MTA16} the employment of \acp{UAV} is discussed to demonstrate their countless reference applications. In particular, public protection and disaster relief operations, but also for commercial and governmental services. Some good examples are surveillance and reconnaissance, public safety, homeland security, forest fire monitoring, environmental monitoring, security and border surveillance, farming, or even Internet delivery, architecture surveillance, goods transportation.

Fortunately, not all the scenarios and applications investigated in the scientific literature are related to dangerous or extreme situations. In particular, network capability could be ineffective/insufficient for temporary high loads. The contribution \cite{WCC17} deals with high-action sports game played on a large field. In such events, the idea of using networked drone cameras to gather real-time data can be considered as a great intuition, still challenging. In the design phase, it clearly emerges that distributed approaches yield sub-optimal solutions, an assumption that is motivated by the lack of coordination. Still, a centralized approach to coordination may imply round-trip latencies of several hundreds of milliseconds. For this reason, the contribution proposes a fog-networking based architecture that aims at coordinating a network of drones to capture and broadcast the sports game, with a tradeoff between coverage and streamed video bitrate. Here, latencies are mitigated by a centralized controller that leverages a predictive approach to upcoming locations in order to re-assign the \acp{UAV} to new locations, when needed. The work shows that relay nodes are able to boost throughput while working in real-time.

On the same theme, the work proposed in \cite{JYK+17}, and already discussed in this Section, can be mentioned once again since the proposal may contribute to highly advanced, and safe, surveillance of hazardous locations.

The paper \cite{KQF+19} proposes a cloud-based system that is able to remotely control and manage all the operations carried out by drones and, more in general, robots.
Since this work proposes an entire system with multiple components, the transport layer of each of them. As for the \ac{UAV}, MAVLink uses \ac{UDP}, \ac{TCP}, \ac{USB} and serial. When \ac{UDP} and \ac{TCP} are used, data are forwarded to client applications through dedicated Websockets.
The framework is able to operate in a wide range of commercial applications, under real-time constraints and below visual \ac{LoS} conditions.
Still, the proposal is facing security problems and future research directions involve both transmission and coordination tasks.

\subsection{Lessons Learnt}
This section is focused on all the technological aspects connected with application layer.
However, several scientific contributions discussed have shown a peculiar influence, and non-negligible impact, on peculiar \ac{QoS} parameters, such as latencies and \ac{PDR} (or, complementary, \ac{PLR}).
These quantities have, in fact, a close link with the \ac{QoE} parameters that can be used in quality assessments on streaming services, especially video.
With respect to path planning, the impact on the application layer is relevant. In fact, optimizing the path that drones are programmed to follow during their mission has a significant impact on the optimization of onboard resources, with a consequent extension of mission longevity.
Similar considerations can referred to the optimization problems connected to resource handling and data exchange capacity.

The dedicated study of the problems connected with the application layer demonstrates that mobility is a key enabler for different applications in several scenarios.
In particular, it is possible to guarantee local and temporary offloading thanks to the repositioning of the drones, both in the case of mobile \ac{BS} deployment and when drones are part of a swarm.
In the case of mobile \acp{BS}, they can be re-deployed even when a replacement is necessary.
Thanks to these functionalities it is possible to think about adaptive, temporary and optimized coverage based on the needs of the application scenario.
Some case studies have shown that the typical \ac{KPIs} that are used to measure \ac{QoE} and \ac{QoS} can become input variables to objective functions and algorithms specifically dedicated to the design and optimization of routing. The latter represents an important frontier in the design of drone networks, especially when large swarms are involved.
At the same time, a softwarization-based approach could be used to create customized solutions tailoring the specific application needs.

%% file: src/8-others.tex
\section{Cross-Layer and Optimization Approaches}\label{sec:others}
This Section is specifically dedicated to the details on the following aspects: (i) Path optimization and collision avoidance, (ii) State estimation and optimization, (iii) Network formation and control, and (iv) Coverage analysis and optimization. All the aforementioned topics are peculiar to the application layer. Nevertheless, the optimization processes include aspects and parameters encompassing other layers of the stack, from channel models, signal energy, energy efficiency, throughput and connectivity/coverage requirements at physical layer \cite{XZZ18,WLZ19,ZZ17,DNN19,HYB+17,ZXZ19,ACD+18}, or resource scheduling at data link layer \cite{LWL+19,CAM+19,DNN19}, and this justifies the cross-layer nature of these studies.
Hence, no survey paper so far investigated these problems as a whole.

\begin{figure}[htbp]
	\centering
	\includegraphics[width=0.7\columnwidth]{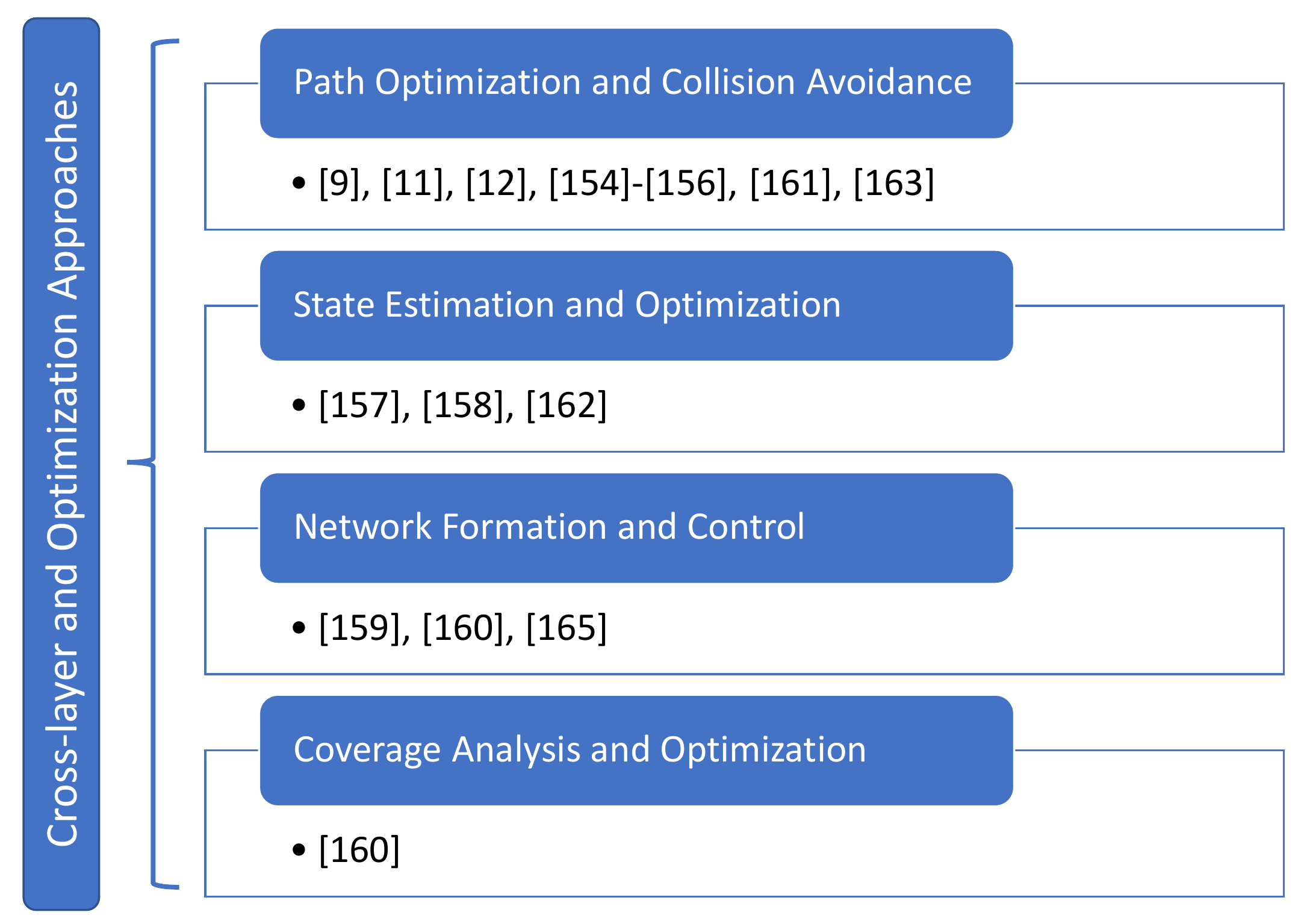}
	\caption{Cross-Layer and optimization section Taxonomy.}
	\label{fig:taxopt}
\end{figure}

The Section closes up highlighting the lessons learnt on the theme.
%In Figure \ref{fig:taxopt}, the overall organization of the present Section is summarized.
%\begin{figure*}[!h]
%	\centering
%	\includegraphics[width=\linewidth]{img/res}
%	\caption{Resource allocation and Optimization Section overview.}
%	\label{fig:sec_res}
%\end{figure*}
Figure \ref{fig:taxopt} summarizes the overall organization of the present Section.

\subsection{Path Optimization and Collision Avoidance}
When discussing drones missions and task list, the specification of the main criteria for path design becomes of critical importance.
Some of the most relevant problems related to path design and optimization criteria have been widely and extensively discussed in \cite{MTA16}, \cite{LWW21}, and \cite{CZX14}.
Since those works are surveys, a deeper insight on the theme of optimizing the path a drone flights over has been discussed in many other works.
For example, the problem of understanding which is the optimal trajectory has been studied in the context of \ac{UAV}-enabled wireless power transfer systems, as proposed in \cite{XZZ18}. In particular, this study discusses the \ac{UAV} maximum speed, which is generally constrained by physical layer aspects, even if the optimization occurs at application layer. This limitation leads to the design of complex hover-and-fly trajectories, which, still, deserve to be optimized. The proposal applies convex programming optimization technique to solve the problem.

Path optimization is a leading trend in research, since power and energy consumption are among the major stakeholders in \acp{UAV} characterization. On this theme, qualitative modeling for energy footprint optimization are widely discussed in several studies, such as \cite{WLZ19,ZZ17,PWW19}, as associated to trajectories design.
Those works are pretty similar since the whole set of parameters involved in all of them is almost the same. Still, they share the wider objective of finding out the minimum energy consumption, the bits allocation (i.e., for each task) and the shortest, yet most efficient, trajectories. Those parameters resulted to be simultaneously optimized. The proposed models are anyway in the need to be verified over wide swarms flying over long distances, while keeping coordinated flight conditions steady.

Another interesting work on the theme of trajectory design and path optimization is presented in \cite{ZZ17}. It assumes that the \ac{UAV} flies horizontally with a fixed altitude while exchanging data with a ground infrastructure.
While on flight, the drones are supposed to communicate with the reference ground infrastructure in an energy-efficient way, that is granted thanks to the compliance to several \ac{KPIs}, including the required throughput and the overall energy footprint. The work derives the contributions due to mechanical propulsion thanks to a theoretical model. The obtained framework is used to design multiple trajectories and iterates the application of the algorithm to demonstrate its versatility.
The theme of complex trajectory design is discussed in \cite{PWW19} too. Here, the mission envisions frequent turnovers between hovering phases and altitude variations. Among the most relevant advantages of the proposed model there is its wide applicability to multi-copters.

In \cite{LWL+19} an energy-efficient scheme is proposed to include multiple variables for the optimization, such as the energy budget of each \ac{UAV}, the number of tasks that each drone has to accomplish, and the type of data exchanged, together with the \acp{UAV} reference speed. Here, both uploading and downloading phases are handled relying on \ac{MEC} solutions.

The study presented in \cite{JSL19} discusses the problem of hybrid delivery employing Truck-drones. in particular, this work frames an optimization problem related with the connection between the payload size and the energy spent while moving over certain trajectory. Moreover the proposed framework includes the possibility that No-Fly zones are within the mission plan. Even though the invasion scenario only include a single truck and a single drum, multiple vehicles are clearly among the future research perspectives. Even more so because the problem can be complicated with different scales or shapes of no fly zones, in which case fleet coordination becomes mandatory. 

\subsection{State Estimation and Optimization}
Several works deal with the drones state estimation with the precise aim of identifying methods to minimize energy consumption, rather than designing trajectories or maximizing area covered during the mission.

The ability to communicate state information is discussed in \cite{CAM+19}, a proposal that aims at achieving the maximization of users throughput in a three-dimensional space where dangerous conditions are verifying. The mathematical formulation of the problem include several parameters, such as overall throughput, remaining battery capacity for each \ac{UAV} in a swarm, and several others. Those variables are modeled on a multi-period graph.
Something similar is done in \cite{DNN19}, a contribution that formulates an optimization problem to specifically address natural disaster scenarios in real-time conditions with relay-assisted \acp{UAV}.

State estimation represents the starting point of the study proposed in \cite{HYB+17}, in which multi-objective optimization algorithms are proposed. Their aim is to allocate tasks and plan paths for multiple \acp{UAV} forming a swarm.
The proposal leverages the \ac{GA} approach to minimize the time in which the mission gets completed and includes area coverage and communication path. As a such, the problem also embraces network connectivity aspects. The solution can be fine-tuned to prioritize coverage or connectivity according to mission commitments.

\subsection{Network Formation and Control}
Two major survey contributions discussed network formation and control-related aspects in the context of the \ac{IoD} \cite{SK17} and \cite{GJV16}.
A detailed analysis of those themes have been extensively discussed in the form of optimization problems in \cite{LDB11, ZXZ19, ACD+18}.
In particular, \cite{LDB11} discusses network formation as the result of an optimal task assignment problem. The formulation leverages the coordinated nature of a swarm of drones that are communicating through \ac{A2A} links. Together with network formation thanks to the cooperation among multiple drones, efficient mission plans are constantly generated and updated to refine the obtained results.

Similarly, \cite{ZXZ19} proposes \acp{UAV}-enabled wireless communications between multiple ground nodes that are subjected to specific requirements, such as throughput, energy consumption optimization, and limitations on communication-related energy expenditure. The proposal models propulsion power in close correlation with physics-related phenomena. The model reaches the energy minimization objective as a results of a non-convex problem.
Therefore, the preliminary fly-hover-communicate assumption is empowered with a more sophisticated design criteria that includes the hovering location and duration, together with the flying trajectory design.

In the same field, another possibility is related to \ac{BS} offloading and coverage area enhancement, as in 5G wireless communication systems \cite{ACD+18}. Another interesting application is aided relaying, with \acp{UAV} providing reliable wireless connectivity between users or user groups in adverse environments.

\subsection{Coverage Analysis and Optimization}
%On coverage analysis, the survey \cite{CZX14}
An interesting research perspective is the \ac{BS} offloading and coverage area enhancement, as in 5G wireless communication systems \cite{ACD+18}.
\acp{UAV} are often involved in information dissemination and/or data collection. In these cases, drones are arranged to fly over a certain area of interest and gather data.
Here, aided relaying envisions \acp{UAV} providing reliable wireless connectivity between users or user groups in adverse environments.
This use greatly reduces the overall energy consumption at the end-node level, thus extending the network lifetime.

\subsection{Lessons Learnt}
Many lessons can be learnt from the cross-layer approaches discussed in this section. All the topics presented (path and trajectory optimization and design, state estimation and optimization, and network formation and control) are of great importance to exchange data among \ac{UAV} and/or between \ac{UAV} and \ac{GS}, while optimizing different metrics like energy consumption, connectivity, throughput, delay in data exchange, task accomplishment, overall mission time, connectivity, and coverage area. Especially the minimization of energy consumption pushes such optimization strategies, because of the positive implications on the increase of the flight time, battery duration and amount of data delivered to destination.

Nevertheless the metrics to be optimized are too many and often in contrast each other, so trade-offs have to be reached especially when the optimization strategies are implemented through multi-objective algorithms which are intrinsically much more computationally complex with respect to their single-objective counterparts. Very often, and as usual, simplifications are introduced to make these algorithms more computationally tractable, but with the risk of oversimplifying real application scenarios.
Another weak point resides in the high mobility of drones, that brings to continuous and frequent changes in the network topology and connectivity. Accordingly, also the state information (position, velocity, etc.) must be continuously updated to recover from unavoidable mismatches between the estimated state and the real one, and this increases the overhead in exchange of control information among drones and/or with \acp{GS}.

%% file: src/9-misc.tex
\section{Security and Privacy Aspects}\label{sec:miscellaneous}
\begin{figure}[htbp]
		\centering
		\includegraphics[width=0.8\columnwidth]{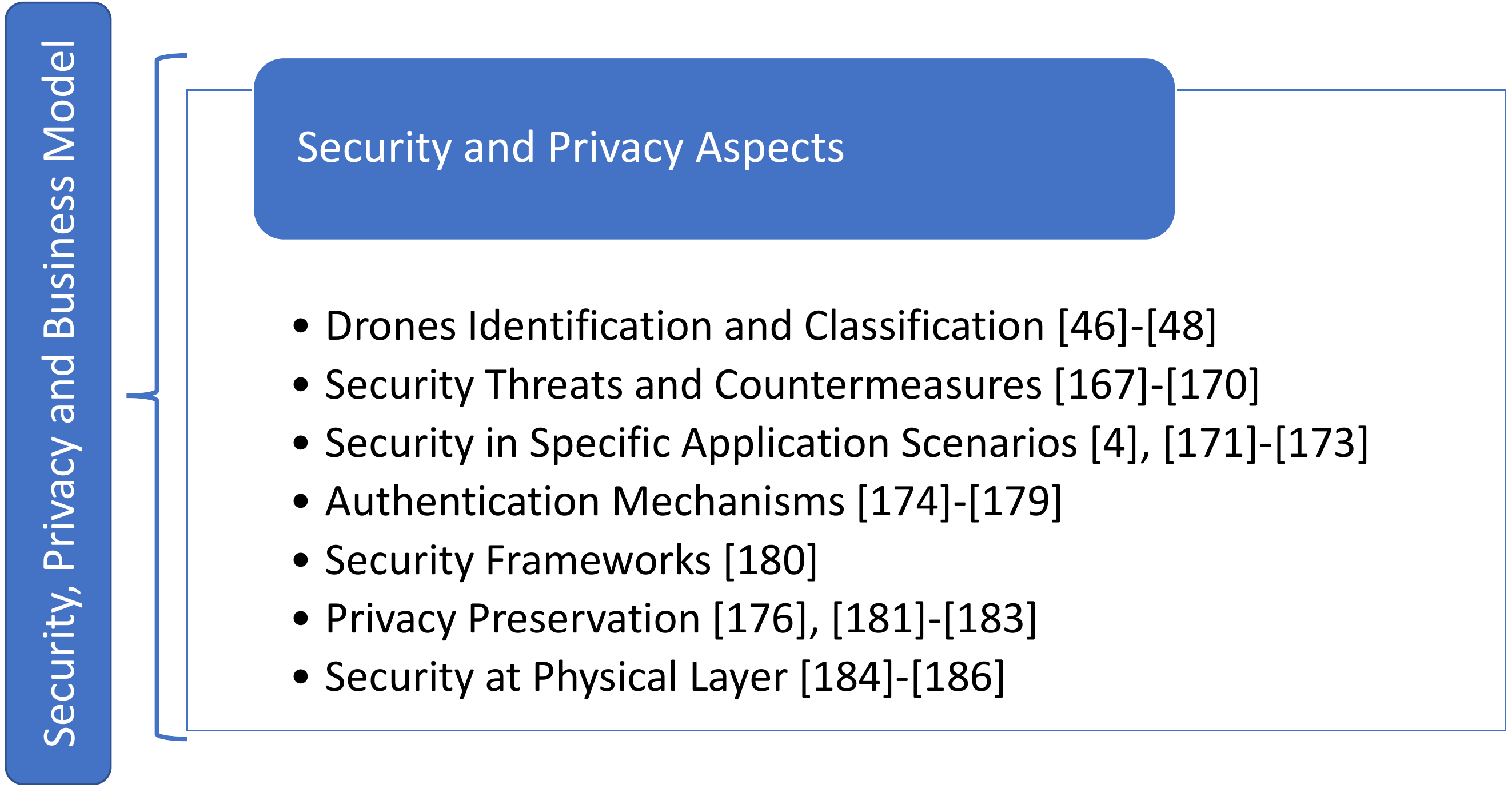}
		\caption{Security, privacy and business models Taxonomy.}
		\label{fig:taxsecurity}
	\end{figure}

%This Section discusses some remaining topics connected to the main economical aspects in the employment of drones, and also addresses security aspects. The Section closes up highlighting the lessons learnt on the theme.
This Section discusses security and privacy related topics that are of relevance in the context of the \ac{IoD}.
In Figure \ref{fig:taxsecurity}, the overall organization of the present Section is summarized.

\acp{UAV} can be considered as complex flight control computer-aided systems with, at least, one sensor. \acp{UAV} is programmed, off-line, before take off, or controlled, on-line, to fly over a specific pattern, for example within a facility. During this period of time, it is supposed to receive sensor data, thanks to on-board units, apply some processing routine to the sensor data, as to detect an unacceptable level of difference in data features, determine a new flight instruction for the vehicle based on the processing. Afterward the new flight instructions are sent for controlling the flight of the vehicle.

According to the reference context, several security aspects may be involved.
Conventionally, commercial surveillance systems are managed by humans, that regularly check in at guard stations and make observations. Surveillance systems have progressed into closed-circuit television monitoring, and more recently integrated systems have been developed to include video cameras installed at strategic locations in a facility. These video cameras are in communication with a centralized remote monitoring facility, and operators visually sweep the facility from the monitoring center.
%These tours are scripted and timed at a user request. 
Upon discovery of suspicious activity, the operator can engage in a response plan.
A surveillance solution has been employed by the military through surveillance unmanned aerial vehicles, commonly referred to as drones. Military surveillance drones are capable of flying over large areas such that video surveillance can be achieved. Military surveillance drones are very expensive and address surveillance of very large outdoor areas. However, they are not practical for businesses that want to maintain security at a single location or small group of locations such as a warehouse or a manufacturing facility.

In \cite{Altawy16,CSG+18}, the leading theme is security in the context of the communications among drones. The main concern on the theme is related to the security and the privacy of the data exchange processes.

%%%%%%%%%%%%%% DRONES IDENTIFICATION AND CLASSIFICATION %%%%%%%%%%%%%%%
\subsubsection{Drones Identification and Classification}
Modulation schemes are studied in some papers to identify and classify drones \cite{EEA+19,SAE18,KPR19}. These contributions aim to identify drones on the basis of modulated RF signals \cite{SAE18,KPR19} and classify them based on their unique features \cite{EEA+19}.
The method proposed in \cite{SAE18} aims at preventing \acp{UAV} from unauthorized flights over reserved areas. The solution proposed to tackle this problem is to use \ac{SDN} to process the \acp{UAV} control signals. On the basis of a detailed analysis of the \acp{UAV} control methods, that rely on different modulation techniques, this study proposes a simple detection technique through measurement campaigns that are conducted on the field. Two different commercially available \acp{UAV} are used to extract the main features of their control signals through a \ac{SDN}.
Another \ac{UAV} detection technique is proposed in \cite{KPR19}, for drones identification purposes. The paper carries out an analysis of the radio communication between the \ac{UAV} and the remote control, as related to a specific type of drone. The radio communication channel is studied with reference to the number of channels adopted, their frequencies, the frequency utilization method and the modulation technique. This last aspect is used to study how the communication is maintained between the \ac{UAV} and the remote controller through a signal analysis.
A classification method for \acp{UAV} based on the analysis of the RF signal is proposed in \cite{EEA+19}, to avoid security and privacy issues and preserve public safety. Drone pilots are identified through the analysis of RF signals captured from the drone controllers, in presence of wireless co-channel interference, namely, WiFi and Bluetooth. The methodology adopted to detect the \acp{UAV} is first described, testifying the robustness against false alarms and missed target detection. The RF features are extracted to classify the controller signals of \acp{UAV}. A RF fingerprinting technique is proposed, able to capture the unique characteristics (the ``fingerprint'') of each \ac{UAV} controller. The proposed approach is then compared with different Machine Learning algorithms, to testify its effectiveness for \ac{UAV} classification.

\subsubsection{Security Threats and Countermeasures}
Some papers on security in drone networks are mainly focused on the description of the security threats and the related countermeasures in drones communication \cite{SA20,BP18,PBC14,TWL+20}.
The importance of cyber-security in \ac{UAV}-based networks is highlighted in \cite{SA20}. The main goal of this work is to describe the main security threats involving networks based on \acp{UAV}. To this end, the main challenges for networks of drones are first presented, among which the need to verify the identity of a drone in some application scenarios, proposing a classification of the drone based on its functionalities, range, altitude, endurance, etc. This classification is the starting point to describe the main attacks and security threats that can be carried out on drones, that testify how easily a \ac{UAV} can be attacked, and its functionalities compromised. The network architecture is described, highlighting the main security threats as referred to each part of the network. Finally, a risk assessment scheme is proposed, that contains integrity, confidentiality and availability. The necessary steps to provide security to drones are considered as concluding considerations of this study.
More focused on secure communications among drones is the study carried out in \cite{BP18}. The security-related risks are addressed in this work, especially focusing on the ``taking over control'' type of attack. It is a software centric type of attack, and exploits the vulnerabilities of the standardized communication and control protocol. To this end, the possible approaches adopted to carry out attacks that exploit the protocol vulnerabilities are discussed in detail, also highlighting that they can be easily exported to other communication protocols. Then, some recommendations are made to improve the security of drones operations. They mainly consist in a strong limitation of the secret information disclosed during  the binding phase, an extension of the binary string that represents the secret information exchanged, and the adoption of cryptographic methods, that are not actually present in the \acp{UAV} communication protocols.
Other vulnerability issues are described in \cite{PBC14}, with reference to a specific type of drone provided with video stream capabilities. This paper raises a warning flag from the privacy and resource accessibility points of view, which should absolutely not be set free in communication networks of any kind. This work focuses on the well known security vulnerabilities like an unencrypted Wi-Fi network connection and the management of the linux-based operating systems running on the drone by a user. This study also shows how the drone can be hacked, illustrating some possible attack scenarios that exploit the port scanning to carry out a combination of attacks that can take the full drone control. Then, some operations are described to secure the WiFi connection of the drone and  its operation with the official smartphone app or third-party software.
The work \cite{TWL+20} discusses security of \acp{UAV}, but considering holistically the interactions among \ac{UAV} systems from the point of view of attack methods and threats. The study is conducted from the perspective of a common security analysis among different systems, and the related security challenges to be tackled. The main goal is to derive a general security model, which can be exploited in different \ac{UAV}-based systems, and that can analyze the security problems. The security analysis at the basis of the model is carried out by addressing both attacks at several layers of the stack, and potential threats, and analyzing security from the point of view of aerial, ground, marine and space unmanned systems. A summary of the security issues for the different systems paves the way to a discussion on the main security challenges for unmanned systems and the future directions that can be taken to tackle them.

\subsubsection{Security in Specific Application Scenarios}
Papers \cite{DCG17,GAK+20,B13,NNB+20} contextualize the \ac{UAV} security analysis to specific application scenarios.
The security aspect is addressed in \cite{DCG17}, specifically referred to drone-assisted public safety networks. Drones can be fruitfully employed in public safety networks, where the supported services aim to protect people and environments from a wide variety of threats, because they enhance the communication capabilities of such wireless networks, since they can reduce coverage gaps and network congestion. Nevertheless, security is of great importance in this context, because sensitive data and/or critical information can be transmitted among drones. This contribution analyzes the most relevant security threats and the related countermeasures, that cover all the aspects of data confidentiality, integrity, availability, authenticity, privacy preservation and non-repudiation \cite{DCG17}.

Another scenario, the smart farming environment, is considered in \cite{GAK+20} to study security and privacy issues related to the use of IoT and communication technologies involving the use of \acp{UAV}, that are subject to cyber-security threats and vulnerabilities. This is mainly an overview paper, whose goal is first to provide an overall description of the communication architecture in the smart farming landscape, at different protocol layers. Then, the security and privacy issues are identified, in terms of security, privacy, authentication and authorization, and compliance with regulations. Cyber-attacks are then presented and classified into data, network, supply chain, and other relevant attacks. In parallel, the current research in cyber-security is discussed, describing also the main contributions and weak points of the works on this topic. Finally, the most interesting research challenges are addressed, that cover different areas ranging from network security to supply chain, artificial intelligence, machine learning, access control and information sharing \cite{GAK+20}.

Security plays a very important role also in military applications. For example, drones are widely used in the context of counter-terrorism \cite{B13}. Here, the concept of security is strictly related to the ability of inspecting active theaters of war. The conventional wisdom on drone warfare holds that drones are considered secure if, and perhaps only if, they have a high effectiveness in striking and disabling terrorist organizations. Nevertheless, the legitimacy of their employment is still under discussion. In fact, there is a diffused willingness to mitigate the strategic use of drones, at least under recognized standards and norms.

\subsubsection{Authentication Mechanisms}
Authentication mechanisms in networks of \acp{UAV} are described in \cite{PL20,GKH+19,DWZ+18,WSB15,GS20,TZA+20}.
The papers \cite{PL20, DWZ+18,GS20,TZA+20} are entirely focused on authentication mechanisms in \ac{UAV}-based networks. In particular, \cite{PL20} leverages a mutual authentication protocol between \acp{UAV} and a ground station. This mechanism aims at overcoming the traditional cryptographic techniques which are inefficient in networks of \acp{UAV}, because of their tight resource constraints, together with the nature of the wireless medium that is ``open'' by its own nature. The proposed mechanism exploits the so-called Physical Unclonable Functions (PUFs), that are the unique physical identity of a device, that depends on its hardware characteristics. Based on PUFs, a challenge-response pair is unique, and is used to authenticate the \ac{UAV} to the ground station, through the well known time-based one-time password algorithm. It is used to exchange the challenge-response pair between the \ac{UAV} and the station in the initial phase of the proposed algorithm, that is developed in such a way to avoid an adversary to overhear, duplicate, corrupt, or alter authentication data.
In a similar way to what previously discussed, the peculiarities of the \ac{IoD} are taken into account by the work presented in \cite{TZA+20}, where data exchange between mobile users, a drone working a server and the remainder \ac{IoD} infrastructure.
Even though the operating context is similar, \cite{GS20} focuses on an efficient privacy-aware scheme specifically designed for authenticated key agreement. The peculiarity of the contribution is that fact that no secret keys are store within the devices, still granting the desired security features.

Another discussion on authentication vulnerability in consumer WiFi drones is found in \cite{GKH+19}. This work proposes a security assessment that illustrates how to manipulate a WiFi drone by a software suite, and without the need for additional hardware equipment that can carry out jamming or eavesdropping attacks, like for example a \ac{SDR}. The assessment procedure show that a deauthentication attack can be effectively carried out; it is a kind of Denial of Service (DoS) attack where a malicious entity launches deauthentication packets to a wireless access point (AP), making the AP believe that hey come from a real client, thus taking down the drone. The drone vulnerabilities are assessed by standardized assessment models. Some prevention strategies are also discussed, like the adoption of encryption protocols and best password practices. The work presented in \cite{WSB15} is an important contribution that tackles both privacy and security from a technological point of view, for example securing transmission and reception activities. More specifically, a mechanism for secure communications in \ac{UAV}-based networks is proposed. It supports key agreement, user authentication, non-repudiation, and user revocation, and all this in one algorithm. This mechanism is then exploited to build a secure communication protocol for \ac{UAV}-based applications, which aims at maximizing the energy efficiency. The performance of the proposed scheme is evaluated in a real testbed.

\subsubsection{Security Frameworks}
A security management framework is proposed in \cite{HZB+20}, which aims at automating the orchestration, configuration and deployment of lightweight Virtual network Security Functions (VSF) in Mobile Edge Computing (MEC) nodes, that are placed on-board \acp{UAV}. Different factors are considered to optimize the security orchestration. The contribution of this work is manyfold. First, a MEC architecture is proposed and implemented. With this background, the orchestration mechanism is provided, to manage security in \ac{UAV} networks, and taking into account different metrics and conditions for the VNF placement on \acp{UAV}, like operating capacity, battery, atmospheric conditions, computing resources (RAM, disk, CPU), and network metrics. The applicability of the proposed solution is then evaluated in a real testbed and use case, testing the performance of the framework from both the MEC node and the \ac{UAV} points of view.

\subsubsection{Privacy Preservation}

Privacy is also a critical issue in the \ac{IoD} environment, as described in \cite{B19,DWZ+18,LHK+18,LS16}.
This aspect is developed in \cite{B19}. This paper discusses the security challenges from the legal point of view, starting from the assumption that the legal framework for \acp{UAV} is very complex in the EU scenario. The main goal of this paper is to point out the key open issues in the legal field, that include public security legislations, telecommunication law, rules on product liability, criminal and insurance law, privacy and data protection. An overview of the new EU rules is first provided, describing the approach followed by the EU to guarantee security and safety for aircraft operations in the sky. The open issues on the security topic are then addressed, synthesized into privacy preservation, telecommunications and cyber-security breaches, registration of \acp{UAV} and the identifiability of their operators and pilots, liability and enforcement regulations. The resulting picture confirm the complexity of the legal framework that necessitates harmonization and clarification of rules \cite{B19}.

A major privacy concern is directly linked to the idea that drones are moving all around urban and suburban areas, impairing privacy of the people, especially in case of amateur drones. This is the theme developed in \cite{DWZ+18}, where different safety, security and privacy threats are discussed in the deployment of amateur \acp{UAV}. The main goal of this review paper is to illustrate the state-of-the-art studies on this topic. A framework is then proposed, that exploits cognitive \ac{IoT} to provide amateur drone surveillance with a higher level of intelligence through cognitive tasks, i.e., sensing, data analytics, knowledge discovery and intelligent decision making \cite{DWZ+18}. A case study is also provided, where detection of both authorized and unauthorized drones is provided, to check the correct operations that the various types of drones can carry out.

The contribution \cite{LHK+18} specifically addresses privacy issues of \ac{IoD} in both civilian and military architectures and a certain number of security and privacy requirements.
The main contribution of this paper is to study the \ac{IoD} architecture and its main security and privacy requirements. To this end, the main security and privacy challenges in the \ac{IoD} scenario are described, that range from the privacy leakage, that occurs when sensitive data are collected by unauthorized drones, to data sharing in the cloud that can be readable by third-parties, and malicious interference in control sites that occurs whenever a drone can be managed, controlled in all its movements by an attacker to compromise a system. For all these issues, solutions are proposed aiming at protecting the privacy of sensitive data. They consist of authentication solution for privacy protection to lightweight cryptography protocols for security and privacy protection in case of stored data. A lightweight data protection protocol is also proposed for data stored in a cloud server. It is based on the well-known Identity-Based Encryption (IBE) protocol, but it is made lightweight to adapt it to the resource-constrained drones. 
Finally, future research directions are also suggested to tackle the main security challenges explained above.

The use of drones drones is widely used also in commercial applications, and this raises concerns about the related privacy and safety issues arising from a domestic employment of drones. This is the main goal of the work \cite{LS16}, that carries out an exhaustive review of the state-of-the-art literature on drones thought for commercial uses, but focusing on the so-called ``tecnoethical'' aspects, i.e., the study of the impact of drones technology on ethics. Social and ethical concerns are studied in different key-areas, also providing insights on the state of the public knowledge on commercial drones, but seen from an ethical point of view. Finally, this paper points out also the need to increase the level of attention towards the potential negative ethical consequences in the use of commercial drones.

%One of the By the way, it is worth noting that, once again, the military and the civil applications must be differentiated. 

%On the other hand, commercial surveillance systems are usually monitoring both regular and sporadic facility access. Surveillance systems usually include \acp{CCTV} monitoring, with cameras installed at strategic observation points. The main drawback of this technique is that these points are fixed by design. Over time, a facility may change, for example as a result of newly installed points of interest for visitors. Moreover, it is not always possible to install cameras in optimal points. For these reasons, drones may be employed as a (re)deployable monitoring units.
%The closed circuit nature of these systems also implies that operator may react upon the discovery of suspicious activity. In fact, in case of emergency, the operator can engage in a custom designed response plan by tracking the suspicious activity switching its point of view from one camera to the other. Such process is fully functional in the hypothesis that no blind-points are left for observation. A drone may effectively participate in the monitoring process as its by-design mobility may ensure an optimal and continuous tracking of suspicious movements throughout the facility. This process can even be automated using tracing routines \cite{LS16}.

\subsubsection{Security at Physical Layer}
Some papers tackle the security aspect in \ac{UAV} networks from the physical layer perspective \cite{SAB18,LXD+18,WMZ19}.
From this point of view, an interesting research topic about security is jamming, as testified in \cite{SAB18,LXD+18}. In \cite{SAB18} jamming is discussed as a kind of attack in optical \ac{UAV} network architectures. The jamming attack is presented as a kind of Denial of service (DoS) attack, where the attacker sends intentionally and interference signal in the communication link between \acp{UAV} exploiting the Free Space Optical (FSO) link. A technique for jamming attack detection is proposed, able to detect the presence of jamming signals in combination with the transmitter signals. A countermeasure approach is also presented, allowing to recover the transmitted signals in presence of a jamming attack.
Another solution to the problem of jamming is presented in \cite{LXD+18} for \ac{UAV}-aided cellular systems. This paper proposes a \ac{UAV} relay scheme based on the Reinforcement Learning (RL) approach, that exploits the RL and transfer learning to optimize the relay signal power against jamming, without knowing the network topology and the models of message generation and jamming signal. The proposed scheme estimates the received jamming power, the condition of the links between jammer, \ac{UAV} and the base stations. From these estimations, the \ac{UAV} chooses the best relay policy that improves the network resistance towards jamming attacks.  neural network is exploited to learn from past \ac{UAV} relay experiences in similar cellular anti-jamming communication systems. Performance bounds, including bit error rate and computational complexity of the proposed relay scheme, are derived through simulation.
A more general description of physical layer related security is carried out in \cite{WMZ19}, where eavesdropping and jamming attacks are taken into account in wireless networks including \acp{UAV}. Eavesdropping and jamming attacks are first described from two different points of view. In the first, UAV-ground communications are analyzed as more subject than terrestrial communications to eavesdropping and jamming attacks by malicious ground nodes. in the second, malicious \acp{UAV} can be more effective in carrying out eavesdropping and jamming attacks to terrestrial communications, if compared to malicious ground nodes. Starting from these premises, countermeasures are proposed to secure \ac{A2G} communications from terrestrial eavesdropping and jamming, and to secure terrestrial communications from \ac{UAV} eavesdropping and jamming.

\subsection{Lessons Learnt}
In the military context, both the role and employment are under discussion nowadays, since a regulation is highly wished from many players.
Turning from the military to the civil context, the considerations connected with privacy are due to lack of legislation. Sometimes, the problems are related to the fact that not all the countries are using same regulations and/or laws on the theme.

%% file: src/10-disc.tex
\section{Discussion on the Main Findings and Future Research Perspectives}\label{sec:disc_future}
The aim of this Section is to discuss the details of the main findings of the analysis carried out so far. At the same time, in this part, the present contribution is specifically focused on the main lessons learnt from the surveyed literature. Particular attention is devoted to the strengths of the main results found and the related issues.

\subsection{Discussion on the Main Findings in the Surveyed Literature}
The analysis of \acp{UAV} connectivity demonstrates an extremely wide set of possible technologies involved for creating both \ac{A2A} and \ac{A2G} links.
In terms of channel modeling and connectivity analysis, several studies show that there is a tight bound between the two aspects.
Several proposals aim at creating reliable mathematical formulations that include as much aspects as possible, in terms of statistical variability and signal fluctuations in the communication channel. They are very useful to analyze the increase in coverage performance, the optimal allocation of resources at different layers of the protocol stack, and the improvements in path reliability for routing, path planning and position optimization strategies.
Although the presented contributions are valuable, connectivity and channel models suffer from multiple problems, mainly connected to real-world operating conditions, such as \ac{NLoS} conditions and/or presence of obstacles.
To solve such issues, the most promising solution seems to be the choice of the most suitable communication technology, that matches the specific application needs.

At data link layer, the current state of the art suggests that drones may benefit from significant modifications to \ac{MAC} layer. For example, cooperative schemes are useful to improve communication efficiency. Synchronization is another important point that arises from the analysis carried out. To reach optimal resource allocation, scheduling is widely considered as the most promising strategy, especially in high densely populated scenarios in which drones are configured as swarms.

Network layer-related problems are, once again, strictly related to enhancements in connectivity; drones are proposed as relaying units to strengthen existing networking solutions.

Turning to application layer considerations, the solutions that have been investigated so far demonstrate that applications may strongly benefit from the adoption of cooperation among drones, as suggested in lower layers solutions. Some of the most interesting works show the suitability of drones for advanced applications in continuous monitoring of widely distributed phenomena, surveillance among others. What emerges as a clear perspective of the employment of drones is the fact that the inspection of difficult access areas is sensibly simplified. Almost all of the referenced works on this topic agree that the quick on-site deployment of inspection system composed by one or more drones generates crucial benefits.

The analysis and optimization of throughput has been discussed as a way to increase the performance of communication systems in the context if both \ac{A2A} and \ac{A2G} links for data exchange, and to optimize different metrics in \ac{IoD} systems (mutual distance, positions and paths of drones, spectrum and energy efficiency, network topologies). In such context, the throughput optimization in most cases translates into solving an optimization problem, whose solution is computationally expensive, or found approximately, or only in simplified scenarios. Furthermore, throughput maximization is counterbalanced by a higher energy consumption, which can become a serious drawback for battery-powered drones.
Given the formulations of the optimization problems proposed for resource allocation, the currently available technological landscape seems to be a sensibly limiting factor, in terms of onboard energy availability and computational capabilities.

Despite the wide agreement on the fact that drones may sensibly enhance all the processes in which they may be involved, some legal issues are clearly arising. In fact, there is too much lack of legislation in many countries to allow a massive employment of networks of drones as enabling technology. Moreover, some of the biggest players in this market argue that legislation should be somehow homogeneous across borders, thus allowing industrial players to act on a worldwide scale.

\subsection{Research Challenges and Possible Future Directions}
In this section, the main research challenges and possible future directions in the \ac{IoD} field are discussed.

As for communications among drones, the main trends are related to the extension of the covered area and the employability of communication protocols to guarantee a target \ac{QoE} in both \ac{A2A} and \ac{A2G}. In the 5G perspective, it could be useful to standardize communications and creating homogeneous frameworks and platforms that provide real integration among heterogeneous technologies, rather than pushing on performance (e.g., increasing the data rate or decreasing latencies). This process is often referred to as \textit{Softwarization}.
It is worth noting that a similar phenomenon is taking place in the \ac{IoT} world., where a number of examples and dedicated studies are specifically focused at integrating the existing \ac{IoT} technologies rather than developing new ones \cite{GBF+16, BDP+16, ZPS+18, MMN+17}. Specific attention is devoted to the security aspects of such interactions \cite{SKP+18}, thus proposing innovative solutions based on Blockchain and smart contracts \cite{RMC+18}.

Another example can be found in the application of \acp{VLC} and \ac{ICN} technologies to drones.
Some works have already contributed on the \ac{VLC} topic \cite{YCL14,CD17,RRL12,SKU16,MBM14,WGP14,LTK+12,WWY14,YTO+14}, especially for what concerns the analysis of the IEEE standards \cite{YCL14}, modulation schema and light patterns at physical layer \cite{RRL12}, MAC layer analysis and performance of network topologies \cite{SKU16,MBM14}, simulation software \cite{MBM14,WGP14}, applications of the \ac{VLC} system architecture \cite{LTK+12,YTO+14} with its strengths and weaknesses  \cite{WWY14}, and related performance evaluation \cite{CD17}. It would be very interesting, as a future research direction, to investigate the possibilities of application of this novel technology to the \ac{IoD} field.

Another research challenge in drone communication concerns security and privacy in data exchange. Strictly related to privacy is the legislation issue. Legislative interventions could also be of importance on the theme of privacy, even if some problems have still to be solved \cite{DWZ+18}. In particular, one of the concerns is related to the idea that drones moving all along urban and suburban areas could impair privacy of the people. To the best of the authors knowledge, very few works analyze these aspects: the work \cite{WSB15} treats these themes from a technological point of view, for example securing transmission and reception activities. The contribution \cite{LHK+18} specifically addresses privacy issues of the \ac{IoD} in both civilian and military architectures and a certain number of security and privacy requirements. The work \cite{PBC14} raises a flag from the privacy and resource accessibility points of view, which should absolutely not be set free in communication networks of any kind. Nevertheless, research efforts could be made in this direction to further analyze and improve these aspects. 

Another interesting research line that definitely deserves attention is the application of \ac{ICN} in the context of the \ac{IoD} \cite{BLP+18}.
In particular, what clearly emerges from the study of the related state-of-the-art, is that the application of the \ac{ICN} in the context of \acp{ITS} seems not to be already as mature as it could \cite{STG16, CM16}.
As a matter of fact, the authors believe that this may represent a limiting factor since the \ac{ICN} architecture proposes one of the best solutions to support communications over users, especially in mobile-by-design scenarios \cite{Kerrche2020,SA18A}.
Among the main reasons that make this of relevance, there is the increasingly emerging limitation of the host-centric nature of the Internet, as known today.
Indeed, there is a continuous request for seamless mobility, coming from users of different nature, and pushing to the limit the Internet capabilities.
The proper support to intrinsic mobile scenarios becomes necessary in those cases in which the handover is frequently verified. The latter is, in fact, an extremely well-known problem in the telecommunications world as, in a mobile scenario, the user remains connected to the network, and does not experience any discontinuity of service, if, and only if, the coverage is acceptable.
Since the global connectivity cannot be granted by using a single access point (i.e., a single cell in the case of cellular networks), the problem of the handover arises when the user moves from one access point to another. In the case of cellular networks, this happens when the user moves from one cell to one of the neighboring ones.
The intrinsic mobility that characterizes \acp{ITS} systems, both from the points of view of users and drones producing data, places serious constraints on sustainability.
It is worth noting that routing problems also arise in these cases, because of the need to realize in real-time which access point is the \ac{UAV} connected to, and both in \ac{DL} and \ac{UL}. If the routes are not known, the \ac{UAV} will experience a loss of connectivity and information exchange.
The \ac{ICN} communication paradigm leveraging the Publish-Subscribe communication scheme represents a good candidate for granting \ac{QoS} design criteria, for instance communication latencies, \ac{PLR}, resilience, and throughput. The joint analysis of even some of these aspects, combined together in \ac{ITS} applications, could be an interesting challenge for the future research.

\subsection{Other Remarks}
Some other considerations are related to technology and to the massive production and widespread diffusion of autonomous vehicles, even more so in industrial applications. It is well known that there is a close relationship between costs and the development of a technology \cite{LALL1992165}. In this context, advanced industrial countries select, and apply without costs, all the innovations that are more immediately useful. As the general level of capital accumulation rises, more capital-intensive technologies become economical.
Since industrialization is driven by technology applicability, and considerably boosted by rapid prototyping, the need for a drone to carry out part of industrial process monitoring or supply chain support/optimization depends on its cost and its accessibility.
The consideration can be completed by reasoning on the legal aspects of the question. Since there is not a unified legislation on the usage of unmanned vehicles, the seamless employability is still far from now. In a global economical context, this means that a firm or an industry may find economical and legal barriers to their welcoming attitude towards drones. In all the analyzed works, these aspects have been neglected, but they surely represent a value added in future research on the \ac{IoD} theme.

%% file: src/11-conclusions.tex
\section{Conclusions}\label{sec:conclusions}
This work aims at providing a thorough overview on the research activities on the \ac{IoD} network architecture. To this end, the available scientific literature has been studied in detail and classified in order to identify all the current research trends.

At a first level of analysis, the proposed classification scheme follows the Internet protocol stack, starting from the physical layer, going up toward the application layer, without neglecting cross-layer approaches.
At a finer description level, for each layer of the stack the papers have been further classified and described, based on the approach proposed.
%, and highlighting the main differences between the classification criteria followed in this survey and the other surveys on the same topics found in literature.

While describing the challenges that threaten the \ac{IoD} diffusion, this work also presents the current open issues, in order to draw future research directions.
%Such an ambitious goal has been achieved through a detailed and comprehensive analysis of the most recent literature on the \ac{IoD} theme, that can be of help to researchers in developing future activities on this topic.
%All the surveyed papers have been classified and described, according to the classification scheme described above.
%From the analysis of the surveyed literature, some important conclusions can be drawn.
%The analysis demonstrated that consistent efforts have been spent in the improvement of communication range. This goal is usually reached through relaying techniques or algorithms for optimal positioning of drones, mainly for coverage extension purposes.
%This is a very challenging task in critical scenarios where communication is severely limited by factors that increases delay and lowers the data rate.
The analysis demonstrates that consistent design efforts have been devoted to specific aspects at the different layers. For instance, several mathematical models/frameworks have been proposed to improve communication range, or reliably model the wireless medium, at physical layer, or even efficiently routing for synchronization. Some works propose cooperation and network formation as an efficient data exchange strategy.
Promising results have been discussed in those works that focus on one of the aforedescribed problems and try to deal with it as an optimization problem. This is even more evident when the \ac{IoD}-related problems are solved together in combined optimization approaches, such as energy consumption and trajectory design, or path planning and routing optimization.

At the same time, novel proposals and design methodologies were surveyed to discuss the effects of the introduction of 5G and 6G-compliant technologies, such as mmWave and \ac{VLC}. Even though the discussion is still ongoing, further efforts could be useful to establish new requirements even in critical operating conditions.

%% file: elsarticle-template-num.bbl
\begin{thebibliography}{100}
\expandafter\ifx\csname url\endcsname\relax
  \def\url#1{\texttt{#1}}\fi
\expandafter\ifx\csname urlprefix\endcsname\relax\def\urlprefix{URL }\fi
\expandafter\ifx\csname href\endcsname\relax
  \def\href#1#2{#2} \def\path#1{#1}\fi

\bibitem{GBW16}
M.~Gharibi, R.~Boutaba, S.~L. Waslander, Internet of drones, IEEE Access 4
  (2016).

\bibitem{KM21}
A.~Kumar, P.~L. Mehta, Internet of drones: An engaging platform for
  iiot-oriented airborne sensors, in: Smart Sensors for Industrial Internet of
  Things, Springer, 2021, pp. 249--270.

\bibitem{SK17}
V.~Sharma, R.~Kumar, Cooperative frameworks and network models for flying ad
  hoc networks: a survey, Concurrency and Computation: Practice And Experience
  29~(4) (2017) 1--36.

\bibitem{NNB+20}
A.~Nayyar, B.-L. Nguyen, N.~G. Nguyen, The internet of drone things (iodt):
  Future envision of smart drones, in: First International Conference on
  Sustainable Technologies for Computational Intelligence, Springer, 2020, pp.
  563--580.

\bibitem{KCZ+18}
A.~A. Khuwaja, Y.~Chen, N.~Zhao, M.~S. Alouini, P.~Dobbins, A survey of channel
  modeling for uav communications, IEEE Communications Surveys Tutorials (2018)
  1--1\href {https://doi.org/10.1109/COMST.2018.2856587}
  {\path{doi:10.1109/COMST.2018.2856587}}.

\bibitem{FQD+18}
A.~Fotouhi, H.~Qiang, M.~Ding, M.~Hassan, L.~G. Giordano, A.~Garcia-Rodriguez,
  J.~Yuan, {Survey on UAV Cellular Communications: Practical Aspects,
  Standardization Advancements, Regulation, and Security Challenges}, IEEE
  Communications Surveys {\&} Tutorials PP~(c) (2018) 1.

\bibitem{IBG20}
G.~Iacovelli, P.~Boccadoro, L.~Grieco, An iterative stochastic approach to
  constrained drones' communications, in: Proc. of IEEE/ACM 24th International
  Symposium on Distributed Simulation and Real Time Applications (DS-RT)
  (DS-RT'20), Prague, Czech Republic, 2020.

\bibitem{GJV16}
L.~Gupta, R.~Jain, G.~Vaszkun, {Survey of Important Issues in UAV Communication
  Networks}, IEEE Communications Surveys and Tutorials 18~(2) (2016)
  1123--1152.
\newblock \href {https://doi.org/10.1109/COMST.2015.2495297}
  {\path{doi:10.1109/COMST.2015.2495297}}.

\bibitem{LWW21}
Y.~Liu, F.~Wu, J.~Wu, Cellular uav-to-device communications: Joint trajectory,
  speed, and power optimisation, IET Communications (2021).

\bibitem{CSG+18}
G.~Choudhary, V.~Sharma, T.~Gupta, J.~Kim, I.~You, Internet of drones (iod):
  Threats, vulnerability, and security perspectives, arXiv preprint
  arXiv:1808.00203 (2018).

\bibitem{MTA16}
N.~H. Motlagh, T.~Taleb, O.~Arouk, Low-altitude unmanned aerial vehicles-based
  internet of things services: Comprehensive survey and future perspectives,
  IEEE Internet of Things Journal 3~(6) (2016) 899--922.

\bibitem{CZX14}
Y.~{Chen}, H.~{Zhang}, M.~{Xu}, The coverage problem in uav network: A survey,
  in: Fifth International Conference on Computing, Communications and
  Networking Technologies (ICCCNT), 2014, pp. 1--5.
\newblock \href {https://doi.org/10.1109/ICCCNT.2014.6963085}
  {\path{doi:10.1109/ICCCNT.2014.6963085}}.

\bibitem{FMT+17}
G.~Ferri, A.~Munafò, A.~Tesei, P.~Braca, F.~Meyer, K.~Pelekanakis,
  R.~Petroccia, J.~Alves, C.~Strode, K.~LePage, Cooperative robotic networks
  for underwater surveillance: an overview, IET Radar, Sonar \& Navigation
  11~(12) (2017) 1740--1761.

\bibitem{CGT+18}
C.~T. Cicek, H.~Gultekin, B.~Tavli, H.~Yanikomeroglu,
  \href{http://arxiv.org/abs/1812.11826}{{UAV Base Station Location
  Optimization for Next Generation Wireless Networks: Overview and Future
  Research Directions}}, CoRR abs/1812.11826 (2018).
\newblock \href {http://arxiv.org/abs/1812.11826} {\path{arXiv:1812.11826}}.
\newline\urlprefix\url{http://arxiv.org/abs/1812.11826}

\bibitem{ZZW+19}
C.~{Zhang}, W.~{Zhang}, W.~{Wang}, L.~{Yang}, W.~{Zhang}, Research challenges
  and opportunities of uav millimeter-wave communications, IEEE Wireless
  Communications 26~(1) (2019) 58--62.
\newblock \href {https://doi.org/10.1109/MWC.2018.1800214}
  {\path{doi:10.1109/MWC.2018.1800214}}.

\bibitem{AA17}
A.~Alnoman, A.~Anpalagan, On d2d communications for public safety applications,
  in: 2017 IEEE Canada International Humanitarian Technology Conference (IHTC),
  2017, pp. 124--127.

\bibitem{ASY+18}
M.~{Alzenad}, M.~Z. {Shakir}, H.~{Yanikomeroglu}, M.~{Alouini}, Fso-based
  vertical backhaul/fronthaul framework for 5g+ wireless networks, IEEE
  Communications Magazine 56~(1) (2018) 218--224.

\bibitem{BA15}
A.~Bader, M.-S. Alouini, An ultra-low-latency geo-routing scheme for team-based
  unmanned vehicular applications, in: 2015 IEEE Global Communications
  Conference (GLOBECOM), 2015, pp. 1--6.

\bibitem{CCB+18}
S.~{Colonnese}, A.~{Carlesimo}, L.~{Brigato}, F.~{Cuomo}, Qoe-aware uav flight
  path design for mobile video streaming in hetnet, in: 2018 IEEE 10th Sensor
  Array and Multichannel Signal Processing Workshop (SAM), 2018, pp. 301--305.
\newblock \href {https://doi.org/10.1109/SAM.2018.8448608}
  {\path{doi:10.1109/SAM.2018.8448608}}.

\bibitem{CSM17}
D.~G. {Cileo}, N.~{Sharma}, M.~{Magarini}, Coverage, capacity and interference
  analysis for an aerial base station in different environments, in: 2017
  International Symposium on Wireless Communication Systems (ISWCS), 2017, pp.
  281--286.

\bibitem{FCC+18}
L.~Ferranti, F.~Cuomo, S.~Colonnese, T.~Melodia,
  \href{http://www.sciencedirect.com/science/article/pii/S157087051830204X}{Drone
  cellular networks: Enhancing the quality of experience of video streaming
  applications}, Ad Hoc Networks 78 (2018) 1 -- 12.
\newblock \href {https://doi.org/https://doi.org/10.1016/j.adhoc.2018.05.003}
  {\path{doi:https://doi.org/10.1016/j.adhoc.2018.05.003}}.
\newline\urlprefix\url{http://www.sciencedirect.com/science/article/pii/S157087051830204X}

\bibitem{KQF+19}
A.~Koubâa, B.~Qureshi, M.-F. Sriti, A.~Allouch, Y.~Javed, M.~Alajlan,
  O.~Cheikhrouhou, M.~Khalgui, E.~Tovar, Dronemap planner: A service-oriented
  cloud-based management system for the internet-of-drones, Ad Hoc Networks 86
  (2019) 46 -- 62.
\newblock \href {https://doi.org/https://doi.org/10.1016/j.adhoc.2018.09.013}
  {\path{doi:https://doi.org/10.1016/j.adhoc.2018.09.013}}.

\bibitem{MGA+17}
H.~Menouar, I.~Guvenc, K.~Akkaya, A.~S. Uluagac, A.~Kadri, A.~Tuncer,
  Uav-enabled intelligent transportation systems for the smart city:
  Applications and challenges, IEEE Communications Magazine 55~(3) (2017)
  22--28.

\bibitem{saad2019vision}
W.~Saad, M.~Bennis, M.~Chen, A vision of 6g wireless systems: Applications,
  trends, technologies, and open research problems, IEEE network 34~(3) (2019)
  134--142.

\bibitem{VCA+19}
D.~{Vasiliev}, A.~{Chunaev}, A.~{Abilov}, I.~{Kaysina}, D.~{Meitis},
  Application layer arq and network coding for qos improving in uav-assisted
  networks, in: 2019 25th Conference of Open Innovations Association (FRUCT),
  2019, pp. 353--360.

\bibitem{VCP17}
B.~Van~den Bergh, A.~Chiumento, S.~Pollin, Ultra-reliable ieee 802.11 for uav
  video streaming: From network to application, in: R.~El-Azouzi, D.~S.
  Menasche, E.~Sabir, F.~De~Pellegrini, M.~Benjillali (Eds.), Advances in
  Ubiquitous Networking 2, Springer Singapore, Singapore, 2017, pp. 637--647.

\bibitem{ZVK+17}
T.~{Zahariadis}, A.~{Voulkidis}, P.~{Karkazis}, P.~{Trakadas}, Preventive
  maintenance of critical infrastructures using 5g networks drones, in: 2017
  14th IEEE International Conference on Advanced Video and Signal Based
  Surveillance (AVSS), 2017, pp. 1--4.
\newblock \href {https://doi.org/10.1109/AVSS.2017.8078465}
  {\path{doi:10.1109/AVSS.2017.8078465}}.

\bibitem{WCC17}
X.~{Wang}, A.~{Chowdhery}, M.~{Chiang}, Networked drone cameras for sports
  streaming, in: 2017 IEEE 37th International Conference on Distributed
  Computing Systems (ICDCS), 2017, pp. 308--318.

\bibitem{EBEID201811}
A survey of open-source uav flight controllers and flight simulators,
  Microprocessors and Microsystems 61 (2018) 11 -- 20.

\bibitem{Deebak2020}
B.~D. Deebak, F.~Al-Turjman, Drone of IoT in 6G Wireless Communications:
  Technology, Challenges, and Future Aspects, Springer International
  Publishing, Cham, 2020, pp. 153--165.
\newblock \href {https://doi.org/{$10.1007/978-3-030-38712-9_9$}}
  {\path{doi:{$10.1007/978-3-030-38712-9_9$}}}.

\bibitem{BMM+20}
L.~{Bariah}, L.~{Mohjazi}, S.~{Muhaidat}, P.~C. {Sofotasios}, G.~K. {Kurt},
  H.~{Yanikomeroglu}, O.~A. {Dobre}, A prospective look: Key enabling
  technologies, applications and open research topics in 6g networks, IEEE
  Access 8 (2020) 174792--174820.
\newblock \href {https://doi.org/10.1109/ACCESS.2020.3019590}
  {\path{doi:10.1109/ACCESS.2020.3019590}}.

\bibitem{deebak2020drone}
B.~Deebak, F.~Al-Turjman, Drone of iot in 6g wireless communications:
  Technology, challenges, and future aspects, in: Unmanned Aerial Vehicles in
  Smart Cities, Springer, 2020, pp. 153--165.

\bibitem{SBC20}
W.~{Saad}, M.~{Bennis}, M.~{Chen}, A vision of 6g wireless systems:
  Applications, trends, technologies, and open research problems, IEEE Network
  34~(3) (2020) 134--142.
\newblock \href {https://doi.org/10.1109/MNET.001.1900287}
  {\path{doi:10.1109/MNET.001.1900287}}.

\bibitem{YXX+19}
P.~{Yang}, Y.~{Xiao}, M.~{Xiao}, S.~{Li}, 6g wireless communications: Vision
  and potential techniques, IEEE Network 33~(4) (2019) 70--75.
\newblock \href {https://doi.org/10.1109/MNET.2019.1800418}
  {\path{doi:10.1109/MNET.2019.1800418}}.

\bibitem{BII18}
{Business Insider Intelligence}, The future of drones for consumers,
  businesses, and the military, DRONES 101 (4 2018).

\bibitem{PV19}
S.~Pollin, E.~Vinogradov, Tut-23: Ieee icc tutorial on wireless communications
  with unmanned aerial vehicles, in: IEEE International Conference on
  Communications, 2019.

\bibitem{castillo2015projecting}
A.~Castillo~O'Sullivan, A.~D. Thierer, Projecting the growth and economic
  impact of the internet of things, Available at SSRN 2618794 (2015).

\bibitem{koiwanit2018analysis}
J.~Koiwanit, Analysis of environmental impacts of drone delivery on an online
  shopping system, Advances in Climate Change Research 9~(3) (2018) 201--207.

\bibitem{PDE+18}
L.~M. {PytlikZillig}, B.~{Duncan}, S.~{Elbaum}, C.~{Detweiler}, A drone by any
  other name: Purposes, end-user trustworthiness, and framing, but not
  terminology, affect public support for drones, IEEE Technology and Society
  Magazine 37~(1) (2018) 80--91.
\newblock \href {https://doi.org/10.1109/MTS.2018.2795121}
  {\path{doi:10.1109/MTS.2018.2795121}}.

\bibitem{jenkins2013economic}
D.~Jenkins, B.~Vasigh, The economic impact of unmanned aircraft systems
  integration in the united states. arlington, va: Association of unmanned
  vehicle systems international, International, 2013.

\bibitem{keaveney2019single}
T.~Keaveney, C.~Stewart, Single european sky atm research joint undertaking,
  SESAR (2019).

\bibitem{clarke2014regulation}
R.~Clarke, L.~B. Moses, The regulation of civilian drones' impacts on public
  safety, Computer law \& security review 30~(3) (2014) 263--285.

\bibitem{andersen2020strategic}
K.~V. Andersen, M.~H. Frederiksen, M.~P. Knudsen, A.~D. Krabbe, The strategic
  responses of start-ups to regulatory constraints in the nascent drone market,
  Research Policy 49~(10) (2020) 104055.

\bibitem{Balafoutis2017}
A.~T. Balafoutis, B.~Beck, S.~Fountas, Z.~Tsiropoulos, J.~Vangeyte, T.~van~der
  Wal, I.~Soto-Embodas, M.~G{\'o}mez-Barbero, S.~M. Pedersen,
  \href{https://doi.org/10.1007/978-3-319-68715-5_2}{Smart Farming Technologies
  -- Description, Taxonomy and Economic Impact}, Springer International
  Publishing, Cham, 2017, pp. 21--77.
\newblock \href {https://doi.org/{$10.1007/978-3-319-68715-5_2$}}
  {\path{doi:{$10.1007/978-3-319-68715-5_2$}}}.
\newline\urlprefix\url{https://doi.org/10.1007/978-3-319-68715-5_2}

\bibitem{pricewaterhousecoopers2018skies}
L.~Pricewaterhousecoopers, Skies without limits--drones-taking the uk’s
  economy to new heights (2018).

\bibitem{EEA+19}
M.~{Ezuma}, F.~{Erden}, C.~{Kumar Anjinappa}, O.~{Ozdemir}, I.~{Guvenc},
  {Detection and Classification of UAVs Using RF Fingerprints in the Presence
  of Wi-Fi and Bluetooth Interference}, IEEE Open Journal of the Communications
  Society 1 (2019) 60--76.
\newblock \href {https://doi.org/10.1109/OJCOMS.2019.2955889}
  {\path{doi:10.1109/OJCOMS.2019.2955889}}.

\bibitem{SAE18}
J.~{Sadovskis}, A.~{Aboltins}, J.~{Eidaks}, Modulation recognition of unmanned
  aerial vehicle control signals using software-defined radio, in: 2018 IEEE
  6th Workshop on Advances in Information, Electronic and Electrical
  Engineering (AIEEE), 2018, pp. 1--5.

\bibitem{KPR19}
P.~{Kozak}, V.~{Platenka}, M.~{Richterova}, {Radio Communication Channel
  Analysis of UAV}, in: 2019 International Conference on Military Technologies
  (ICMT), 2019, pp. 1--4.

\bibitem{AKL+18}
S.~{Atoev}, O.-H. {Kwon}, S.~H. {Lee}, K.~R. {Kwon}, {An efficient SC-FDM
  modulation technique for a UAV communication link}, Electronics 7~(12) (2018)
  1--18.
\newblock \href {https://doi.org/10.3390/electronics7120352}
  {\path{doi:10.3390/electronics7120352}}.

\bibitem{JTY20}
Y.~{Jia}, X.~{Tu}, W.~{Yan}, {An UAV Wireless Communication Noise Suppression
  Method Based on OFDM Modulation and Demodulation}, Radio Science 55~(2)
  (2020) 1--14.
\newblock \href {https://doi.org/10.1029/2019RS006959}
  {\path{doi:10.1029/2019RS006959}}.

\bibitem{GKD17}
B.~Galkin, J.~Kibiłda, L.~A. DaSilva, Coverage analysis for low-altitude uav
  networks in urban environments, in: 2017 IEEE Global Communications
  Conference (GLOBECOM), 2017, pp. 1--6.

\bibitem{RD16}
V.~V.~C. Ravi, H.~S. Dhillon, Downlink coverage probability in a finite network
  of unmanned aerial vehicle (uav) base stations, in: Proc. 2016 IEEE 17th
  International Workshop on Signal Processing Advances in Wireless
  Communications (SPAWC), Edinburgh, UK, 2016, pp. 1--5.

\bibitem{LC17}
Y.~Li, L.~Cai, Uav-assisted dynamic coverage in a heterogeneous cellular
  system, IEEE Network 31~(4) (2017) 56--61.

\bibitem{MSB+15}
M.~Mozaffari, W.~Saad, M.~Bennis, M.~Debbah, Drone small cells in the clouds:
  Design, deployment and performance analysis, in: 2015 IEEE Global
  Communications Conference (GLOBECOM), 2015, pp. 1--6.

\bibitem{CSY17}
M.~Chen, W.~Saad, C.~Yin, Liquid state machine learning for resource allocation
  in a network of cache-enabled lte-u uavs, in: 2017 IEEE Global Communications
  Conference (GLOBECOM), 2017, pp. 1--6.

\bibitem{MSB+17}
M.~Mozaffari, W.~Saad, M.~Bennis, M.~Debbah, Performance optimization for
  uav-enabled wireless communications under flight time constraints, in: 2017
  IEEE Global Communications Conference (GLOBECOM), 2017, pp. 1--6.

\bibitem{AGS+16}
D.~Athukoralage, I.~Guvenc, W.~Saad, M.~Bennis, Regret based learning for uav
  assisted lte-u/wifi public safety networks, in: 2016 IEEE Global
  Communications Conference (GLOBECOM), 2016, pp. 1--7.

\bibitem{NXL+17}
M.~Narang, S.~Xiang, W.~Liu, J.~Gutierrez, L.~Chiaraviglio, A.~Sathiaseelan,
  A.~Merwaday, Uav-assisted edge infrastructure for challenged networks, in:
  2017 IEEE Conference on Computer Communications Workshops (INFOCOM WKSHPS),
  2017, pp. 60--65.

\bibitem{KYW+17}
L.~Kong, L.~Ye, F.~Wu, M.~Tao, G.~Chen, A.~V. Vasilakos, Autonomous relay for
  millimeter-wave wireless communications, IEEE Journal on Selected Areas in
  Communications 35~(9) (2017) 2127--2136.

\bibitem{QHG+16}
W.~Qi, W.~Hou, L.~Guo, Q.~Song, A.~Jamalipour, A unified routing framework for
  integrated space/air information networks, IEEE Access 4 (2016) 7084--7103.

\bibitem{GSY17}
G.~Gankhuyag, A.~P. Shrestha, S.-J. Yoo, Robust and reliable predictive routing
  strategy for flying ad-hoc networks, IEEE Access 5 (2017) 643--654.

\bibitem{PJS+17}
S.-Y. Park, D.~Jeong, C.~S. Shin, H.~Lee, Dronenet+: Adaptive route recovery
  using path stitching of uavs in ad-hoc networks, in: 2017 IEEE Global
  Communications Conference (GLOBECOM), 2017, pp. 1--7.

\bibitem{OLL+16}
O.~S. Oubbati, A.~Lakas, N.~Lagraa, M.~B. Yagoubi, Uvar: An intersection
  uav-assisted vanet routing protocol, in: 2016 IEEE Wireless Communications
  and Networking Conference (WCNC 2016), Doha, Qatar, 2016, pp. 1--6.

\bibitem{OLZ+17}
O.~S. Oubbati, A.~{Lakas}, F.~{Zhou}, M.~{Gunes}, N.~{Lagraa}, M.~B. {Yagoubi},
  Intelligent uav-assisted routing protocol for urban vanets, Computer
  Communications 107~(C) (2017) 93--111.

\bibitem{ZJF+18}
O.~S. Oubbati, A.~Lakas, F.~Zhou, M.~Gunes, N.~Lagraa, M.~B. Yagoubi,
  Intelligent uav-assisted routing protocol for urban vanets, Computer
  Communications 107~(C) (2017) 93--111.

\bibitem{GPM+18}
M.~Gapeyenko, V.~Petrov, D.~Moltchanov, S.~Andreev, N.~Himayat, Y.~Koucheryavy,
  Flexible and reliable uav-assisted backhaul operation in 5g mmwave cellular
  networks, IEEE Journal on Selected Areas in Communications 36~(11) (2018)
  2486--2496.

\bibitem{MKM+15}
R.~Motooka, T.~Katagiri, S.~Murayama, J.~Takahashi, Y.~Tobe, R.~Nishikawa,
  Distance control between multiple drones for stable communication, in: 2015
  IEEE SENSORS, 2015, pp. 1--3.

\bibitem{RDC17}
D.~Rautu, R.~Dhaou, E.~Chaput, Maintaining a permanent connectivity between
  nodes of an air-to-ground communication network, in: 2017 13th International
  Wireless Communications and Mobile Computing Conference (IWCMC), 2017, pp.
  681--686.

\bibitem{SKS+17}
S.~A.~W. Shah, T.~Khattab, M.~Z. Shakir, M.~O. Hasna, A distributed approach
  for networked flying platform association with small cells in 5g+ networks,
  in: 2017 IEEE Global Communications Conference (GLOBECOM), 2017, pp. 1--7.

\bibitem{RIG16}
N.~Rupasinghe, A.~S. Ibrahim, I.~Guvenc, Optimum hovering locations with
  angular domain user separation for cooperative uav networks, in: 2016 IEEE
  Global Communications Conference (GLOBECOM), 2016, pp. 1--6.

\bibitem{AS15}
F.~Ahdi, S.~Subramaniam, Using unmanned aerial vehicles as relays in wireless
  balloon networks, in: 2015 IEEE International Conference on Communications
  (ICC), 2015, pp. 3795--3800.

\bibitem{CSB19}
U.~Challita, W.~Saad, C.~Bettstetter, Interference management for
  cellular-connected uavs: A deep reinforcement learning approach, IEEE
  Transactions on Wireless Communications 18~(4) (2019) 2125--2140.

\bibitem{LXN+17}
X.~Liu, T.~Xi, E.~Ngai, W.~Wang, Path planning for aerial sensor networks with
  connectivity constraints, in: 2017 IEEE International Conference on
  Communications (ICC), 2017, pp. 1--6.

\bibitem{MSS16}
M.-A. Messous, S.-M. Senouci, H.~Sedjelmaci, Network connectivity and area
  coverage for uav fleet mobility model with energy constraint, in: Proc. 2016
  IEEE Wireless Communications and Networking Conference, Doha, Qatar, 2016,
  pp. 1--6.

\bibitem{MSB+16}
M.~Mozaffari, W.~Saad, M.~Bennis, M.~Debbah, Unmanned aerial vehicle with
  underlaid device-to-device communications: Performance and tradeoffs, IEEE
  Transactions on Wireless Communications 15~(6) (2016) 3949--3963.

\bibitem{ARC+16B}
M.~M. Azari, F.~Rosas, K.~Chen, S.~Pollin, Optimal uav positioning for
  terrestrial-aerial communication in presence of fading, in: 2016 IEEE Global
  Communications Conference (GLOBECOM), 2016, pp. 1--7.

\bibitem{CS17_02}
U.~Challita, W.~Saad, Network formation in the sky: Unmanned aerial vehicles
  for multi-hop wireless backhauling, in: 2017 IEEE Global Communications
  Conference (GLOBECOM), 2017, pp. 1--6.

\bibitem{CMS+17}
M.~Chen, M.~Mozaffari, W.~Saad, C.~Yin, M.~Debbah, C.~S. Hong, Caching in the
  sky: Proactive deployment of cache-enabled unmanned aerial vehicles for
  optimized quality-of-experience, IEEE Journal on Selected Areas in
  Communications 35~(5) (2017) 1046--1061.

\bibitem{WZZ17}
Q.~Wu, Y.~Zeng, R.~Zhang, Joint trajectory and communication design for
  uav-enabled multiple access, in: 2017 IEEE Global Communications Conference
  (GLOBECOM), 2017, pp. 1--6.

\bibitem{LZZ17}
J.~Lyu, Y.~Zeng, R.~Zhang, Spectrum sharing and cyclical multiple access in
  uav-aided cellular offloading, in: 2017 IEEE Global Communications Conference
  (GLOBECOM), 2017, pp. 1--6.

\bibitem{CG17}
J.~Chen, D.~Gesbert, Optimal positioning of flying relays for wireless
  networks: A los map approach, in: 2017 IEEE International Conference on
  Communications (ICC), 2017, pp. 1--6.

\bibitem{WDZ19}
Z.~Wang, L.~Duan, R.~Zhang, Traffic-aware adaptive deployment for uav-aided
  communication networks, in: 2018 IEEE Global Communications Conference
  (GLOBECOM), Abu Dhabi, United Arab Emirates, United Arab Emirates, 2018, pp.
  1--6.

\bibitem{HNK+16}
M.~Horiuchi, H.~Nishiyama, N.~Kato, F.~Ono, R.~Miura, Throughput maximization
  for long-distance real-time data transmission over multiple uavs, in: 2016
  IEEE International Conference on Communications (ICC), 2016, pp. 1--6.

\bibitem{ZZZ17}
J.~Zhang, Y.~Zeng, R.~Zhang, Spectrum and energy efficiency maximization in
  uav-enabled mobile relaying, in: 2017 IEEE International Conference on
  Communications (ICC), 2017, pp. 1--6.

\bibitem{AFN+14}
A.~Abdulla, Z.~M. Fadlullah, H.~Nishiyama, N.~Kato, F.~Ono, R.~Miura, An
  optimal data collection technique for improved utility in uas-aided networks,
  in: 2014 IEEE International Conference on Computer Communications (INFOCOM),
  2014, pp. 736--744.

\bibitem{YKB11}
E.~Yanmaz, R.~Kuschnig, C.~Bettstetter, Channel measurements over 802.11a-based
  uav-to-ground links, in: 2011 IEEE GLOBECOM Workshops (GC Wkshps), Houston,
  TX, USA, 2011, pp. 1280--1284.

\bibitem{YCL14}
H.-F. Yu, X.-F. Chi, J.~Liu, An integrated phy-mac analytical model for ieee
  802.15.7 vlc network with mpr capability, Optoelectronics Letters 10~(5)
  (2014) 365--368.

\bibitem{MBM14}
A.~Musa, M.~D. Baba, H.~M. A.~H. Mansor, The design and implementation of ieee
  802.15.7 module with ns-2 simulator, in: 2014 International Conference on
  Computer, Communications, and Control Technology (I4CT), 2014, pp. 111--115.

\bibitem{SKU16}
P.~Shams, M.~Erol-Kantarci, M.~Uysal, Mac layer performance of the ieee
  802.15.7 visible light communication standard, Transactions on Emerging
  Telecommunications Technologies 27~(5) (2016) 662--674.

\bibitem{WGP14}
Q.~Wang, D.~Giustiniano, D.~Puccinelli, Openvlc: software-defined visible light
  embedded networks, in: 1st ACM MobiCom Workshop on Visible Light
  Communication Systems(VLCS'14), 2014, pp. 15--20.

\bibitem{TKN+17}
Y.~Takahashiy, Y.~Kawamotoy, H.~Nishiyamay, N.~Katoy, F.~F. Onoz, R.~Miura, A
  td-lte-a based efficient radio access scheme for real-time data transmission
  over relay unmanned aerial vehicle networks, in: 2017 IEEE 86th Vehicular
  Technology Conference (VTC-Fall), 2017, pp. 1--5.

\bibitem{WCL+18}
L.~Wang, Y.~L. Che, J.~Long, L.~Duan, K.~Wu, Multiple access mmwave design for
  uav-aided 5g communications, IEEE Wireless Communications 26~(1) (2019)
  64--71.

\bibitem{SIL+16}
S.~Say, H.~Inata, J.~Liu, S.~Shimamoto, Priority-based data gathering framework
  in uav-assisted wireless sensor networks, IEEE Sensors Journal 16~(14) (2016)
  5785--5794.

\bibitem{KNK+17}
Y.~Kawamoto, H.~Nishiyama, N.~Kato, F.~Ono, R.~Miura, An efficient
  throughput-aware resource allocation technique for data transmission in
  unmanned aircraft systems, in: 2017 IEEE International Conference on
  Communications (ICC), 2017, pp. 1--6.

\bibitem{FC12}
H.~Feng, L.~J. Cimini, On optimum relay deployment in a multi-hop linear
  network with cooperation, in: MILCOM 2012 - 2012 IEEE Military Communications
  Conference, 2012, pp. 1--6.

\bibitem{LNW+16}
K.~Li, W.~Ni, X.~Wang, R.~P. Liu, S.~S. Kanhere, S.~Jha, Energy-efficient
  cooperative relaying for unmanned aerial vehicles, IEEE Transactions on
  Mobile Computing 15~(6) (2016) 1377--1386.

\bibitem{LKY+16}
J.~Lee, K.~Kim, S.~Yoo, A.~Y. Chung, J.~Y. Lee, S.~J. Park, H.~Kim,
  Constructing a reliable and fast recoverable network for drones, in: 2016
  IEEE International Conference on Communications (ICC), 2017, pp. 1--6.

\bibitem{QHS+16}
W.~Qi, W.~Hou, Q.~Song, L.~Guo, A.~Jamalipour, Topology control and routing
  based on adaptive rf/fso switching in space-air integrated networks, in: 2016
  IEEE Global Communications Conference (GLOBECOM), 2016, pp. 1--6.

\bibitem{LWC+17}
J.~Lu, S.~Wan, X.~Chen, P.~Fan, Energy-efficient 3d uav-bs placement versus
  mobile users’ density and circuit power, in: 2017 IEEE Global
  Communications Conference (GLOBECOM), 2017, pp. 1--6.

\bibitem{GGK+17}
H.~Ghazzai, M.~B. Ghorbel, A.~Kadri, M.~J. Hossain, H.~Menouar,
  Energy-efficient management of unmanned aerial vehicles for underlay
  cognitive radio systems, IEEE Transactions on Green Communications and
  Networking 1~(4) (2017) 434--443.

\bibitem{MSB+16A}
M.~{Mozaffari}, W.~{Saad}, M.~{Bennis}, M.~{Debbah}, Efficient deployment of
  multiple unmanned aerial vehicles for optimal wireless coverage, IEEE
  Communications Letters 20~(8) (2016) 1647--1650.

\bibitem{BEY18}
I.~{Bor-Yaliniz}, A.~{El-Keyi}, H.~{Yanikomeroglu}, {Spatial Configuration of
  Agile Wireless Networks With Drone-BSs and User-in-the-loop}, IEEE
  Transactions on Wireless Communications 18~(2) (2019) 753--768.

\bibitem{LH17}
J.~Li, Y.~Han, Optimal resource allocation for packet delay minimization in
  multi-layer uav networks, IEEE Communications Letters 21~(3) (2017) 580--583.

\bibitem{ARC+16A}
M.~M. Azari, F.~Rosas, K.-C. Chen, S.~Pollin, Joint sum-rate and power gain
  analysis of an aerial base station, in: 2016 IEEE Global Communications
  Conference (GLOBECOM), 2016, pp. 1--6.

\bibitem{GW15}
N.~Goddemeier, C.~Wietfeld, Investigation of air-to-air channel characteristics
  and a uav specific extension to the rice model, in: 2015 IEEE Global
  Communications Conference (GLOBECOM), 2015, pp. 1--5.

\bibitem{CPA+17}
X.~{Cai}, A.~{Gonzalez-Plaza}, D.~{Alonso}, L.~{Zhang}, C.~B. {Rodríguez},
  A.~P. {Yuste}, X.~{Yin}, Low altitude uav propagation channel modelling, in:
  2017 11th European Conference on Antennas and Propagation (EUCAP), 2017, pp.
  1443--1447.

\bibitem{ANM+17}
R.~{Amorim}, H.~{Nguyen}, P.~{Mogensen}, I.~Z. {Kovacs}, J.~{Wigard}, T.~B.
  {Sorensen}, Radio channel modeling for uav communication over cellular
  networks, IEEE Wireless Communications Letters 6~(4) (2017) 514--517.

\bibitem{KOG17}
W.~{Khawaja}, O.~{Ozdemir}, I.~{Guvenc}, Uav air-to-ground channel
  characterization for mmwave systems, in: 2017 IEEE 86th Vehicular Technology
  Conference (VTC-Fall), 2017, pp. 1--5.

\bibitem{XXX16}
Z.~Xiao, P.~Xia, X.-G. Xia, Enabling uav cellular with millimeter-wave
  communication: Potentials and approaches, IEEE Communications Magazine 54~(5)
  (2016) 66--73.

\bibitem{RYR+19}
A.~{Rahmati}, Y.~{Yap{\i}c{\i}}, N.~{Rupasinghe}, I.~{Guvenc}, H.~{Dai},
  A.~{Bhuyany}, {Energy Efficiency of RSMA and NOMA in Cellular-Connected
  mmWave UAV Networks}, arXiv e-prints (2019) arXiv:1902.04721.

\bibitem{WTZ+18}
H.~{Wu}, X.~{Tao}, N.~{Zhang}, X.~{Shen}, Cooperative uav cluster-assisted
  terrestrial cellular networks for ubiquitous coverage, IEEE Journal on
  Selected Areas in Communications 36~(9) (2018) 2045--2058.

\bibitem{ZJ18}
J.~{Zhao}, W.~{Jia}, Efficient channel tracking strategy for mmwave uav
  communications, Electronics Letters 54~(21) (2018) 1218--1220.

\bibitem{CEG+17}
J.~Chen, O.~Esrafilian, D.~Gesbert, U.~Mitra, Efficient algorithms for
  air-to-ground channel reconstruction in uav-aided communications, in: 2017
  IEEE Global Communications Conference (GLOBECOM), 2017, pp. 1--6.

\bibitem{BSY18}
I.~{Bor-Yaliniz}, S.~S. {Szyszkowicz}, H.~{Yanikomeroglu}, Environment-aware
  drone-base-station placements in modern metropolitans, IEEE Wireless
  Communications Letters 7~(3) (2018) 372--375.

\bibitem{VYS+16}
V.~Vahidi, A.~P. Yazdanpanah, E.~Saberinia, E.~E. Regentova, Channel
  estimation, equalisation, and evaluation for high-mobility airborne
  hyperspectral data transmission, IET Communications 10~(8) (2016) 2656--2662.

\bibitem{DGI+18}
D.~Darsena, G.~Gelli, I.~Iudice, F.~Verde, Equalization techniques of control
  and non-payload communication links for unmanned aerial vehicles, IEEE Access
  6 (2018) 4485--4496.

\bibitem{BDL13}
C.~C. Baseca, J.~R. Díaz, J.~Lloret, Communication ad hoc protocol for
  intelligent video sensing using ar drones, in: 2013 IEEE 9th International
  Conference on Mobile Ad-hoc and Sensor Networks, 2013, pp. 449--453.

\bibitem{OOF+17}
A.~Orsino, A.~Ometov, G.~Fodor, D.~Moltchanov, L.~Militano, S.~Andreev,
  O.~N.~C. Yilmaz, T.~Tirronen, J.~Torsner, G.~Araniti, A.~Iera, M.~Dohler,
  Y.~Koucheryavy, Effects of heterogeneous mobility on d2d- and drone-assisted
  mission-critical mtc in 5g, IEEE Communications Magazine 55~(2) (2018)
  79--87.

\bibitem{ASA17}
A.~Ali, G.~Shah, M.~Aslam, Model for autonomous agents in machine-to-machine
  navigation networks, International Journal of Communication Systems 31~(4)
  (2017) 1--13.

\bibitem{FAA18}
W.~Fawaz, C.~Abou-Rjeily, C.~Assi, Uav-aided cooperation for fso communication
  systems, IEEE Communications Magazine 56~(1) (2018) 70--75.

\bibitem{MNP+17}
S.~Manfredi, E.~Natalizio, C.~Pascariello, N.~R. Zema, A packet loss tolerant
  rendezvous algorithm for wireless networked robot systems, Asian Journal of
  Control 19~(4) (2017) 1413--1423.

\bibitem{M16}
S.~Mori, Cooperative sensing data collecting framework by using unmanned
  aircraft vehicle in wireless sensor network, in: 2016 IEEE International
  Conference on Communications (ICC), 2016, pp. 1--6.

\bibitem{WDZ+17}
W.~Wang, C.~Dong, S.~Zhu, H.~Wang, Dfra: Demodulation-free random access for
  uav ad hoc networks, in: 2017 IEEE International Conference on Communications
  (ICC), 2017, pp. 1--6.

\bibitem{ZSW18}
Z.~Zheng, A.~K. Sangaiah, T.~Wang, Adaptive communication protocols in flying
  ad hoc network, IEEE Communications Magazine 56~(1) (2018) 136--142.

\bibitem{KSD+19B}
P.~{Kumar}, P.~{Singh}, S.~{Darshi}, S.~{Shailendra}, Drone assisted network
  coded co-operation, in: TENCON 2019 - 2019 IEEE Region 10 Conference
  (TENCON), 2019, pp. 1174--1179.

\bibitem{KSD+19}
P.~Kumar, P.~{Singh}, S.~{Darshi}, S.~{Shailendra}, Analysis of drone assisted
  network coded cooperation for next generation wireless network, IEEE
  Transactions on Mobile Computing (2019) 1--11.

\bibitem{ZHM+17}
M.~{Zayene}, O.~{Habachi}, V.~{Meghdadi}, T.~{Ezzeddine}, J.~{Cances}, {Joint
  delay and energy minimization for instantly decodable network coding}, in:
  2017 IEEE International Conference on Communications (ICC), 2017, pp. 1--6.

\bibitem{ZHM+18}
M.~Zayene, O.~{Habachi}, V.~{Meghdadi}, T.~{Ezzeddine}, J.~{Cances}, Delay and
  energy aware instantly decodable network coding for multi-hop cooperative
  data exchange, in: 2018 IEEE Wireless Communications and Networking
  Conference (WCNC), 2018, pp. 1--6.

\bibitem{CF19}
G.~{Cocco}, D.~{Floreano}, Cross-packet coding for delay-constrained streaming
  applications, IEEE Communications Letters 23~(11) (2019) 1962--1966.

\bibitem{CJK+19}
C.~{Chun}, K.~M. {Jeon}, T.~{Kim}, W.~{Choi}, Drone noise reduction using deep
  convolutional autoencoder for uav acoustic sensor networks, in: 2019 IEEE
  16th International Conference on Mobile Ad Hoc and Sensor Systems Workshops
  (MASSW), 2019, pp. 168--169.

\bibitem{LNW+15}
K.~Li, W.~Ni, X.~Wang, R.~P. Liu, S.~S. Kanhere, S.~Jha, Epla: Energy-balancing
  packets scheduling for airborne relaying networks, in: 2015 IEEE
  International Conference on Communications (ICC), 2015, pp. 6246--6251.

\bibitem{PC20}
C.~{Pu}, L.~{Carpenter}, $psched$: A priority-based service scheduling scheme
  for the internet of drones, IEEE Systems Journal (2020) 1--10\href
  {https://doi.org/10.1109/JSYST.2020.2998010}
  {\path{doi:10.1109/JSYST.2020.2998010}}.

\bibitem{MKD16}
R.~K. X.~Ma, R.~Dhaou, Fairness-aware uav-assisted data collection in mobile
  wireless sensor networks, in: 2016 International Wireless Communications and
  Mobile Computing Conference (IWCMC), 2016, pp. 995--1001.

\bibitem{NSA10}
H.~H. Nejad, D.~Sauter, S.~Aberkane, On-line scheduling and fault detection in
  ncs with communication constraints in drone application, in: 2010 Conference
  on Control and Fault-Tolerant Systems (SysTol), 2010, pp. 867--872.

\bibitem{KAM16}
T.~Kitagawa, S.~Ata, M.~Murata, Retrieving information with autonomously-flying
  routers in information-centric network, in: 2016 IEEE International
  Conference on Communications (ICC), 2016, pp. 1--6.

\bibitem{MCK+17}
X.~Ma, S.~Chisiu, R.~Kacimi, R.~Dhaou, Opportunistic communications in wsn
  using uav, in: 2017 14th IEEE Annual Consumer Communications \& Networking
  Conference (CCNC), 2017, pp. 510--515.

\bibitem{SBI+16}
B.~Sliwa, D.~Behnke, C.~Ide, C.~Wietfeld, B.a.t.mobile: Leveraging mobility
  control knowledge for efficient routing in mobile robotic networks, in: 2016
  IEEE Global Communications Conference (GLOBECOM), 2016, pp. 1--6.

\bibitem{ZZJ+17}
P.~{Zhang}, Q.~{Zhang}, M.~{Jiang}, Z.~{Feng}, Cube based space region
  partition routing algorithm in uav networks, in: 2017 23rd Asia-Pacific
  Conference on Communications (APCC), 2017, pp. 1--6.

\bibitem{LGY+15}
X.~Li, D.~Guo, H.~Yin, G.~Wei, Drone-assisted public safety wireless broadband
  network, in: 2015 IEEE Wireless Communications and Networking Conference
  Workshops (WCNCW), 2015, pp. 323--328.

\bibitem{LXX+17}
X.~Lu, D.~Xu, L.~Xiao, L.~Wang, W.~Zhuang, Anti-jamming communication game for
  uav-aided vanets, in: 2017 IEEE Global Communications Conference (GLOBECOM),
  2017, pp. 1--6.

\bibitem{SDZ17}
K.~Scott, R.~Dai, J.~Zhang, Online-relaying-based image communication in
  unmanned aerial vehicle networks, in: 2017 IEEE International Conference on
  Communications (ICC), 2017, pp. 1--6.

\bibitem{KSJ+19}
Z.~Kaleem, J.~Souza, J.~Jailton, T.~Carvalho, J.~Araújo, R.~Francês, A
  proposal for routing protocol for fanet: A fuzzy system approach with qoe/qos
  guarantee, Wireless Communications and Mobile Computing (2019).

\bibitem{RWSK+18}
R.~Valentino, W.-S. Jung, Y.-B. Ko, A design and simulation of the
  opportunistic computation offloading with learning-based prediction for
  unmanned aerial vehicle (uav) clustering networks, Sensors 18~(11) (2018).

\bibitem{VJK18}
R.~{Valentino}, W.~{Jung}, Y.~{Ko}, Opportunistic computational offloading
  system for clusters of drones, in: 2018 20th International Conference on
  Advanced Communication Technology (ICACT), 2018, pp. 303--306.

\bibitem{MAA+17}
M.~{Messous}, A.~{Arfaoui}, A.~{Alioua}, S.~{Senouci}, A sequential game
  approach for computation-offloading in an uav network, in: GLOBECOM 2017 -
  2017 IEEE Global Communications Conference, 2017, pp. 1--7.

\bibitem{CWC+20}
X.~{Chen}, C.~{Wu}, T.~{Chen}, Z.~{Liu}, M.~{Bennis}, Y.~{Ji}, Age of
  information-aware resource management in uav-assisted mobile-edge computing
  systems, in: GLOBECOM 2020 - 2020 IEEE Global Communications Conference,
  2020, pp. 1--6.
\newblock \href {https://doi.org/10.1109/GLOBECOM42002.2020.9322632}
  {\path{doi:10.1109/GLOBECOM42002.2020.9322632}}.

\bibitem{JYK+17}
W.~{Jung}, J.~{Yim}, Y.~{Ko}, S.~{Singh}, Acods: adaptive computation
  offloading for drone surveillance system, in: 2017 16th Annual Mediterranean
  Ad Hoc Networking Workshop (Med-Hoc-Net), 2017, pp. 1--6.

\bibitem{MSH+17}
M.~A. Messous, H.~Sedjelmaci, N.~Houari, S.~M. Senouci, Computation offloading
  game for an uav network in mobile edge computing, in: 2017 IEEE International
  Conference on Communications (ICC), 2017, pp. 1--6.

\bibitem{YUA+21}
S.~Yaqoob, A.~Ullah, M.~Awais, I.~Katib, A.~Albeshri, R.~Mehmood, M.~Raza,
  S.~ul~Islam, J.~J. Rodrigues, Novel congestion avoidance scheme for internet
  of drones, Computer Communications (2021).

\bibitem{HRC+19}
X.~{Hou}, Z.~{Ren}, W.~{Cheng}, C.~{Chen}, H.~{Zhang}, Fog based computation
  offloading for swarm of drones, in: ICC 2019 - 2019 IEEE International
  Conference on Communications (ICC), 2019, pp. 1--7.

\bibitem{WLS+17}
F.~Wamser, F.~Loh, M.~Seufert, P.~Tran-Gia, R.~Bruschi, P.~Lago, Dynamic cloud
  service placement for live video streaming with a remote-controlled drone,
  in: 2017 IFIP/IEEE Symposium on Integrated Network and Service Management
  (IM), 2017, pp. 893--894.

\bibitem{LWH+18}
F.~{Loh}, F.~{Wamser}, T.~{Hoßfeld}, P.~{Tran-Gia}, Quality of service
  assessment of live video streaming with a remote-controlled drone, in: 2018
  4th IEEE Conference on Network Softwarization and Workshops (NetSoft), 2018,
  pp. 462--469.

\bibitem{NHP+18}
S.~A.~R. Naqvi, S.~A. Hassan, H.~Pervaiz, Q.~Ni, Drone-aided communication as a
  key enabler for 5g and resilient public safety networks, IEEE Communications
  Magazine 56~(1) (2018) 36--42.

\bibitem{XZZ18}
J.~{Xu}, Y.~{Zeng}, R.~{Zhang}, Uav-enabled wireless power transfer: Trajectory
  design and energy optimization, IEEE Transactions on Wireless Communications
  17~(8) (2018) 5092--5106.

\bibitem{WLZ19}
Q.~{Wu}, L.~{Liu}, R.~{Zhang}, Fundamental trade-offs in communication and
  trajectory design for uav-enabled wireless network, IEEE Wireless
  Communications 26~(1) (2019) 36--44.

\bibitem{ZZ17}
Y.~{Zeng}, R.~{Zhang}, Energy-efficient uav communication with trajectory
  optimization, IEEE Transactions on Wireless Communications 16~(6) (2017)
  3747--3760.

\bibitem{DNN19}
T.~Q. {Duong}, L.~D. {Nguyen}, L.~K. {Nguyen}, Practical optimisation of path
  planning and completion time of data collection for uav-enabled disaster
  communications, in: 2019 15th International Wireless Communications Mobile
  Computing Conference (IWCMC), 2019, pp. 372--377.

\bibitem{HYB+17}
S.~{Hayat}, E.~{Yanmaz}, T.~X. {Brown}, C.~{Bettstetter}, Multi-objective uav
  path planning for search and rescue, in: 2017 IEEE International Conference
  on Robotics and Automation (ICRA), 2017, pp. 5569--5574.

\bibitem{ZXZ19}
Y.~{Zeng}, J.~{Xu}, R.~{Zhang}, Energy minimization for wireless communication
  with rotary-wing uav, IEEE Transactions on Wireless Communications 18~(4)
  (2019) 2329--2345.

\bibitem{ACD+18}
L.~{Amorosi}, L.~{Chiaraviglio}, F.~{D'Andreagiovanni}, N.~{Blefari-Melazzi},
  Energy-efficient mission planning of uavs for 5g coverage in rural zones, in:
  2018 IEEE International Conference on Environmental Engineering (EE), 2018,
  pp. 1--9.

\bibitem{LWL+19}
L.~{Li}, X.~{Wen}, Z.~{Lu}, Q.~{Pan}, W.~{Hu}, Energy-efficient uav-enabled mec
  system: Bits allocation optimization and trajectory design, Sensors 19 (2019)
  1148--1162.

\bibitem{CAM+19}
L.~{Chiaraviglio}, L.~{Amorosi}, F.~{Malandrino}, C.~F. {Chiasserini},
  P.~{Dell'Olmo}, C.~{Casetti}, Optimal throughput management in uav-based
  networks during disasters, in: IEEE INFOCOM 2019 - IEEE Conference on
  Computer Communications Workshops (INFOCOM WKSHPS), 2019, pp. 307--312.

\bibitem{PWW19}
A.~S. {Prasetia}, R.~{Wai}, Y.~{Wen}, Y.~{Wang}, Mission-based energy
  consumption prediction of multirotor uav, IEEE Access 7 (2019) 33055--33063.

\bibitem{JSL19}
H.~Y. Jeong, B.~D. Song, S.~Lee, Truck-drone hybrid delivery routing:
  Payload-energy dependency and no-fly zones, International Journal of
  Production Economics 214 (2019) 220--233.

\bibitem{LDB11}
S.~{Leary}, M.~{Deittert}, J.~{Bookless}, Constrained uav mission planning: A
  comparison of approaches, in: 2011 IEEE International Conference on Computer
  Vision Workshops (ICCV Workshops), 2011, pp. 2002--2009.

\bibitem{Altawy16}
R.~Altawy, A.~M. Youssef, \href{http://doi.acm.org/10.1145/3001836}{Security,
  privacy, and safety aspects of civilian drones: A survey}, ACM Trans.
  Cyber-Phys. Syst. 1~(2) (2016) 7:1--7:25.
\newblock \href {https://doi.org/10.1145/3001836} {\path{doi:10.1145/3001836}}.
\newline\urlprefix\url{http://doi.acm.org/10.1145/3001836}

\bibitem{SA20}
B.~Siddappaji, K.~B. Akhilesh, Role of Cyber Security in Drone Technology,
  Springer Singapore, Singapore, 2020, pp. 169--178.

\bibitem{BP18}
C.~Bunse, S.~Plotz, Security analysis of drone communication protocols, in:
  Engineering Secure Software and Systems, Springer International Publishing,
  2018, pp. 96--107.
\newblock \href {https://doi.org/10.1109/ICCCNT.2014.6963085}
  {\path{doi:10.1109/ICCCNT.2014.6963085}}.

\bibitem{PBC14}
J.-S. Pleban, R.~Band, R.~Creutzburg,
  \href{https://doi.org/10.1117/12.2044868}{Hacking and securing the ar.drone
  2.0 quadcopter: investigations for improving the security of a toy}, in:
  Proceedings of SPIE - The International Society for Optical Engineering, Vol.
  9030, 2014, pp. 9030 -- 9030 -- 12.
\newblock \href {https://doi.org/10.1117/12.2044868}
  {\path{doi:10.1117/12.2044868}}.
\newline\urlprefix\url{https://doi.org/10.1117/12.2044868}

\bibitem{TWL+20}
Y.~{Tan}, J.~{Wang}, J.~{Liu}, Y.~{Zhang}, Unmanned systems security: Models,
  challenges, and future directions, IEEE Network 34~(4) (2020) 291--297.

\bibitem{DCG17}
D.~He, S.~Chan, M.~Guizani, {Drone-Assisted Public Safety Networks: The
  Security Aspect}, IEEE Communications Magazine 55~(8) (2017) 218--223.

\bibitem{GAK+20}
M.~{Gupta}, M.~{Abdelsalam}, S.~{Khorsandroo}, S.~{Mittal}, Security and
  privacy in smart farming: Challenges and opportunities, IEEE Access 8 (2020)
  34564--34584.

\bibitem{B13}
M.~J. BOYLE, \href{https://doi.org/10.1111/1468-2346.12002}{{The costs and
  consequences of drone warfare}}, International Affairs 89~(1) (2013) 1--29.
\newblock \href {https://doi.org/10.1111/1468-2346.12002}
  {\path{doi:10.1111/1468-2346.12002}}.
\newline\urlprefix\url{https://doi.org/10.1111/1468-2346.12002}

\bibitem{PL20}
C.~{Pu}, Y.~{Li}, Lightweight authentication protocol for unmanned aerial
  vehicles using physical unclonable function and chaotic system, in: 2020 IEEE
  International Symposium on Local and Metropolitan Area Networks (LANMAN),
  2020, pp. 1--6.

\bibitem{GKH+19}
J.~{Gordon}, V.~{Kraj}, J.~H. {Hwang}, A.~{Raja}, A security assessment for
  consumer wifi drones, in: 2019 IEEE International Conference on Industrial
  Internet (ICII), 2019, pp. 1--5.

\bibitem{DWZ+18}
G.~Ding, Q.~Wu, L.~Zhang, Y.~Lin, T.~A. Tsiftsis, Y.~D. Yao, An amateur drone
  surveillance system based on the cognitive internet of things, IEEE
  Communications Magazine 56~(1) (2018) 29--35.

\bibitem{WSB15}
J.~Won, S.~H. Seo, E.~Bertino, A secure communication protocol for drones and
  smart objects, in: 2015 10th ACM Symposium on Information, Computer and
  Communications Security (ASIA CCS '15), 2015, pp. 249--260.

\bibitem{GS20}
P.~{Gope}, B.~{Sikdar}, An efficient privacy-preserving authenticated key
  agreement scheme for edge-assisted internet of drones, IEEE Transactions on
  Vehicular Technology 69~(11) (2020) 13621--13630.
\newblock \href {https://doi.org/10.1109/TVT.2020.3018778}
  {\path{doi:10.1109/TVT.2020.3018778}}.

\bibitem{TZA+20}
M.~{Tanveer}, A.~H. {Zahid}, M.~{Ahmad}, A.~{Baz}, H.~{Alhakami}, Lake-iod:
  Lightweight authenticated key exchange protocol for the internet of drone
  environment, IEEE Access 8 (2020) 155645--155659.
\newblock \href {https://doi.org/10.1109/ACCESS.2020.3019367}
  {\path{doi:10.1109/ACCESS.2020.3019367}}.

\bibitem{HZB+20}
A.~{Hermosilla}, A.~M. {Zarca}, J.~B. {Bernabe}, J.~{Ortiz}, A.~{Skarmeta},
  Security orchestration and enforcement in nfv/sdn-aware uav deployments, IEEE
  Access 8 (2020) 131779--131795.

\bibitem{B19}
E.~{Bassi}, European drones regulation: Today’s legal challenges, in: 2019
  International Conference on Unmanned Aircraft Systems (ICUAS), 2019, pp.
  443--450.

\bibitem{LHK+18}
C.~Lin, D.~He, N.~Kumar, K.~K.~R. Choo, A.~Vinel, X.~Huang, Security and
  privacy for the internet of drones: Challenges and solutions, IEEE
  Communications Magazine 56~(1) (2018) 64--69.

\bibitem{LS16}
R.~Luppicini, A.~So,
  \href{http://www.sciencedirect.com/science/article/pii/S0160791X16300033}{A
  technoethical review of commercial drone use in the context of governance,
  ethics, and privacy}, Technology in Society 46 (2016) 109 -- 119.
\newblock \href {https://doi.org/https://doi.org/10.1016/j.techsoc.2016.03.003}
  {\path{doi:https://doi.org/10.1016/j.techsoc.2016.03.003}}.
\newline\urlprefix\url{http://www.sciencedirect.com/science/article/pii/S0160791X16300033}

\bibitem{SAB18}
M.~{Sliti}, W.~{Abdallah}, N.~{Boudriga}, Jamming attack detection in optical
  uav networks, in: 2018 20th International Conference on Transparent Optical
  Networks (ICTON), 2018, pp. 1--5.

\bibitem{LXD+18}
X.~{Lu}, L.~{Xiao}, C.~{Dai}, H.~{Dai}, {UAV-Aided Cellular Communications with
  Deep Reinforcement Learning Against Jamming}, arXiv e-prints (2018)
  arXiv:1805.06628\href {http://arxiv.org/abs/1805.06628}
  {\path{arXiv:1805.06628}}.

\bibitem{WMZ19}
Q.~{Wu}, W.~{Mei}, R.~{Zhang}, {Safeguarding Wireless Network with UAVs: A
  Physical Layer Security Perspective}, arXiv e-prints (2019)
  arXiv:1902.02472\href {http://arxiv.org/abs/1902.02472}
  {\path{arXiv:1902.02472}}.

\bibitem{GBF+16}
J.~{Guth}, U.~{Breitenbücher}, M.~{Falkenthal}, F.~{Leymann}, L.~{Reinfurt},
  Comparison of iot platform architectures: A field study based on a reference
  architecture, in: Cloudification of the Internet of Things (CIoT), 2016, pp.
  1--6.
\newblock \href {https://doi.org/10.1109/CIOT.2016.7872918}
  {\path{doi:10.1109/CIOT.2016.7872918}}.

\bibitem{BDP+16}
A.~Botta, W.~de~Donato, V.~Persico, A.~Pescapé,
  \href{http://www.sciencedirect.com/science/article/pii/S0167739X15003015}{Integration
  of cloud computing and internet of things: A survey}, Future Generation
  Computer Systems 56 (2016) 684 -- 700.
\newline\urlprefix\url{http://www.sciencedirect.com/science/article/pii/S0167739X15003015}

\bibitem{ZPS+18}
Y.~Zhang, L.~Peng, Y.~Sun, H.~Lu,
  \href{https://doi.org/10.1007/s11036-017-0939-1}{Editorial: Intelligent
  industrial iot integration with cognitive computing}, Mobile Networks and
  Applications 23~(2) (2018) 185--187.
\newline\urlprefix\url{https://doi.org/10.1007/s11036-017-0939-1}

\bibitem{MMN+17}
F.~{Marino}, L.~{Maggiani}, L.~{Nao}, P.~{Pagano}, M.~{Petracca}, Towards
  softwarization in the iot: Integration and evaluation of t-res in the onem2m
  architecture, in: IEEE Conference on Network Softwarization (NetSoft), 2017,
  pp. 1--5.

\bibitem{SKP+18}
C.~Stergiou, K.~E. Psannis, B.-G. Kim, B.~Gupta,
  \href{http://www.sciencedirect.com/science/article/pii/S0167739X1630694X}{Secure
  integration of iot and cloud computing}, Future Generation Computer Systems
  78 (2018) 964 -- 975.
\newline\urlprefix\url{http://www.sciencedirect.com/science/article/pii/S0167739X1630694X}

\bibitem{RMC+18}
A.~Reyna, C.~Martín, J.~Chen, E.~Soler, M.~Díaz,
  \href{http://www.sciencedirect.com/science/article/pii/S0167739X17329205}{On
  blockchain and its integration with iot. challenges and opportunities},
  Future Generation Computer Systems 88 (2018) 173 -- 190.
\newline\urlprefix\url{http://www.sciencedirect.com/science/article/pii/S0167739X17329205}

\bibitem{CD17}
A.~M. Cailean, M.~Dimian, Impact of ieee 802.15.7 standard on visible light
  communications usage in automotive applications, IEEE Communications Magazine
  55~(4) (2017) 169--175.

\bibitem{RRL12}
S.~Rajagopal, R.~D. Roberts, S.~K. Lim, Ieee 802.15.7 visible light
  communication: modulation schemes and dimming support, IEEE Communications
  Magazine 50~(3) (2012) 72--82.

\bibitem{LTK+12}
N.~Lourenco, D.~Terra, N.~Kumar, L.~N. Alves, R.~L. Aguiar, Visible light
  communication system for outdoor applications, in: 2012 8th International
  Symposium on Communication Systems, Networks \& Digital Signal Processing
  (CSNDSP), 2012, pp. 1--6.

\bibitem{WWY14}
S.~Wu, H.~Wang, C.~H. Youn, Visible light communications for 5g wireless
  networking systems: from fixed to mobile communications, IEEE Network 28~(6)
  (2014) 41--45.

\bibitem{YTO+14}
T.~Yamazato, I.~Takai, H.~Okada, T.~Fujii, T.~Yendo, S.~Arai, M.~Andoh,
  T.~Harada, K.~Yasutomi, K.~Kagawa, S.~Kawahito, Image-sensor-based visible
  light communication for automotive applications, IEEE Communications Magazine
  52~(7) (2014) 88--97.

\bibitem{BLP+18}
P.~Boccadoro, M.~Losciale, G.~Piro, L.~A. Grieco, A standard-compliant and
  information-centric communication platform for the internet of drones, in:
  Proc. of European Wireless (EW), Catania, Italy, 2018.
\newblock \href
  {http://arxiv.org/abs/https://telematics.poliba.it/publications/2018/BoccadoroEW18.pdf}
  {\path{arXiv:https://telematics.poliba.it/publications/2018/BoccadoroEW18.pdf}}.

\bibitem{STG16}
A.~Shariat, A.~Tizghadam, A.~Leon-Garcia, An icn-based publish-subscribe
  platform to deliver uav service in smart cities, in: 2016 IEEE Conference on
  Computer Communications Workshops (INFOCOM WKSHPS)), 2016, pp. 698--703.

\bibitem{CM16}
S.~{Colucci}, M.~{Mongiello}, Pushing the role of information in icn, in: 2016
  23rd International Conference on Telecommunications (ICT), 2016, pp. 1--5.

\bibitem{Kerrche2020}
C.~A. Kerrche, F.~Ahmad, M.~Elhoseny, A.~Adnane, Z.~Ahmad, B.~Nour, Internet of
  Vehicles Over Named Data Networking: Current Status and Future Challenges,
  Springer International Publishing, Cham, 2020, pp. 83--99.

\bibitem{SA18A}
Z.~Sabir, A.~Amine, Connected vehicles using ndn for intelligent transportation
  systems, in: Proceedings of the international conference on industrial
  engineering and operations management, Vol. 2018, 2018, pp. 2433--2441.

\bibitem{LALL1992165}
S.~Lall,
  \href{http://www.sciencedirect.com/science/article/pii/0305750X9290097F}{Technological
  capabilities and industrialization}, World Development 20~(2) (1992) 165 --
  186.
\newblock \href {https://doi.org/https://doi.org/10.1016/0305-750X(92)90097-F}
  {\path{doi:https://doi.org/10.1016/0305-750X(92)90097-F}}.
\newline\urlprefix\url{http://www.sciencedirect.com/science/article/pii/0305750X9290097F}

\end{thebibliography}
